\documentclass[a4paper, 11pt]{article}
\usepackage{jheppub}

\usepackage{wrapfig}
\usepackage{float}

\hypersetup{
    colorlinks=true,
    linkcolor=blue,
    citecolor=cyan,
    urlcolor=blue,
    }
    
\usepackage{amsthm}
\usepackage{bm}
\usepackage{multirow,booktabs}
\usepackage{longtable} 
\usepackage[table,svgnames,dvipsnames]{xcolor}
\usepackage{enumitem}
\usepackage{mathrsfs}
\usepackage{slashed}
\usepackage{mathtools}
\usepackage{bbm}

\edef\restoreparindent{\parindent=\the\parindent\relax}
\usepackage{parskip}
\restoreparindent

\renewcommand{\arraystretch}{1.6}

\usepackage{tikz}
\usepackage{tikz-3dplot}
\usetikzlibrary{arrows,arrows.meta,intersections, calc,positioning,decorations.pathreplacing,decorations.pathmorphing,shapes}
\usetikzlibrary{patterns}
\usetikzlibrary{decorations.markings}

\usepackage{lmodern} 
\usepackage{physics}  
\usepackage{subcaption} 
\usepackage{caption}    
\usepackage{tabularx} 

\def\i{{\rm i}}
\def\d{{\rm d}}

\def\CD{\mathcal{D}}

\def\CH{\mathcal{H}}
\def\CI{\mathcal{I}}
\def\CK{\mathcal{K}}

\def\CN{\mathcal{N}}
\def\CO{\mathcal{O}}

\def\CS{\mathcal{S}}
\def\CT{\mathcal{T}}

\def\BH{\mathbb{H}}

\def\BR{\mathbb{R}}
\def\BS{\mathbb{S}}
\def\BT{\mathbb{T}}
\def\BZ{\mathbb{Z}}

\def\Dp{\mathcal{D}^{(p)}}
\def\oap{\overleftrightarrow{\partial}}
\def\TR{\mathsf{TR}}
\def\sfs{\mathsf{S}}
\def\hO{\hat{\CO}}
\def\hD{\hat{\Delta}}
\def\hDO{\hat{\Delta}_{\hat{\CO}}}

\def\hx{\hat{x}}
\def\hy{\hat{y}}
\def\ha{\hat{a}}
\def\hb{\hat{b}}

\newcommand{\hphi}{\hat\phi}
\newcommand{\hatt}{\hat t}
\newcommand{\Cphi}{C_\phi}
\newcommand{\normord}[1]{{:}\,#1\,{:}}   

\def\SO{\mathrm{SO}}

\def\O{\mathrm{O}}

\renewcommand{\normord}[1]{#1}

\title{Conformal defects of general dimensions at finite temperature}
\author{Yucheng Li,}
\author{Haruki Nakayama}
\author{and Tatsuma Nishioka}

\affiliation{Department of Physics, The University of Osaka,\\
Machikaneyama-Cho 1-1, Toyonaka 560-0043, Japan}

\preprint{OU-HET-1327}

\abstract{
We consider defect conformal field theories (DCFTs) at finite temperature, where a $p$-dimensional conformal defect wraps the thermal circle. 
Extending the previous study of line defects to defects of general dimensions, we determine the general forms of the thermal one-point functions of bulk scalars, conserved currents and the stress tensor by imposing the residual symmetry and the conservation laws.
The low temperature expansion of the thermal one-point functions is shown to be reproduced by the bulk-defect operator expansion, allowing us to read off the thermal one-point coefficients of defect operators.
We examine the general results in four classes of examples with free bulk theories: the trivial defect, free scalar and fermion theories with a boundary, the free scalar theory with a localized $\phi$ deformation in $d=2\,p+2$ dimensions and the free $\mathrm{O}(N)$ model with a localized $\phi^2$ deformation in $d=p+2-\epsilon$ dimensions.
In the last example with $\epsilon = 1$, where the defect becomes an interface, we find that the thermal one-point functions coincide with those of the free scalar theory with the Dirichlet boundary condition, in accordance with the  conjectured factorization of an interface CFT into two decoupled boundary CFTs.
}

\begin{document}

\maketitle

\newpage

\section{Introduction}
Conformal field theories (CFTs) provide a universal and practical description of various physical systems at criticality.
They have played a pivotal role in both high energy theory and condensed matter physics, ranging from hard scattering processes in QCD \cite{Braun:2003rp,Hofman:2008ar} to quantum phase transitions at zero temperature \cite{Sachdev:2011fcc}.
In realistic situations, however, most of the physical systems described by CFTs are at finite temperature, where a part of conformal symmetries is broken by the scale $\beta$ for the temperature. 
Since the scale is introduced not by the theory itself but by placing the theory on the thermal manifold $\BS^1_\beta \times \BR^{d-1}$, the local data of the CFT such as the operator spectrum and the operator product expansion (OPE) remain intact \cite{El-Showk:2011yvt,Iliesiu:2018fao}.
Nevertheless, the residual symmetry is not sufficient to constrain the forms of the correlation functions, and new structures which are absent at zero temperature appear in the thermal correlation functions.
In two dimensions, the modular invariance relates the thermal effects to the finite-size effects at zero temperature \cite{Cardy:1986ie}.
In higher dimensions, such an invariance does not exist in general,\footnote{A generalization of the modular invariance on $\BS^1_\beta \times \BT^{d-1}$ has been discussed in \cite{Shaghoulian:2015kta,Belin:2016yll,Allameh:2024qqp}.} and the thermal correlators provide additional data that characterizes the CFT \cite{Iliesiu:2018fao}.

On the thermal manifold $\BS_\beta^1 \times \BR^{d-1}$, a CFT does not have a phase transition as the temperature varies since the theory has no scale other than $\beta$.
Thus, a global symmetry is either preserved or broken at all temperatures in contrast to the symmetry restoration at high temperature in ordinary QFTs.
Examples of CFTs with thermal order have been constructed recently in \cite{Chai:2020zgq,Chai:2020onq,Chaudhuri:2020xxb,Chai:2021djc,Chai:2021tpt,Komargodski:2024zmt}, where the thermal one-point function $\langle\,\CO\,\rangle_\beta$ of a charged operator $\CO$ is used as an order parameter.
For a scalar primary $\CO$ with dimension $\Delta_{\CO}$, the one-point function is fixed to be
\begin{align}
    \langle\,\CO\,\rangle_\beta
        =
            \frac{b_\CO}{\beta^{\Delta_\CO}} \ ,
\end{align}
by the scale and translation symmetries.
The coefficient $b_\CO$ is an example of the additional data at finite temperature.
Using the KMS condition for the correlation functions with the OPE, one can constrain $b_\CO$ by bootstrap methods \cite{Iliesiu:2018fao,Iliesiu:2018zlz,Petkou:2018ynm,David:2023uya,Marchetto:2023fcw,David:2024naf,Barrat:2024fwq,Barrat:2025nvu,Kumar:2025txh}.
Also, the asymptotic behavior of the density of states for operators with large dimensions is characterized by the coefficient $b_T$ for the stress tensor through the thermal effective action \cite{Benjamin:2023qsc,Allameh:2024qqp,Kusuki:2025pgx}.

Given these developments in CFTs at finite temperature, it is natural to incorporate conformal defects into thermal CFTs.
Conformal defects are extended objects such as boundaries and line operators that preserve a part of the conformal symmetry.
A CFT with a $p$-dimensional conformal defect $\CD^{(p)}$ in $d=p+q$ dimensions is called a defect CFT (DCFT).
Besides the bulk operators $\CO$, there are defect operators $\hO$ localized on $\CD^{(p)}$.
In addition to the bulk CFT data, the DCFT data consist of the spectrum and the OPE coefficients of the defect primaries, the bulk one-point coefficients $a_\CO$, and the bulk-defect operator expansion (DOE) coefficients $b_{\CO\hO}$ \cite{McAvity:1993ue,McAvity:1995zd,Liendo:2012hy,Gadde:2016fbj,Billo:2016cpy,Lauria:2018klo,Kobayashi:2018okw,Guha:2018snh}.
When we place a DCFT on the thermal manifold, a defect may or may not wrap the thermal circle.
The simplest example of the former is the Polyakov loop in gauge theories, which is used as the order parameter of the confinement/deconfinement phase transition \cite{Polyakov:1978vu}.
Another example is a boundary CFT (BCFT) at finite temperature, where the boundary extends along the thermal circle, and the thermal one-point functions in free field theories with boundaries were studied in \cite{Brown:1969na,Kennedy:1979ar,Kennedy:1981yi}.
In the latter case, a defect extends to the spatial directions and acts on the thermal states.
The most familiar example is a topological defect associated with a global symmetry \cite{Gaiotto:2014kfa}.

In this paper, we consider DCFTs at finite temperature with a $p$-dimensional defect $\CD^{(p)}$ which wraps the thermal circle.
For a line case ($p=1$), the structure of the thermal correlation functions has been worked out in \cite{Barrat:2024aoa} (see also \cite{Giombi:2026kdz}).
In a thermal DCFT, the coefficients $\hat b_{\hO}$ for the thermal one-point functions of defect operators $\hO$ appear as additional data.
The purpose of this paper is two-fold.
First, we generalize the previous study of \cite{Barrat:2024aoa} for line defects to the case with defects of arbitrary $p\ge 1$ dimensions, including surface defects, boundaries and interfaces.
Second, we examine the thermal data $\hat b_{\hO}$ in several concrete DCFT models whose bulk theory is free, and check the validity of the general formulation by comparing with the exact thermal one-point functions.

\paragraph{Summary of the results.}
When a $p$-dimensional defect $\CD^{(p)}$ wraps the thermal circle, the residual symmetry of the DCFT becomes the translations along $\CD^{(p)}$ and the rotations $\O(p-1)\times \SO(q)$ around it when $p \ge 2$.
Following the strategy for thermal CFTs in \cite{Iliesiu:2018fao}, we do not assume the theory to have a parity, but be invariant under the $\TR^a$ transformation which flips both the Euclidean time $\tau \to - \tau$ and one of the spatial coordinates $x^a_\parallel \to - x^a_\parallel$ on the defect.
The residual symmetry and modified conformal Ward identity for dilatation do not fix the thermal correlation functions $\langle\,\CO_1(x_1)\cdots \CO_n(x_n)\,\rangle_{\CD,\,\beta}$ completely, but constrain their forms as functions of the ratios of the coordinates to $\beta$.
For a bulk operator $\CO$, we decompose the thermal one-point function $\langle\,\CO\,\rangle_{\CD,\,\beta}$ to the bulk part $\langle\,\CO\,\rangle_{\beta}$ (the one-point function in the absence of the defect) and the defect-induced part $\langle\,\CO\,\rangle_{\CD,\,\beta}^\text{def}$ that depends on the transverse distance $|x_\perp|$ to the defect through the dimensionless distance $z :=|x_\perp|/\beta$.
We determine the general forms of the thermal one-point function for bulk scalar $\CO$, conserved current $J^\mu$ and the stress tensor $T^{\mu\nu}$.
The defect-induced part of a bulk scalar is represented by a function $f(z)$ while it is fixed up to a constant for a conserved current due to the conservation law.
For the stress tensor, we find that the defect-induced part is characterized by three functions $F_i(z)~(i=1,2,3)$ associated with three traceless symmetric tensor structures, where $F_1(z)$ and $F_2(z)$ are related by the conservation law.
For a line defect $(p=1)$ or a boundary $(q=1)$, some of the three functions vanish due to the reduction of the tensor structures.
Since the $\TR^a$ transformation is absent for a line case, the thermal one-point function of a conserved current and the stress tensor can have an additional term when the time-reversal symmetry is broken on the defect.
For defects without parity, the thermal one-point function of a bulk spin-two operator can have a parity odd structure only when $d=3$ and $p=2$.

We also examine the high and low temperature behaviors of the thermal one-point functions.
Since the one-point function depends on the distance $|x_\perp|$ to the defect through $z = |x_\perp|/\beta$, the high temperature limit corresponds to the $|x_\perp|\to \infty$ limit where the effect of the defect is negligible while the low temperature limit corresponds to the $|x_\perp|\to 0$ limit where the DOE provides a good approximation.
In the high temperature expansion, the defect-induced part of the thermal one-point function is controlled by the dimensionally-reduced theory on the thermal circle, and whether the effective theory is gapped or not may be read off from the decay in the large $z$ limit.
On the other hand, the low temperature expansion of the thermal one-point function is obtained by taking the thermal expectation value of the DOE.
The coefficients of the expansion are given by the DOE coefficients $b_{\CO\hO}$ times the thermal one-point coefficients $\hat b_{\hO}$ of defect operators $\hO$.
This relation can be used to read off $\hat b_{\hO}$ from the low temperature expansion given the DOE coefficients $b_{\CO\hO}$ at zero temperature.
We demonstrate this strategy with a concrete example.

We apply the above general results to four classes of examples of DCFTs whose bulk theories are free.
The first is the trivial defect in the free scalar theory, where $\CD^{(p)}$ is a $p$-dimensional hyperplane without any localized deformation on it.
This example is motivated by the more non-trivial ones considered later, which reduce to the trivial defect when the defect couplings are turned off.
While the defect is trivial at first sight, the defect theory is described by a generalized free field theory (GFF) of defect primaries $\hatt_s$ with transverse spin $s\ge 0$.
The bilinear defect operators of $\hatt_s$ may have non-vanishing thermal one-point functions, and one can perform a non-trivial test for the general relation between the low temperature expansion and the DOE.

The second deals with the free scalar and fermion theories in the presence of a boundary.
We calculate the exact thermal one-point functions of $\phi^2$, $\bar\psi \psi$ and $T^{\mu\nu}$ by using the method of images, and reproduce the results of \cite{Kennedy:1979ar} when $d=4$.
For the scalar case, we determine the thermal one-point coefficients $\hb_{\hO}$ of the boundary operators from the low temperature expansion as a demonstration of the strategy mentioned above.
For the fermionic case, the conformal boundary condition breaks parity.
We find a parity odd term in the thermal one-point function of the stress tensor in $d=3$ as predicted by the general argument.

The third one is the free scalar theory with the localized $\phi$ deformation.
This model is the generalization of the scalar Wilson line in $d=4$ dimensions \cite{Kapustin:2005py} to a $p$-dimensional defect in $d=2\,p+2$ dimensions \cite{Lauria:2020emq,Nishioka:2021uef}.
Since the deformation is linear in $\phi$, it can be absorbed by the shift of $\phi$ by the classical solution $\phi_0$ that the defect induces.
After the redefinition, the theory reduces to the free scalar theory with the trivial defect, and one can employ the results of the first example to compute the thermal one-point functions of $\phi$, $\phi^2$ and $T^{\mu\nu}$ exactly.

The fourth is the free $\O(N)$ model with the localized $\O(N)$-invariant $\phi^2$ deformation in $d=p+2-\epsilon$ dimensions.
The localized $\phi^2$ deformation has been studied at zero temperature in both the free and interacting $\O(N)$ models \cite{Shachar:2022fqk,Trepanier:2023tvb,Raviv-Moshe:2023yvq,Giombi:2023dqs,deSabbata:2024xwn,Kim:2025tvu,Ge:2025fsm}.
In the free $\O(N)$ case, the non-trivial IR fixed point for the defect coupling can be obtained to all orders in $\epsilon$ \cite{Giombi:2023dqs,Ge:2025fsm}.
At zero temperature, we determine the exact DCFT data such as the spectrum of the bilinear defect primaries with dimension up to $d+2\,\epsilon$ and the DOE coefficients for $\phi^2$ and $T^{\mu\nu}$.
We then calculate the exact thermal one-point functions of the bulk operators at the fixed point.
The consistency between the low temperature expansions and the DOE is verified for defect operators with dimension up to $d+2\,\epsilon$.
In the high temperature limit, we find that the defect-induced part decays as a power law, which is consistent with the fact that the dimensionally reduced theory is described by the massless free scalar in one dimension lower.
For $p=1$, the high temperature expansion diverges in $z\to \infty$, and we explain that the breakdown is caused by the IR divergence arising from the zero mode.
We also examine the case with $\epsilon = 1$, where the defect becomes an interface.
We find that the thermal one-point functions coincide with those of two copies of the scalar BCFT with Dirichlet boundary conditions.
Moreover, we determine the precise relation between the boundary operators in the Dirichlet BCFT and defect operators in the free $\O(N)$ model with the localized $\phi^2$ deformation with $\epsilon = 1$, which provides further support for the factorization conjecture in \cite{Bray:1977fvl,Krishnan:2023cff,Popov:2025cha} (see also \cite{Raviv-Moshe:2023yvq,Diatlyk:2024ngd, deSabbata:2024xwn,Ge:2025fsm}).

\paragraph{Organization of the paper.}
Section \ref{ss:DCFT_review} reviews CFTs at finite temperature and DCFTs at zero temperature, and fixes our conventions and notation for later use,
which are summarized in table \ref{tab:notation}.
In section \ref{sec:thermal_DCFT}, we examine the general structures of DCFTs at finite temperature, including the thermal one-point functions, the high and low temperature expansions, and the relation to the DOE.
Then, the four classes of examples for thermal DCFTs, the trivial defect, the free theories with a boundary, the localized $\phi$ and $\phi^2$ deformations, are dealt with in sections \ref{sec:GFF_on_defect}, \ref{sec:free_BCFT}, \ref{sec:scalar_Wilson} and \ref{sec:O(N)_phi2_bulk_free}, respectively.
Section \ref{sec:discussion} is devoted to discussion and future directions.
Appendix \ref{app:DCFT_zetoT_details} lists the details of the two-point functions and the DOE in DCFTs at zero temperature.
Appendix \ref{app:Ward_identity} derives the modified conformal Ward identities in the presence of defects by using the topological operator method, which generalizes those of \cite{Marchetto:2023fcw,Barrat:2024aoa}.
Appendix \ref{sec:app_useful} collects useful formulas and appendix \ref{sec:app_Inu} provides the properties of the function $\CI_\nu[a,b](z)$ that is used in the thermal one-point functions of the free $\O(N)$ model in section \ref{sec:O(N)_phi2_bulk_free}.

\paragraph{Note added:}
After the submission of this manuscript, we became aware of the independent work \cite{Artico:2026rye}, which appeared on arXiv after the second version of this paper.
It overlaps with our results in section \ref{sec:thermal_DCFT} and \ref{sec:free_BCFT}.
We thank the authors for bringing their work to our attention.

\section{Preliminaries}\label{ss:DCFT_review}
In this section, we begin with reviewing CFTs at finite temperature in section \ref{sec:review_CFT}, and then summarize the DCFT data at zero temperature in section \ref{sec:DCFT_zeroT_review}.
Those who are familiar with these subjects may skim table \ref{tab:notation} for our conventions and notation, and skip to section \ref{sec:thermal_DCFT}.

\subsection{CFTs at finite temperature}\label{sec:review_CFT}
We review the basic facts about CFTs at finite temperature.
Our presentation follows \cite{Iliesiu:2018fao,Miscioscia:2025pjh}, to which we refer for more details.

By a thermal CFT, we mean a CFT on the thermal manifold $\BS^1_\beta\times\BR^{d-1}$ without changing any conformal data at zero temperature.
Due to the periodicity along the thermal circle $\BS^1_\beta$, correlation functions are subject to the KMS condition 
\begin{align}\label{KMS condition}
        \langle\,\phi_1(\tau_1)\,\phi_2(\tau_2)\,\rangle_{\beta}
            =
                \pm \langle\, \phi_1(\tau_1)\,\phi_2(\tau_2-\beta)\, \rangle_{\beta}\ .
\end{align}
For free fields, the thermal propagator is obtained by the method of images from the propagators at zero temperature:
\begin{align}
    G_\beta(\tau, \vec{x})=\sum_{n\in \BZ}\,(\pm 1)^n\,G_0(\tau+n\,\beta, \vec{x})\ .
\end{align}
In the above equations, the $+$ sign is for bosonic theories and the $-$ sign for fermionic theories.

\paragraph{Symmetries and Ward identities.}
On the thermal manifold, the conformal symmetry group $\SO(d+1,1)$ is broken to the rotation group $\SO(d-1)$ and the translation group along the $d$ directions.
Note that the symmetry breaking is brought by the background geometry and the stress tensor is still traceless and conserved.
Thus, the local conformal Ward identities still hold.
On the thermal manifold, however, the conformal Ward identities are modified due to the non-invariance of the manifold under the dilatation (the modified one is termed ``broken'' Ward identities in \cite{Marchetto:2023fcw}).
The identity concerning scale transformation is of particular importance among the other Ward identities.
In the convention where the stress tensor being defined as
\begin{align}
    T^{\mu\nu}
        =
            -\frac{2}{\sqrt{g}}\frac{\delta I_E}{\delta g_{\mu\nu}}\ ,
\end{align}
where $I_E$ denotes the Euclidean action, it reads \cite{Marchetto:2023fcw}
\begin{align}\label{scale ward}
    \left[\sum_i\left(\Delta_i+x^{\mu}_i\frac{\partial}{\partial x^{\mu}_i}\right)+\beta\,\frac{\partial}{\partial\beta}\right]\langle \,\CO_1(x_1)\cdots\CO_n(x_n)\,\rangle_{\beta}
        =
            0\ .
\end{align}
The last term is the consequence of a scale transformation which amounts to changing the temperature on $\BS^1_\beta\times\BR^{d-1}$ (see appendix \ref{app:Ward_identity} for the derivation using topological operators).
Since the one-point function of any operator is a constant due to translational invariance, we conclude from the above identity that $\langle\, \CO\,\rangle_\beta \propto \beta^{-\Delta_\CO}$, where $\Delta_\CO$ is the scaling dimension of $\CO$.

As in \cite{Iliesiu:2018fao}, we do not require the parity invariance, which is defined as the reflection in one or several spatial directions.
Instead, we require the invariance under the $\TR^a$ transformation, which is defined by combining the reflection in one of the spatial directions with the reflection in time direction:
\begin{align}\label{reflection}
    \TR^a:~\, (\tau, x^1, \cdots, x^a, \cdots, x^{d-1}) ~\to~(-\tau, x^1, \cdots, -x^a, \cdots, x^{d-1}) \ .
\end{align}
In a theory without defects, the choice of $a$ can be arbitrary.
The invariance under such transformation is guaranteed by the $\SO(d)$ invariance of the theory at zero temperature as the $\TR^a$ transformation is a rotation by $\pi$ in the $\tau$-$x^a$ plane. 
This together with the $\SO(d-1)$ invariance gives us the $\O(d-1)$ group.

\paragraph{Thermal one-point function.}
The Ward identity \eqref{scale ward} for scale transformation and the $\O(d-1)$ invariance determine the structure of the thermal one-point functions.
For scalar primary operators, 
\begin{align}\label{thermal 1-pt scalar}
    \langle\, \CO(x)\, \rangle_\beta
        =
            \frac{b_\CO}{\beta^{\Delta_\CO}}\ ,
\end{align}
and for a symmetric and traceless primary spin-$l$ operator, 
\begin{align}\label{thermal 1-pt with spin}
    \langle\, \CO^{\mu_1\cdots \mu_l}(x)\,\rangle_\beta
        =
            \frac{b_\CO}{\beta^{\Delta_\CO}}\,\left(e^{\mu_1}\cdots e^{\mu_l}-\text{traces}\right),
\end{align}
where $e^\mu$ is the unit vector pointing toward increasing $\tau$-direction.
Since the $\TR^a$ symmetry acts on $e^\mu$ as the sign flip, the reflection invariance forces $b_\CO=0$ for odd $l$.
The thermal one-point coefficients $b_\CO$ are new CFT data that are independent of those at zero temperature.

\paragraph{OPE at finite temperature.}
Since the temperature does not change the UV property of a CFT, the OPE relation continues to hold at finite temperature as long as a pair of operators are close enough.
The periodicity $\beta$ of the thermal manifold puts an upper bound on the convergence radius of OPE.
For a pair of scalar operators $\phi(x),\phi(0)$ satisfying $|x|:=\sqrt{\tau^2+\vec{x}^2}<\beta$, the thermal two-point function can be expanded as \cite{Iliesiu:2018fao} 
\begin{align}\label{eq:thermal_2pt}
    \langle\,\phi(x)\,\phi(0)\,\rangle_{\beta}
        =
            \sum_{\CO^{\mu_1\cdots \mu_l}}\frac{f_{\phi\phi \CO}}{C_\CO}\,|x|^{\Delta_\CO-2\Delta_{\phi}-l}\,x_{\mu_1}\cdots x_{\mu_l}\,\langle\, \CO^{\mu_1\cdots \mu_l}(0)\,\rangle_\beta \ ,
\end{align}
where $C_\CO$ and $f_{\phi\phi \CO}$ are coefficients of the two- and three-point functions at zero temperature.\footnote{We use the following normalization for the three-point function:
\begin{align}\label{eq:3pt}
\begin{aligned}
    \langle\,\phi(x_1)\,\phi(x_2)\,\CO^{\mu_1\cdots\mu_l}(x_3)\rangle
        &=
            f_{\phi\phi \CO}\,\frac{Z^{\mu_1}\cdots Z^{\mu_l} - \text{traces}}{|x_1-x_2|^{2\Delta_{\phi}-\Delta_{\CO} + l}\,|x_2-x_3|^{\Delta_{\CO}-l}\,|x_3-x_1|^{\Delta_{\CO}-l}} \ ,\\
    Z^{\mu}
        &=
            \frac{(x_1-x_3)^{\mu}}{|x_1-x_3|^2} - \frac{(x_2-x_3)^{\mu}}{|x_2-x_3|^2} \ .
\end{aligned}
\end{align}
}
The OPE converges only for $|x|<\beta$ while the KMS condition relates the OPE around $\tau=0$ to the one around $\tau = \beta$.
Combining the KMS condition \eqref{KMS condition} and the OPE relation \eqref{eq:thermal_2pt}, one can derive a set of non-trivial constraints on the thermal one-point coefficients $b_\CO$ \cite{El-Showk:2011yvt,Iliesiu:2018fao}.

\subsection{Defect CFTs at zero temperature}\label{sec:DCFT_zeroT_review}
We consider a $d$-dimensional defect CFTs with a $p$-dimensional defect $\CD^{(p)}$ and collect the DCFT data at zero temperature that will be used in later sections.
We set our conventions in defect CFTs in section \ref{ss:DCFT_convention}, and summarize the general forms of the correlation functions of the bulk and defect primary operators in section \ref{ss:DCFT_correlators_zeroT}.

\subsubsection{Conventions in defect CFT}\label{ss:DCFT_convention}
In the presence of a $p$-dimensional conformal defect $\CD^{(p)}$ in a $d$-dimensional CFT, the conformal symmetry $\SO(d+1, 1)$ is broken to the subgroup $\SO(p+1, 1)\times \SO(q)$ where $q:= d-p$.
We denote the bulk coordinates $x^\mu$ and the coordinates parallel and transverse to $\CD^{(p)}$ by $\hat x$ and $x_\perp$, respectively, i.e.,
\begin{align}
    x^\mu
        =
            (\hat x^{\hat a}, x_\perp^i) \ ,\qquad
    \hat a = 0, 1, \cdots, p-1\ , \quad 
    i=p, \cdots, d-1 \ .
\end{align}
The defect is always located at $x_\perp^i = 0$, and we choose the bulk extending to $x_\perp \ge 0$ in a boundary CFT ($q=1$).

We also denote the coordinate along the thermal circle by $\tau:=x^0$ and write
\begin{align}
    \hat x^{\hat a}
        =
            (\tau, x_\parallel^a) \ , \qquad a = 1, \cdots, p-1 \ .
\end{align}
The metrics on the defect and in the transverse direction are 
\begin{align}
    h_{\mu\nu}
        :=
            \delta_{\hat a\,\hat b}\,\delta_\mu^{\hat a}\,\delta_\nu^{\hat b} \ , \qquad
    h_{\perp\,\mu\nu}
        :=
            \delta_{\mu\nu} - h_{\mu\nu} \ ,
\end{align}
respectively.

Correlation functions in the presence of the defect are covariant under the residual conformal symmetry, and their tensor structures can be built from the invariant tensors $\delta_{\mu\nu}$, $h_{\mu\nu}$, $h_{\perp\,\mu\nu}$ and the vector 
\begin{align}
    n^\mu
        :=
            \delta^\mu_{i}\,\frac{x_\perp^i}{|x_\perp|} \ .
\end{align}
At finite temperature, they also depend on the vector $e^\mu$.
When we discuss the one-point function of a symmetric traceless spin-two operator, we find it convenient to use the following tensor structures:
\begin{align}\label{tensor basis}
    \sfs_1^{\mu\nu}
        =
            p\,h^{\mu\nu}_\perp-q\,h^{\mu\nu}\ ,\qquad
    \sfs_2^{\mu\nu}
        =
            h^{\mu\nu}_\perp-q\,n^\mu n^\nu\ ,\qquad
    \sfs_3^{\mu\nu}
        =
        h^{\mu\nu}-p\,e^\mu e^\nu\ .
\end{align}
These tensors are traceless and symmetric, and satisfy the following orthogonality relations:
\begin{align}
    \sfs_i^{\mu\nu}\,\sfs_{j\,\mu\nu}
        =
            \CN_i\,\delta_{ij} \ ,
\end{align}
where
\begin{align}
    \left( \CN_1, \CN_2, \CN_3\right)
        =
        \left(\,p\,q\,d,\ q\,(q-1),\ p\,(p-1)\,\right) \ .
\end{align}
We note that $\sfs_2^{\mu\nu}$ and $\sfs_3^{\mu\nu}$ vanish for a boundary $(q=1)$ and a line defect $(p=1)$, respectively.

The conservation law of the stress tensor in DCFT takes the form\footnote{Our convention of the displacement operator differs from the one in \cite{Billo:2013jda} by the overall sign.}
\begin{align}
    \partial_\mu T^{\mu\nu} = \delta^{(q)}(x_\perp)\,\delta^\nu_i\,\hat D^i \ ,
\end{align}
where $\hat D^i$ is the displacement operator with dimension $\hD_{\hat D} = p+1$.
In a boundary CFT ($q=1$), the displacement operator is given by
\begin{align}\label{displacement_BCFT_definition}
    \hat D = \lim_{x_\perp \to 0}\,n^\mu\,n^\nu\,T_{\mu\nu} \ .
\end{align}

\begin{table}[H]
    \centering
    \renewcommand{\arraystretch}{1.15}
    \begin{tabular}{lp{8.2cm}}
        \toprule
         Symbol & Meaning \\
         \midrule
            $d,~p,~q:=d-p$ & dimensions of bulk, defect and transverse coordinates \\
            $\CD^{(p)}$ & $p$-dimensional defect \\
            $x^\mu$ ~ $(\mu=0,1,\cdots,d-1)$  &    bulk coordinates \\
            $\tau := x^0$ & thermal circle \\
            $\hat x^{\hat a}:=x^{\hat a}$ ~ $(\hat a =0,1,\cdots,p-1)$  &  $p$-dimensional defect coordinates \\
            $x_{\parallel}^a :=x^a$ ~ $(a = 1,\cdots,p-1)$  &  spatial coordinates on defect \\
            $x_{\perp}^i:=x^i$ ~ $(i = p,\cdots,d-1)$  &  bulk coordinates transverse to defect \\
            $\vec{x} := (x_\parallel^a, x_\perp^i)$ & coordinates except thermal circle\\
            $e^\mu$ & future-pointing unit vector in $\tau$ direction \\
            $z := \frac{|x_\perp|}{\beta}$ & dimensionless distance to defect \\
            $n^\mu:= \delta^{\mu}_i\,\frac{x_\perp^i}{|x_\perp|}$ & unit vector normal to defect \\
            $h_{\mu\nu}:= \delta_{\hat a \hat b}\,\delta^{\hat a}_\mu\,\delta^{\hat b}_\nu$ & induced metric on defect \\
            $h_{\perp\,\mu\nu}:= \delta_{\mu\nu} - h_{\mu\nu}$ &  metric transverse to defect \\
            $\sfs_{1,2,3}^{\mu\nu}$ & orthogonal basis of symmetric traceless tensors, defined in \eqref{tensor basis}\\
            $(ab)$,~$[ab]$,~$\langle ab\rangle$ & symmetrization, anti-symmetrization, traceless symmetrization of indices\\
            \midrule
            $\CO$,~$\Delta_\CO$,~$l$ & bulk primary, its dimension and $\SO(d)$ spin \\
            $\hO_{(s,j)}$,~$\hD_{\hO}$ & defect primary with transverse spin $s$ and parallel spin $j$, its dimension \\
            $A,B=1,\cdots, N$ & $\O(N)$ indices \\
            \midrule
            $\langle\cdots\rangle_\beta$,~$\langle\cdots\rangle_\CD$,~$\langle\cdots\rangle_{\CD,\beta}$ & thermal, defect, and thermal defect correlators \\
            $C_\CO$,~$C_{\hO}$ & two-point normalizations of bulk and defect primaries \\
            $a_\CO$ & one-point coefficient of $\CO$ in DCFT at zero temperature \\
            $b_\CO$,~$\hat b_{\hO}$ & thermal one-point coefficients of $\CO$ and $\hO$ \\
            $b_{\CO\hO}$ & DOE coefficients \\
        \bottomrule
    \end{tabular}
    \caption{List of notations used in this paper.}
    \label{tab:notation}
\end{table}

\subsubsection{Correlation functions}\label{ss:DCFT_correlators_zeroT}
In a defect CFT, there are two types of operators.
Bulk primaries $\CO$ are labeled by the conformal dimension $\Delta_\CO$ and the $\SO(d)$ spin $l$ while defect primaries $\hO$ are labeled by the conformal dimension $\hD_{\hO}$, the parallel $\SO(p)$ spin $j$ and the transverse $\SO(q)$ spin $s$.

We denote an $n$-point correlation function in the presence of the defect $\CD^{(p)}$ as
\begin{align}
    \langle\,\CO_1(x_1)\cdots \CO_n(x_n)\,\rangle_{\CD}
        :=
            \frac{\langle\,\CO_1(x_1)\cdots \CO_n(x_n)\,\CD^{(p)}\,\rangle}{\langle\,\CD^{(p)}\,\rangle} \ ,
\end{align}
where we regard $\CD^{(p)}$ as an operator in the bulk CFT without the defect on the right hand side.

\paragraph{One-point functions.}
The one-point function of a defect primary $\hO(\hat x)$ vanishes due to the conformal symmetry $\SO(p+1,1)$ on the defect.
In contrast, a bulk primary $\CO(x)$ can have a non-vanishing one-point function in the presence of the defect, which only depends on $|x_\perp|$ due to the translation symmetry on the defect.

For a bulk scalar primary $\CO$, the one-point function is fixed up to a constant by the residual conformal symmetry as
\begin{align}\label{bulk 1-pt scalar}
    \langle\,\CO(x)\,\rangle_{\CD}
        =
        \frac{a_{\CO}}{|x_\perp|^{\Delta_\CO}} \ .
\end{align}

The one-point function of a bulk spin-one operator vanishes unless $q=2$ and the theory is parity odd or parity violating (see section 3.4 of \cite{Billo:2016cpy}).

For the stress tensor, the one-point function is also fixed up to a constant $a_T$:
\begin{align}\label{bulk 1-pt stress}
    \langle\, T^{\mu\nu}(x)\,\rangle_{\CD}
        =
            \frac{a_T}{q\,|x_\perp|^{d}}\left(\frac{q-1}{d}\,\sfs^{\mu\nu}_1 + \sfs^{\mu\nu}_2\right) \ ,
\end{align}
which is automatically conserved.

\paragraph{Two-point functions.}
The two-point function of a pair of defect operators in a defect CFT take the same forms as those of a pair of operators in CFTs.
For defect scalar primaries,
\begin{align}\label{defect 2-pt scalar}
    \langle\,\hO(\hx)\,\hO'(\hx')\,\rangle_{\CD}
        =
            \delta_{\hO,\hO'}\,\frac{C_{\hO}}{|\hx-\hx'|^{2\hDO}} \ ,
\end{align}
where $C_{\hO}$ is the coefficient that fixes the normalization of $\hO$.
The two-point functions of defect primaries with parallel or transverse spin are given in appendix \ref{app:defect-defect}.

The two-point function of bulk and defect scalar primaries takes the form:
\begin{align}\label{bulk-defect 2-pt scalar}
    \langle\,\CO(x)\,\hO(\hx')\,\rangle_{\CD}
        =
            \frac{c_{\CO\hO}}{|x_\perp|^{\Delta_\CO-\hDO}\,\left(|\hat x-\hx'|^2 + |x_\perp|^2 \right)^{\hDO}} \ .
\end{align}
For primaries with spin, the two-point function is given by the sum of several tensor structures with independent coefficients.
The bulk-defect two-point functions for operators of spin up to two are listed in appendix \ref{app:bulk-defect}.

\subsubsection{Bulk-defect operator expansion}
Bulk operators can be expanded in terms of defect operators, whose coefficients depend on the distance to the defect schematically as
\begin{align}\label{DOE_formal}
    \CO(\hx,x_\perp)
        =
            \sum_{\hO_{(s,j)}}\,|x_\perp|^{\hDO-\Delta_\CO}\,\left[\sum_I\,b^{(I)}_{\CO\hO}\,\CS_I\right]\, \hO_{(s,j)}(\hx)+ \text{(descendants)} \ ,
\end{align}
where $\hO_{(s,j)}$ stands for a defect primary with transverse spin $s$ and parallel spin $j$, and the sum is over all defect primaries including the defect identity $\hat{\bm{1}}$.
The tensor structures $\CS_I$ are constructed from $n^\mu$, $h_{\mu\nu}$ and $h_{\perp\,\mu\nu}$, and labeled by the indices $I$.
The descendant part is fixed by the residual conformal symmetry, and 
the DOE is convergent if there are no operators closer to the defect than the bulk operator.
The DOE coefficient $b_{\CO\hO}^{(I)}$ is fixed by taking the two-point function with a defect operator on both sides of \eqref{DOE_formal}.
See appendix \ref{app:DOE} for details.

Using the bulk-defect two-point functions in appendix \ref{app:bulk-defect}, the DOE for a bulk scalar $\CO$ is determined as
\begin{align}\label{scalar DOE}
    \begin{aligned}
        \CO(\hx,x_\perp)
            &= 
                \frac{1}{|x_\perp|^{\Delta_\CO}}\left[a_{\CO}+\sum_{\hO}b_{\CO\hO}\,|x_\perp|^{\hDO}\,\hO(\hx)+\sum_{\hO_{(s,0)}}b_{\CO\hO}\,|x_\perp|^{\hD_{\hO}}\,n_{i_1}\cdots n_{i_s}\,\hO_{(s,0)}^{i_1\cdots i_s} (\hx)\right]\\
            &\qquad\qquad +
                \text{(descendants)} \ .
    \end{aligned}
\end{align}

The DOE for a bulk operator with spin can be obtained in a similar manner, and given for the spin-one and spin-two cases in appendix \ref{app:DOE}.
Here, we only record the DOE for the conserved current and stress tensor.

For a conserved current $J^\mu$ with $\Delta_J = d-1$, it reads
\begin{align}\label{current OE}
    \begin{aligned}
        J^\mu(\hx,x_\perp)
            =
                \frac{1}{|x_\perp|^{d-1}}
                \Big[&\,n^\mu \sum_{\hO:\,\hDO=p}\,b_{J\hO}\,|x_\perp|^{p}\,\hO(\hx)
                +
                \sum_{\hO_{(0,1)}}b_{J\hO}\,|x_\perp|^{\hD_{\hO}}\,\delta^\mu_{\ha}\,\hO_{(0,1)}^{\ha}(\hx)\\
                &+
                    \sum_{\hO_{(s,0)}}|x_\perp|^{\hD_{\hO}}\,\left(
                        b^{(1)}_{J\hO}\,\delta^\mu_{i_1}\,n\,_{i_2}\cdots n_{i_s}
                        +
                            b^{(2)}_{J\hO}\,n^\mu n_{i_1}\cdots n_{i_s}\right)\,\hO_{(s,0)}^{i_1\cdots i_s}(\hx)\\
                &+
                \sum_{\hO_{(s,1)}}b_{J\hO}\,|x_\perp|^{\hD_{\hO}}\,\delta^\mu_{\ha}\,n_{i_1}\cdots n_{i_s}\,\hO_{(s,1)}^{i_1\cdots i_s,\ha}(\hx)\Big]
                +
                \text{(descendants)} \ ,
    \end{aligned}
\end{align}
where only the defect scalars of $\hD_{\hO} = p$ appear due to the conservation law $\partial_\mu J^\mu = 0$, and the summations in the second and third lines are taken over $s \ge 1$.

For the stress tensor with $\Delta_T = d$, it reads
\begin{align}\label{stress_DOE}
    \begin{aligned}
        T^{\mu\nu}(\hx,x_\perp)
            &=
                \frac{a_T}{q\,|x_\perp|^{d}}\left(
                    \frac{q-1}{d}\,\sfs^{\mu\nu}_1 +
                    \sfs^{\mu\nu}_2\right)\\
            &\qquad \quad 
                    +
                    \sum_{ (s,\,j),\, j\le 2}\,\sum_{\hO_{(s,j)}}|x_\perp|^{\hDO-d}\,\left[\sum_{I}\,b^{(I)}_{T\hO}\,\CS^{\mu\nu}_I\right]\,\hO_{(s,j)}(\hx)+\text{(descendants)} \ ,
    \end{aligned}
\end{align}
where the first term is the contribution from the defect identity.
The defect operators and tensor structures that appear in the second line are listed in table \ref{tab:stress OE}.

\begin{table}[t]
    \centering
    \renewcommand{\arraystretch}{1.7}
    \begin{tabular}{ccl}
        \toprule
        $(s,j)$ & defect operator & $\sum_I b^{(I)}_{T\hO}\,\CS^{\mu\nu}_I$ \\
        \midrule
        $(0,0)$ & $\hO$ & $b^{(1)}\,\sfs_1^{\mu\nu} + b^{(2)}\,\sfs_2^{\mu\nu}$ \\
        $(s\ge 1,0)$ & $\hO_{(s,0)}^{i_1\cdots i_s}$ & \begin{tabular}[t]{@{}l@{}}
        $\left(b^{(1)}\,\sfs_1^{\mu\nu}
            +
                b^{(2)}\,\sfs_2^{\mu\nu}\right)\,n_{i_1}\cdots n_{i_s}$\\[2pt] $\quad
            +
                \,b^{(3)}\Big(n^{(\mu}\delta^{\nu)}_{i_1}-\dfrac{h^{\mu\nu}_\perp}{q}\, n_{i_1}\Big)n_{i_2}\cdots n_{i_s}+b^{(4)}\delta^\mu_{i_1}\delta^\nu_{i_2}n_{i_3}\cdots n_{i_s}$\end{tabular} \\[4pt]
        $(0,1)$ & 
        \parbox[c]{2.6cm}{\centering
            $\hO_{(0,1)}^{\ha}$\\[2pt] $(\hD = p+1)$
        }
        & $b\,n^{(\mu}\delta^{\nu)}_{\ha}$ \\
        $(s\ge 1,1)$ & $\hO_{(s,1)}^{i_1\cdots i_s,\ha}$ & $b^{(1)}\delta^{(\mu}_{\ha}n^{\nu)}n_{i_1}\cdots n_{i_s}+b^{(2)}\delta^{(\mu}_{\ha}\delta^{\nu)}_{i_1}n_{i_2}\cdots n_{i_s}$ \\
        $(s,2)$ & $\hO_{(s,2)}^{i_1\cdots i_s,\ha\hb}$ & $b\,\delta^\mu_{\ha}\delta^\nu_{\hb}\,n_{i_1}\cdots n_{i_s}$ \\
        \bottomrule
    \end{tabular}
    \caption{The defect primaries and the associated tensor structures that appear in the DOE \eqref{stress_DOE} of the stress tensor. For $(s=1,0)$, $b^{(4)}$ is absent.}
    \label{tab:stress OE}
\end{table}

\paragraph{Constraints on $b_{T\hO}$ from the conservation law.}
The DOE coefficients $b_{T\hO}$ for the stress tensor in \eqref{stress_DOE} are not independent due to the conservation law $\partial_\mu T^{\mu\nu} = 0$ that holds when $T^{\mu\nu}$ is away from the defect.
For the defect scalar channel, one finds
\begin{align}\label{stress_DOE_coefficient_scalar_relation}
    p\,(\hDO-d)\,b^{(1)}_{T\hO}
        =
            (q-1)(\hDO-p)\,b^{(2)}_{T\hO} \ ,
\end{align}
which implies that $b^{(1)}_{T\hO} = 0$ for $\hO$ with $\hD_{\hO} = p$.
There are similar relations for the other channels, whose derivations are relegated to appendix \ref{app:DOE}.

In a boundary (or interface) CFT ($q=1$), the structure $\sfs_2^{\mu\nu}$ vanishes and \eqref{stress_DOE_coefficient_scalar_relation} forces $b^{(1)}_{T\hO} = 0$ unless $\hD_{\hO} = d$.
In other words, only defect scalars with $\hD_{\hO} = d$ contribute to the scalar channel in the DOE \eqref{stress_DOE}.
The identity channel automatically vanishes when $q=1$, and contracting both sides of \eqref{stress_DOE} with $n_\mu\,n_\nu$, 
\begin{align}
    T^{nn} 
        :=
            n_\mu\,n_\nu\,T^{\mu\nu}
        \supset
            (d-1)\,\sum_{\hD_{\hO}=d}\,b^{(1)}_{T\hO}\,\hO \ .
\end{align}
In a boundary CFT, the displacement operator $\hat D$ is defined by \eqref{displacement_BCFT_definition}, which is a boundary scalar with $\hD = d$.
Thus we find the bulk-boundary operator expansion (BOE) for the stress tensor:
\begin{align}\label{stress_OE_boundary}
    T^{\mu\nu}(\hx,x_\perp)
        \sim
            \frac{1}{d-1}\,\sfs_1^{\mu\nu}\,\hat D (\hx)
            +
            \text{(non-scalar channel)} \ ,
\end{align}
from which we read off the BOE coefficient $b^{(1)}_{T\hat D} = 1/(d-1)$.

\section{DCFTs at finite temperature}\label{sec:thermal_DCFT}
In this section, we consider a defect CFT at finite temperature.
We focus on the case where the defect wraps the thermal circle parametrized by the coordinate $\tau$.
The residual symmetries and the general structures of the correlation functions in a defect CFT at finite temperature are discussed in section \ref{ss:residual_symmetry}.
The thermal one-point functions of bulk and defect operators are determined in section \ref{ss:general_thermal_1pt}, which reduce to those of \cite{Barrat:2024aoa,Giombi:2026kdz} when $p=1$.
Finally, the high and low temperature limits of the one-point functions are examined in section \ref{ss:general_high_lowT}.

\subsection{Residual symmetries and thermal correlation functions}\label{ss:residual_symmetry}

In the presence of a defect at finite temperature, the unbroken symmetries are translations along the directions parallel to the defect, and rotational invariance, $\SO(p-1)\times \SO(d-p)$. 

To define the $\TR^a$ transformation as in \eqref{reflection}, we have to make sure that it is an element of the $\SO(p)\times \SO(d-p)$ group at zero temperature.
We thus restrict to $p\geq2$, and let the reflected spatial direction be along the defect, $a\in 1,\cdots, p-1$.
We do not choose $a\in p,\cdots ,d-1$ for $\TR^a$, because this invariance holds only when there is parity invariance along the defect at zero temperature, which is too strong. Thus the total invariance should be 
\begin{align}
    (\text{translational invariance along the defect}) \, \times \, \O(p-1)\times \SO(q)\ .
\end{align}
For a line defect $(p=1)$, there are no $\TR^a$ symmetries.
This case will be discussed separately below.

\paragraph{Thermal correlation functions.}
As in a thermal CFT, we can also derive the modified version of the conformal Ward identities such as \eqref{scale ward}, in a similar way to the $p=1$ case \cite{Marchetto:2023fcw,Barrat:2024aoa,Miscioscia:2025pjh}. 
See appendix \ref{app:Ward_identity} for a derivation of the Ward identities using topological operator.
Using the translational invariance along the defect and taking the scaling dimensions of operators into account, an $n$-point correlation function takes the general form:
\begin{align}\label{general structure of correlation}
    \langle\,\CO_1(x_1)\cdots \CO_n(x_n)\,\rangle_{\CD,\,\beta}
        =
            \frac{1}{\beta^{\sum_k\Delta_k}} \,
            F\left(\frac{\hx_{k}-\hx_{l}}{\beta},\frac{x_{k,\perp}}{\beta}\right)
            \ ,
\end{align}
where $F$ denotes a function that depends on $(\hx_{k}-\hx_{l})/\beta$ and $x_{k,\perp}/\beta$ for every $k$ and $l$, and we have omitted the tensor structures for operators with spin.
This form automatically satisfies the scale Ward identity \eqref{scale ward}.

\subsection{Thermal one-point functions}\label{ss:general_thermal_1pt}
We determine the general structure of the one-point function of the bulk scalar, conserved current $J^\mu$ and the stress tensor $T^{\mu\nu}$ in the presence of a $p$-dimensional defect.
Following \cite{Kennedy:1979ar}, we decompose the one-point functions to the bulk and defect-induced parts, and assume the defect-induced part is given by a combination of local geometrical tensors on the defect that respect the residual symmetry.
We then parallel transport this combination to the bulk point where the one-point function is evaluated.
In flat space, the parallel transport is trivial.

\subsubsection{Defect operators}
Since the defect operators form a $p$-dimensional CFT by themselves, the thermal one-point function of defect operators take the same forms as \eqref{thermal 1-pt scalar} and \eqref{thermal 1-pt with spin},
\begin{align}
    \begin{aligned}
        \langle\,\hat{\CO}(\hat{x})\,\rangle_{\CD,\,\beta}
            &=
                \frac{\hat{b}_{\hat{O}}}{\beta^{\hat{\Delta}_{\hat{O}}}}\ ,\\
        \langle\,\hO_{(0,j)}^{\hat{a}_1\cdots\hat{a}_j}(\hx)\,\rangle_{\CD,\,\beta}
            &=
                \frac{\hat{b}_{\hat{O}}}{\beta^{\hat{\Delta}_{\hat{O}}}}\left(e^{\hat{a}_1}\cdots e^{\hat{a}_j}-\text{traces}\right)\ ,\\
        \langle\,\hO_{(s\ge 1,j)}^{i_1\cdots i_s, \hat a_1\cdots \hat a_j}(\hx)\,\rangle_{\CD,\,\beta}
            &=
                0 \ .
    \end{aligned}
\end{align}
Among defect primaries with spin, only operators with parallel spin can have non-zero one-point functions.
Operators with transverse spin have vanishing one-point function because there are no symmetric traceless tensor structures available that are invariant under the $\SO(q)$ transformation.
Also, the $\TR^a$ invariance forces $\hat b_{\hO} = 0$ for defect operators with odd parallel spin.
When the defect does not respect a parity,
the one-point function of a parallel spin operator may develop a parity violating part.

\subsubsection{Bulk scalars}
For a general bulk operator $\CO$, with or without spin, we define the \textit{defect-induced thermal one-point function} by subtracting the thermal one-point function without the presence of the defect from the total one-point function, 
\begin{align}
    \langle\,\CO(x)\,\rangle_{\CD,\,\beta}
        =
            \langle\,\CO(x)\,\rangle_\beta
            +
            \langle\,\CO(x)\,\rangle_{\CD,\,\beta}^\text{def}\ ,
\end{align} 
where the first term is the one-point function in the absence of the defect, which is already given in \eqref{thermal 1-pt scalar} or \eqref{thermal 1-pt with spin}.
By the translational invariance along the defect, the second term depends only on $x_\perp$, and we require it to approach zero as $|x_\perp|\to\infty$, where the effect of the defect becomes negligible. 

In what follows, we will use the dimensionless distance to the defect
\begin{align}
    z
        :=
            \frac{|x_\perp|}{\beta} \ .
\end{align}
For a bulk scalar operator, it follows from the general form \eqref{general structure of correlation} that the thermal one-point function becomes
\begin{align}\label{scalar bulk one-point}
    \langle\,\CO(x)\,\rangle_{\CD,\,\beta}
        =
            \frac{1}{\beta^{\Delta_\CO}}\left(\,b_\CO + f(z)\,\right) \ ,
\end{align}
where the defect-induced part is represented by the function $f(z)$.

\subsubsection{Conserved current}
For a conserved current $J^\mu$ with $\Delta_J = d-1$, the only invariant structure that can be used for the one-point function is $n^\mu$ when $p\geq2$ as terms proportional to $e^\mu$ are forbidden by the $\TR^a$ invariance.
By taking into account the conservation law $\partial_\mu J^\mu=0$, we find
\begin{align}
    \langle \, J^\mu(x)\,\rangle_{\CD,\,\beta}
        =
            \frac{f_J}{\beta^{d-1}}\,n^\mu\, z^{-(q-1)} \ ,
\end{align}
where $f_J$ is a constant.
In a boundary CFT $(q=1)$, the one-point function must vanish in $z\to \infty$, thus $f_J=0$.

\subsubsection{Stress tensor}

For the stress tensor $T^{\mu\nu}$ with $\Delta_T = d$, the available tensor structures are 
\begin{align}
    h^{\mu\nu},\, e^\mu e^\nu,\, h^{\mu\nu}_{\perp},\, n^\mu n^\nu,
\end{align}
where we assume $p\geq2$ and used the $\TR^a$ invariance to remove $e^{(\mu}n^{\nu)}$ term.
By using the tracelessness condition and the tensor basis \eqref{tensor basis}, we have 
\begin{align}\label{general_stress_tensor_thermal_1pt}
        \langle\,T^{\mu\nu}(x)\,\rangle_{\CD,\,\beta}
            =
                \langle\,T^{\mu\nu}(x)\,\rangle_{\beta}
                +
                \langle\,T^{\mu\nu}(x)\,\rangle_{\CD,\,\beta}^\text{def} \ ,
\end{align}
where the bulk part is 
\begin{align}
    \langle\,T^{\mu\nu}(x)\,\rangle_{\beta}
        =
            \frac{b_T}{\beta^{d}} \,\left(e^\mu e^\nu-\frac{\delta^{\mu\nu}}{d}\right) \ ,
\end{align}
and the defect-induced part is
\begin{align}\label{general_stress_tensor_defect}
    \langle\,T^{\mu\nu}(x)\,\rangle_{\CD,\,\beta}^\text{def}
        =
         \frac{1}{\beta^{d}}\,
            \left[ \,                    
                \sfs^{\mu\nu}_1\,F_1(z)
                +
                \sfs^{\mu\nu}_2\,F_2(z)
                +
                \sfs^{\mu\nu}_3\,F_3(z) \,
            \right] \ .
\end{align} 
The functions $F_i(z)~(i=1,2,3)$ approach zero in the $z\to \infty$ limit, and are subject to the following condition that follows from the conservation law in the transverse directions:
\begin{align}\label{stress conservation}
    -(q-1)\,F_2' - q\,(q-1)\,\frac{F_2}{z} + p\,F'_1 = 0 \ .
\end{align}

For $q=1$, the equation \eqref{stress conservation} reduces to $F_1'(z)=0$, and 
the falloff condition at $z=\infty$ forces $F_1 = 0$.
Since $\sfs_2^{\mu\nu}$ vanishes for $q=1$, the one-point function simplifies to
\begin{align}\label{eq:BCFT_stress_1pt}
    \langle\, T^{\mu\nu}(x)\,\rangle_{\CD,\,\beta}
        =
            \frac{1}{\beta^d}\left[\,
                b_T\,\left(e^\mu e^\nu-\frac{\delta^{\mu\nu}}{d}\right)
            +
                \sfs^{\mu\nu}_3\,F_3(z)\,\right]\ .
\end{align}

\subsubsection{$p=1$ case}
In a defect CFT with a line defect ($p=1$), the tensor structure $\sfs_3^{\mu\nu}$ vanishes as $h^{\mu\nu} = e^\mu e^\nu$.
Also, the $\TR^a$ transformation cannot be defined as there is no spatial direction along the defect.
For a theory without time reversal symmetry on the defect, there exist tensor structures proportional to $e^\mu$.
The most general forms of the one-point functions for the conserved current and stress tensor become
\begin{align}
    \begin{aligned}
        \langle\,J^\mu(x)\,\rangle_{\CD,\,\beta}
            &=
                \frac{1}{\beta^{d-1}}\,\left[
                    f_J\,n^\mu\,z^{-(d-2)}
                    +
                    e^\mu\,g_1(z)
                \right] \ , \\
        \langle\, T^{\mu\nu}(x)\,\rangle_{\CD,\,\beta}^\text{def}
            &=
                \frac{1}{\beta^d}\left[\,
                    \sfs^{\mu\nu}_1\,F_1(z)
                    +
                    \sfs^{\mu\nu}_2\,F_2(z)
                    +
                    e^{(\mu}n^{\nu)}\,F_4(z)\right]\ . \label{line defect spin-2}
    \end{aligned}
\end{align}
There are no constraints on $g_1(z)$ other than the falloff condition at $z=\infty$ while the conservation law $\partial_\mu T^{\mu\nu} = 0$ along the defect yields
\begin{align}
    F_4(z)
        =
            f_4\,z^{-(d-2)},\quad f_4=\text{const}\ .
\end{align}

In a defect CFT with time reversal symmetry, both $g_1$ and $F_4$ vanish.

\subsubsection{Parity odd case}\label{ss:parity odd}

We briefly comment on the case where parity is violated.
We do not presume the invariance under parity transformation as in \cite{McAvity:1993ue,Herzog:2017xha}.
We restrict our attention to defects preserving the transverse reflections, and study possible breaking of the spatial reflections parallel to the defect with $p\geq2$.
For $p=1$, the reflection of the defect coordinate corresponds to the Euclidean time reflection and should be treated separately.
In this case, the one-point functions of bulk operators can develop parity violating terms.
As a result, the Levi-Civita tensor, which is odd under the parity transformation, must be included in  constructing tensor structures for operators with spin.
Since tensor structures that can be contracted with the Levi-Civita tensor are $e^\mu, n^\mu$, such parity violating terms can exist only in $d=3$ (thus $p=2$) and only for spin-two bulk operators such as the stress tensor:
\begin{align}\label{parity violating of tensors}
    \langle\, T^{\mu\nu}(x)\,\rangle^\text{def}_{\CD,\,\beta}
        &\supset \frac{1}{\beta^d}\,V^{(\mu}\,e^{\nu)}\,F_5(z)\ ,
\end{align}
where $V^{\mu} = \epsilon^{\mu\nu\rho}\,e_\nu\, n_\rho $.
The function $F_5$ is not constrained by the conservation law.
For spin-one operators, $V^\mu$ cannot appear in the one-point function as it is odd under reflection.
Thus, among all thermal one-point functions, the one-point function of spin-two operator in $d=3$, $p=2$ is the only observable from which the parity violating can be directly observed.
We will see such parity violating terms in the free fermion theory with a boundary in section \ref{sec:fermion_BCFT}.

\subsection{High and low temperature expansions}\label{ss:general_high_lowT}
We discuss the high and low temperature expansions of the thermal one-point functions determined in section \ref{ss:general_thermal_1pt}.

\paragraph{High temperature expansion.}
The high temperature limit $(\beta\to 0)$ is equivalent to the large distance limit $|x_\perp| \to \infty$ as the one-point functions depend on the latter through $z=|x_\perp|/\beta$.
The falloff condition in the large $z$ limit implies
\begin{align}
    \lim_{z\to \infty}\,\beta^{\Delta_\CO}\,\langle\,\CO(x)\,\rangle_{\CD,\,\beta}^\text{def}
        =
            0 \ .
\end{align} 
The decay rate is governed by the effective theory on $\BR^{d-1}$ obtained by the dimensional reduction on the thermal circle.
The defect part $\langle\,\CO(x)\,\rangle_{\CD,\,\beta}^\text{def}$ falls off exponentially fast in $z$ if the effective bulk theory is gapped, while it follows a power law if there are massless modes.
We will observe these behaviors in the examples in later sections.

\paragraph{Low temperature expansion and bulk-defect operator expansion.}
The low temperature expansion is equivalent to the DOE, which is convergent as long as the distance to the defect is smaller than $\beta/2$ \cite{Miscioscia:2025pjh} . 
The thermal one-point function can be obtained by taking the thermal expectation value of the DOE. 
Among defect operators that appear in the DOE, those with transverse spin cannot survive due to the $\SO(q)$ invariance.
For a bulk operator with spin-$l$, only defect operators with parallel spin $j\le l$ can appear in the DOE.
Also, defect operators with odd parallel spin drop out for $p\geq 2$ due to the $\TR^a$ invariance, while there are no parallel spin operators for $p=1$.
Hence, for an arbitrary bulk operator $\CO$, only defect scalars and operators with even parallel spin $j\le l$ contribute to the thermal one-point function of $\CO$.

For a bulk scalar, the DOE \eqref{scalar DOE} yields
\begin{align}
    \begin{aligned}
        \langle\,\CO(x)\,\rangle_{\CD,\,\beta}
            =
                \frac{1}{\beta^{\Delta_\CO}}\left(\,b_\CO + f(z)\,\right)
            =
                \frac{1}{\beta^{\Delta_\CO}}\,\left(
                    \frac{a_\CO}{z^{\Delta_\CO}}
                    +
                    \sum_{\hat{\CO}}\,b_{\CO\hat{\CO}}\hat{b}_{\hat{\CO}}\,z^{\hat{\Delta}_{\hat{\CO}}-\Delta_\CO}
                    \right)\ ,
    \end{aligned}
\end{align}
where the summation is over defect scalars and we use \eqref{thermal 1-pt scalar}. 

For a conserved current, we use \eqref{current OE} to obtain
\begin{align}\label{current_thermal_1pt}
    \begin{aligned}
        \langle\,J^\mu(x)\,\rangle_{\CD,\,\beta}
            &=
                \frac{1}{\beta^{d-1}}
                    \,n^\mu\, g(z) 
                    \\
            &=
                \frac{1}{\beta^{d-1}}\,
                        n^\mu\,\sum_{\hat{\CO}}\,b_{J\hO}\,\hat{b}_{\hO}\,z^{-(q-1)}\ ,
    \end{aligned}
\end{align}
where the summation over defect scalars is restricted to those with $\hD = p$.
The linear combination of these operators is called the tilt operator.
It follows that 
\begin{align}\label{general_lowT_fJ}
    f_J
        =
        \sum_{\hDO=p}\,b_{J\hat{\CO}}\,\hat{b}_{\hO} \ .
\end{align} 
In a boundary CFT, the right hand side of \eqref{general_lowT_fJ} must vanish.
In the examples studied in this paper, there is no tilt operator, hence $\langle\,J^\mu(x)\,\rangle_{\CD,\,\beta} = 0$.

For the stress tensor, we obtain from the DOE \eqref{stress_DOE}
\begin{align}
    \begin{aligned}
        \langle\, T^{\mu\nu}(x)\,\rangle_{\CD,\,\beta}
            &=
                \frac{a_T}{q\,|x_\perp|^{d}}\,\left(           \frac{q-1}{d}\,\sfs^{\mu\nu}_1
                    +
                    \sfs^{\mu\nu}_2 \right)\\
            &\quad +
                \frac{1}{\beta^{d}}
                \left[
                    \sum_{\hat{\CO}}\,\hat{b}_{\hat{\CO}}\,z^{\hDO-d}\, \left( b_{T\hat{\CO}}^{(1)}\,\sfs^{\mu\nu}_1
                    +
                    b_{T\hat{\CO}}^{(2)}\,\sfs^{\mu\nu}_2
                    \right)
                    -
                    \frac{\sfs^{\mu\nu}_3}{p}\,\sum_{\hO_{(0,2)}}\,b_{T\hO}\,\hat{b}_{\hO}\,z^{\hD_{\hO} -d}
                    \right] \ .
    \end{aligned}
\end{align}
Compared with \eqref{general_stress_tensor_defect}, we find the low temperature expansions of the functions $F_i$\,:
\begin{align}\label{F low temp}
    \begin{aligned}
        F_1(z)
            &=\frac{b_T}{pd}+
                \frac{(q-1)\,a_T}{q\,d}\,z^{-d}+\sum_{\hO}b^{(1)}_{T\hO}\,\hat{b}_{\hO}\,z^{\hDO-d}\ ,\\
        F_2(z)
            &=
                \frac{a_T}{q}\,z^{-d}+\sum_{\hO}b^{(2)}_{T\hO}\,\hat{b}_{\hO}\,z^{\hDO-d}\ ,\\
        F_3(z)
            &=\frac{b_T}{p}
                -\frac{1}{p}\sum_{\hO_{(0,2)}}b_{T\hO}\,\hat{b}_{\hO}\,z^{\hDO-d}\ .
    \end{aligned}
\end{align}
Since the tensor basis $\sfs_i$ are orthogonal to each other, one can distinguish which part of the low temperature expansion is associated with defect scalars and which with defect parallel spin-two operators.

\section{Generalized free field theory on defect}\label{sec:GFF_on_defect}

The simplest defect one can consider in a free massless scalar theory in $d$ dimensions is the trivial defect, which is a $p$-dimensional hyperplane where nothing is inserted.
We will see, however, that the DCFT data are non-trivial as the defect theory is a generalized free field theory in $p$ dimensions.

Having later applications in mind, we consider the bulk theory consisting of $N$ free massless scalars $\phi_A~(A=1, \cdots, N)$:
\begin{align}\label{eq:bulk_free_scalar_action}
    I
        =
            \frac{1}{2}\int\d^d x\,(\partial \phi_A)^2
            \ ,
\end{align}
and the two-point function is normalized as
\begin{align}\label{eq:GFF_bulk_two_point}
    \langle\,\phi_A(x)\,\phi_B(y)\,\rangle
        =
            \delta_{AB}\,\frac{C_\phi}{|x-y|^{2\Delta_\phi}} \ ,
\end{align}
where $\Delta_\phi = \frac{d}{2} - 1$ and
\begin{align}\label{eq:Cphi}
    C_{\phi}
        :=
            \frac{1}{(d-2)\,\Omega_{d-1}}
        :=
            \frac{\Gamma\left({d\over2}-1\right)}{4\pi^{d\over2}} \ ,
\end{align}
where $\Omega_{d-1}$ is the volume of $\BS^{d-1}$:
\begin{align}
    \Omega_{d-1}
        :=
            \frac{2\,\pi^\frac{d}{2}}{\Gamma(\frac{d}{2})} \ .
\end{align}

\subsection{Bulk local operators}
\label{sec:GFF_bulk_operators}

Besides $\phi_A$, the bulk primaries of lower dimensions are bilinear in $\phi$.
The scalar bilinear primaries $\normord{\phi_A\phi_B}$ decompose into the singlet (S) and the symmetric traceless representation (T) of $\O(N)$:
\begin{align}
    \normord{\phi_A\phi_B}
        =
            \frac{\delta_{AB}}{N}\,\phi^2 + \normord{\phi_{\langle A}\phi_{B\rangle}} \ .
\end{align}
Both have dimension $\Delta = d-2$.
The spin-one bilinear primary is the $\O(N)$ conserved current of dimension $\Delta = d-1$ that is in the anti-symmetric (adjoint) representation (A):
\begin{align}\label{O(N)_current}
    J^\mu_{AB}
        =
            \phi_A \partial^\mu\phi_B - \phi_B \partial^\mu\phi_A \ .
\end{align}
There are two spin-two bilinear primaries of dimension $\Delta = d$.
One of them is the stress tensor
\begin{align}\label{O(N)_scalar_stress_tensor}
    T^{\mu\nu}
        =
            \partial^\mu\phi_A\,\partial^\nu\phi^A - \frac{\delta^{\mu\nu}}{2}(\partial\phi_A)^2
            -
           \xi\,\left(\partial^\mu\partial^\nu - \delta^{\mu\nu}\partial^2\right)\phi_A\phi^A \ ,
\end{align}
where $\xi :=  \frac{d-2}{4(d-1)}$.
The stress tensor is $\O(N)$ singlet, and the other is the symmetric traceless $\phi_{\langle A}\partial^{\langle \mu}\partial^{\nu\rangle} \phi_{B\rangle}$.
Table \ref{tab:GFF_bulk_primary} lists the bulk bilinear primaries of the free $\O(N)$ model with dimension $\Delta \le d$.
There are also multi-linear composite primaries such as $\normord{\phi^3}$.
The operator $\normord{\phi^n}$ has dimension $\Delta = n\,\Delta_\phi$, which is at most $d$ when $n\le \frac{2d}{d-2}$.
In lower dimensions, these operators may be heavier than the bilinear primaries in table \ref{tab:GFF_bulk_primary}.
In this paper, we focus on the bilinear primaries of lower dimensions.
Extending the analysis to multi-linear operators is straightforward.

\begin{table}[ht]
    \centering
    \begin{tabular}{ccccc}
    \toprule
       primary & $\O(N)$ rep & spin & $\Delta$ & $b_{\CO}/C_\phi$ \\
    \midrule
    $\phi_A$ & V & $0$ & $\Delta_\phi = \frac{d-2}{2}$ & $0$ \\
    \cmidrule(lr){1-5}
    $\normord{\phi^2} = \normord{\phi_A\phi^A}$ & S & $0$ & $d-2$ & $2\,N\,\zeta(d-2)$ \\
    $\normord{\phi_{\langle A}\phi_{B\rangle}}$ & T & $0$ & $d-2$ & $0$ \\
    \cmidrule(lr){1-5}
    $J^{\mu}_{AB}$ & A & $1$ & $d-1$ & $0$ \\
    \cmidrule(lr){1-5}
    $T^{\mu\nu}$ & S & $2$ & $d$ & $-2\,N\,d\,(d-2)\,\zeta(d)$ \\
    $\normord{\phi_{\langle A}\,\partial^{\langle\mu}\partial^{\nu\rangle}\phi_{B\rangle}}$ & T & $2$ & $d$ & $0$ \\
    \bottomrule
    \end{tabular}
    \caption{The bulk primary $\phi_A$ and the bilinear bulk primaries with $\Delta\le d$, in the $\O(N)$ singlet (S), vector (V), symmetric traceless (T) and adjoint (A) representations, together with their thermal one-point coefficients $b_{\CO}$ calculated in section \ref{sec:GFF_thermal_one_point}.
    At $N=1$ the representations T and A are absent and the list reduces to $\phi$, $\normord{\phi^2}$ and $T^{\mu\nu}$.
    }
    \label{tab:GFF_bulk_primary}
\end{table}

\subsection{Defect local operators}
\label{sec:GFF_defect_operators}

The basic defect local primary operator is
\begin{align}\label{eq:GFF_hphi}
    \hphi_A(\hat x)
        :=
            \lim_{|x_\perp|\to 0}\,\phi_A(\hat x, x_\perp)
            \ .
\end{align}
These defect primaries have the same dimension as $\phi_A$, $\hD_{\hphi} = \Delta_\phi = \frac{d}{2}-1$, and the two-point function is
\begin{align}\label{eq:GFF_hphi_two_point}
    \langle\,\hphi_A(\hat x)\,\hphi_B(\hat y)\,\rangle
        =
            \delta_{AB}\,\frac{\Cphi}{|\hat x - \hat y|^{2\Delta_\phi}} \ .
\end{align}
Thus, $\hphi_A$ behaves as the generalized free field of dimension $\Delta_\phi$ in $p$ dimensions.

Defect primaries with transverse spin can be constructed by acting with $\partial_{\perp\,i}$ on $\phi_A$ and projecting to the traceless parts:\footnote{The trace parts are not defect primaries as $\partial_\perp^2\phi_A = - \hat\partial^2\phi_A$ due to the equation of motion.}
\begin{align}
    \hatt_{s,A}^{i_1\cdots i_s}(\hat x)
        :=
            \lim_{|x_\perp|\to 0}\Pi_\perp^{i_1\cdots i_s,\, j_1\cdots j_s}\partial_{\perp\,j_1}\cdots\partial_{\perp\,j_s}\,\phi_A(\hat x, x_\perp)  \ ,
\end{align}
where $\Pi_\perp$ is the projector to symmetric traceless tensor in $q$ dimensions.\footnote{For $s=2$, it becomes $\displaystyle \Pi^{i_1 i_2,\,j_1 j_2} = \frac{1}{2}\left(\delta^{i_1 j_1}\,\delta^{i_2 j_2} + \delta^{i_1 j_2}\,\delta^{i_2 j_1}\right) - \frac{1}{q}\,\delta^{i_1 i_2}\,\delta^{j_1 j_2}$.}
By introducing a transverse null vector $w^i$ satisfying $w^2 = 0$, it can be written more compactly as
\begin{align}\label{eq:GFF_tower}
    \hatt_{s,A}(\hat x; w)
    :=
        \hatt_{s, A}^{i_1\cdots i_s}(\hat x) \,w_{\perp\,i_1}\cdots w_{\perp\,i_s}
    =
            \lim_{|x_\perp|\to 0}(w\cdot\partial_{\perp})^s\,\phi_A(\hat x, x_\perp)  \ .
\end{align}
The transverse spin-$s$ field $\hatt_{s,A}$ has dimension $\hD_{\hatt_s} = \Delta_\phi + s$ and the two-point function
\begin{align}\label{eq:GFF_tower_two_point}
    \langle\, \hatt_{s, A}(\hat x; w)\,\hatt_{s, B}(\hat y; w')\,\rangle
        =
                \delta_{AB}\,2^s\,\Gamma(s+1)\,(\Delta_\phi)_s\,\Cphi\,\frac{(w\cdot w')^s}{|\hat x - \hat y|^{2(\Delta_\phi + s)}} \ ,
\end{align}
where $(x)_s := \Gamma(x+s)/\Gamma(x)$, and $\hatt_{0, A} = \hphi_A$.
There are no more defect primaries linear in $\phi_A$ as acting with the parallel derivative $\hat\partial^{\hat a}$ yields the defect descendants.

Since $\phi_A$ are free fields, the two-point functions \eqref{eq:GFF_tower_two_point} are the only connected correlators of the tower $\{\hatt_{s, A}\}$, and all the higher-point functions are obtained by the Wick contraction.
In other words, the defect theory is a generalized free field theory in $p$ dimensions whose elementary fields are the tower $\hatt_{s, A}$ of dimension $\Delta_\phi + s$ transforming in the rank-$s$ symmetric traceless representation of the transverse rotation $\SO(q)$ and the vector representation of $\O(N)$.
Its operator content is therefore exhausted by the multi-twist operators built out of $\hatt_{s, A}$.

\subsection{Bilinear defect primaries}
\label{sec:GFF_bilinears}

Since the defect theory is a generalized free field theory of the defect primaries $\hatt_{s, A}$, defect bilinear operators are given by the double-twist operators $[\hatt_{s_1}\hatt_{s_2}]_{n, j}$ of dimension $\hD = 2\,\Delta_\phi + s_1 + s_2 + 2\,n + j$ with parallel spin-$j$ \cite{Fitzpatrick:2011dm}.
For a pair of defect scalar primaries $\hO_1$ and $\hO_2$ of dimension $\hD_1$ and $\hD_2$, the double-twist operators with $n=0$ and $j=1,2$ are given by \cite{Braun:2003rp}
\begin{align}
    [\hO_1\,\hO_2]_{0,1}^{\hat a}
        =
            \hD_1\,\normord{\hO_1(\hat\partial^{\hat a}\hO_2)}
            -
            \hD_2\,\normord{(\hat\partial^{\hat a}\hO_1)\hO_2} \ ,
\end{align}
and
\begin{align}
    [\hO_1\,\hO_2]_{0,2}^{\hat a \hat b}
        =
            \frac{\hD_1}{2(\hD_2+1)}\,\normord{\hO_1(\hat\partial^{\langle\hat a}\hat\partial^{\hat b\rangle}\hO_2)}
            +
            \frac{\hD_2}{2(\hD_1+1)}\,\normord{(\hat\partial^{\langle\hat a}\hat\partial^{\hat b\rangle}\hO_1)\hO_2}
            -
            \normord{(\hat\partial^{\langle\hat a}\hO_1)(\hat\partial^{\hat b\rangle}\hO_2)}
            \ ,
\end{align}
while the operator with $n=1, j=0$ is
\begin{align}
    [\hO_1\,\hO_2]_{1,0}
        =
            \frac{\hD_1}{2\hD_2 + 2-p}\normord{\hO_1(\hat\partial^2\hO_2)} + \frac{\hD_2}{2\hD_1 + 2-p}\normord{(\hat\partial^2\hO_1)\hO_2} - \normord{\hat\partial_{\hat a}\hO_1\, \hat\partial^{\hat a}\hO_2} \ .
\end{align}

\begin{table}[ht]
    \centering
    \setlength{\tabcolsep}{5pt} 
    \begin{tabular}{ccccc}
    \toprule
       primary  & $\O(N)$ & $\hat \Delta_{\hO}$ &
       $C_{\hO}/C_{\phi}^2$ & $\hat b_{\hO}/C_\phi$\\
    \midrule
    $\normord{\hphi\,\hphi} = [\hphi\,\hphi]_{0,0}$
      & S\,$\oplus$\,T & $d-2$ & $2\,N$
      & $2\,N\,\zeta(d-2)$ \\
    \cmidrule(lr){1-5}
    $\normord{\hphi\,\hatt_1^{\,i}} = [\hphi\,\hatt_1]^{i}_{0,0}$
      & S\,$\oplus$\,T\,$\oplus$\,A & \multirow{2}{*}{$d-1$} & $2\,N\,\Delta_\phi$ & $0$ \\
    $\hat J^{\,\hat a} = [\hphi\,\hphi]^{\hat a}_{0,1}$
      & A &  & $8\,\Delta_\phi^3$ & $0$ \\
    \cmidrule(lr){1-5}
    $[\hphi\,\hphi]_{1,0}$
      & S\,$\oplus$\,T & \multirow{7}{*}{$d$}  & $\displaystyle 8N\,p\,\Delta_\phi^2\left(1+\frac{2\Delta_\phi}{q}\right)$ & $\displaystyle 4N\,\Delta_\phi\,(2\Delta_\phi +q)\,\zeta(d)$ \\
    $\normord{\hatt_1\!\cdot\!\hatt_1}
      = \delta_{ij}[\hatt_1\,\hatt_1]^{ij}_{0,0}$
      & S\,$\oplus$\,T & & $8N\,q\,\Delta_\phi^2$
      & $4N\,q\,\Delta_\phi\,\zeta(d)$ \\
    $\normord{\hatt_1^{\langle i}\hatt_1^{\,j\rangle}}
      = [\hatt_1\hatt_1]^{\langle ij\rangle}_{0,0}$
     & S\,$\oplus$\,T & & $8N\,\Delta_\phi^2$ & $0$ \\
    $\normord{\hatt_1^{[i}\hatt_1^{\,j]}}
      = [\hatt_1\hatt_1]^{[ij]}_{0,0}$
     & A & & $8\,\Delta_\phi^2$ & $0$ \\
    $\normord{\hphi\,\hatt_2^{\,ij}} = [\hphi\,\hatt_2]^{ij}_{0,0}$
     & S\,$\oplus$\,T\,$\oplus$\,A & & $8N\,\Delta_\phi(\Delta_\phi+1)$ & $0$ \\
    $[\hphi\,\hatt_1]^{\hat a i}_{0,1}$
     & S\,$\oplus$\,T\,$\oplus$\,A & & $4N\,\Delta_\phi^2(\Delta_\phi+1)(2\Delta_\phi+1)$ & $0$ \\
    $[\hphi\,\hphi]^{\hat a\hat b}_{0,2}$
     & S\,$\oplus$\,T & & $\displaystyle\frac{8N\,\Delta_\phi^2(2\Delta_\phi+1)}{\Delta_\phi + 1}$ & $8N\,\Delta_\phi(2\Delta_\phi+1)\,\zeta(d)$ \\
    \bottomrule
    \end{tabular}
    \caption{The defect primaries bilinear in $\hphi_A$ with $\hD \le d$.
    The second column lists the decomposition to the $\O(N)$ irreducible representations S, T and A.
    For the rows containing the singlet, the primary column shows the flavor-singlet operators, and the normalizations $C_{\hO}$ of the two-point functions and the thermal one-point coefficients $\hat b_{\hO}$ are those of the singlet, equal to $N$ times the single-scalar values.
    The T and A representations of the same row have the coefficient $C_{\hO}/N$ in the convention \eqref{eq:GFF_nonsinglet_two_point} and have $\hat b_{\hO} = 0$ by the $\O(N)$ symmetry.
    The purely antisymmetric primaries $\hat J^{\,\hat a}$ and $\normord{\hatt_1^{[i}\hatt_1^{\,j]}}$ exist only for $N\ge2$, and their $C_{\hO}$ is the coefficient of $P_{\rm A}$.
    At $p=1$, the last row is absent due to $\hat\partial^{\langle\hat a}\hat\partial^{\hat b\rangle} = 0$.}
    \label{tab:bilinear_defect_primary}
\end{table}

A bilinear operator of two vector representations of $\O(N)$ decomposes into the three irreducible representations S, T and A as
\begin{align}
    \hO_{AB}
        =
            \frac{\delta_{AB}}{N}\,\hO_\text{S} + (\hO_\text{T})_{AB} + (\hO_\text{A})_{AB} \ ,
\end{align}
where $\hO_\text{S}$, $\hO_\text{T}$ and $\hO_\text{A}$ are the singlet, symmetric traceless and anti-symmetric operators defined by
\begin{align}
    \hO_\text{S}
        &:=
            \hO_C^{~C} \ ,\\
    (\hO_\text{T})_{AB}
        &:=
            \frac{\hO_{AB} + \hO_{BA}}{2} - \frac{\delta_{AB}}{N}\,\hO_C^{~C} \ , \\
    (\hO_\text{A})_{AB}
        &:=
            \frac{\hO_{AB} - \hO_{BA}}{2} \ .
\end{align}
For the two-point functions in the T and A representations, we adopt the following convention
\begin{align}\label{eq:GFF_nonsinglet_two_point}
    \langle\,(\hO_\text{T/A})_{AB}(\hat x)\,(\hO_\text{T/A})_{CD}(0)\,\rangle
        =
            C_{\hO_\text{T/A}}\,\frac{(P_\text{T/A})_{AB, CD}}{|\hat x|^{2\hD_{\hO}}} \ ,
\end{align}
up to the tensor structures associated with the parallel and transverse spins.
The flavor projectors $P_\text{T/A}$ are defined by
\begin{align}
    (P_\text{T})_{AB, CD}
        &=
            \frac{\delta_{AC}\,\delta_{BD} + \delta_{AD}\,\delta_{BC}}{2} - \frac{\delta_{AB}\,\delta_{CD}}{N}\ , \\
    (P_\text{A})_{AB, CD}
        &=
            \frac{\delta_{AC}\,\delta_{BD} - \delta_{AD}\,\delta_{BC}}{2} \ .
\end{align}
For instance, the Wick contraction gives
\begin{align}\label{eq:GFF_phiphi_flavor}
    \langle\,\normord{\hphi_A\hphi_B}(\hat x)\,\normord{\hphi_C\hphi_D}(0)\,\rangle
        =
            \frac{\Cphi^2}{|\hat x|^{4\Delta_\phi}}\left(\delta_{AC}\,\delta_{BD}+\delta_{AD}\,\delta_{BC}\right) ,
\end{align}
so that the singlet $\hphi^2$ has $C_{\hO}  = 2\,N\,\Cphi^2$, while the symmetric traceless channel $\normord{\hphi_{A}\hphi_{B}} - \delta_{AB}\,\hphi^2/N$ has $C_{\hO} = 2\,\Cphi^2$.

From the bulk conserved current $J^\mu_{AB}$ given in \eqref{O(N)_current}, one finds two defect primaries in the A representation:
\begin{align}\label{eq:GFF_defect_current}
    \hat J^{\,\hat a}_{AB}
        &:=
             \Delta_\phi\,\lim_{|x_\perp|\to 0} J^{\hat a}_{AB}
        =
            [\hphi\,\hphi]^{\hat a}_{[AB],\,0,1} \ , \\
    ([\hphi\,\hatt_1]^i_{0,0})_{[AB]}
        &=
            \frac{1}{2}\lim_{|x_\perp|\to0} J^{i}_{AB} \ ,
\end{align}
both of which have dimension $\hD_{\hO} = 2\Delta_\phi + 1 = d-1$.

Table \ref{tab:bilinear_defect_primary} lists the double-twist defect operators of dimension $\hat\Delta\le d$ with their $\O(N)$ decompositions to the irreducible representations.
At dimension $d$ one finds five defect scalars $[\hphi\,\hphi]_{1,0},\, \normord{\hatt_1\cdot\hatt_1}, \, \normord{\hatt_1^{\langle i}\hatt_1^{\,j\rangle}},\, \normord{\hatt_1^{[ i}\hatt_1^{\,j]}},\, \normord{\hphi\,\hatt_2^{\,ij}}$, one defect vector $[\hphi\,\hatt_1]^{\hat a i}_{0,1}$, and one defect symmetric traceless tensor $[\hphi\,\hphi]^{\hat a\hat b}_{0,2}$ for $p\ge 2$.
Note that $\normord{\hatt_1^{[ i}\hatt_1^{\,j]}}$ that is anti-symmetric both in the transverse and flavor indices exists only for $N\ge 2$. 
For $p=1$, the defect tensor $[\hphi\,\hphi]^{\hat a\hat b}_{0,2}$ is absent as $\hat\partial^{\langle\hat a}\hat\partial^{\hat b\rangle} = 0$.
The two-point normalizations $C_{\hO}$ in the table follow from \eqref{eq:GFF_tower_two_point} by the Wick contraction, and the thermal one-point coefficients $\hat b_{\hO}$ will be discussed in section \ref{sec:GFF_thermal_one_point}.
Multi-twist operators of $\hatt_{s, A}$ also exist, and some of them may have dimension $\hD \le d$ for lower $d$.
For example, $\hphi^n$ with $n\ge 3$ have $\hD_{\hphi^n} = n\,\Delta_\phi \le d$ when $n \le \frac{2d}{d-2}$.
We do not include them in table \ref{tab:bilinear_defect_primary} as they do not appear in the DOE of the bulk operators in table \ref{tab:GFF_bulk_primary}.

\subsection{Defect operators from the bulk stress tensor}
\label{sec:GFF_defect_stress_tensor}

The bulk stress tensor $T^{\mu\nu}$ yields defect local operators of dimension $d$ that are irreducible under the $\SO(p)\times \SO(q)$ symmetry:
\begin{align}\label{defect_stress_tensor_primaries}
    \hat{\CT}(\hat{x})
        &:=
            \lim_{|x_\perp|\to 0}\,T_i^{~i}(\hat{x}, x_\perp)\ ,\\
    \hat{\CT}^{\hat{a}\hat{b}}(\hat{x})
        &:=
            \lim_{|x_\perp|\to 0}\left[\,T^{\hat{a}\hat{b}}(\hat{x}, x_\perp) + \frac{\delta^{\hat{a}\hat{b}}}{p}\,T_i^{~i}(\hat{x}, x_\perp)\,\right] \ ,\\
    \hat{\CT}^{ij}(\hat{x})
        &:=
            \lim_{|x_\perp|\to 0}\left[\,T^{ij}(\hat{x}, x_\perp) - \frac{\delta^{ij}}{q}\,T_k^{~k}(\hat{x}, x_\perp)\,\right] \ ,\\
    \hat{\CT}^{\hat{a} i}(\hat{x})
        &:=
            \lim_{|x_\perp|\to 0}\,T^{\hat{a} i}(\hat{x}, x_\perp)\ ,
\end{align}
where $T^{\mu\nu}$ is the stress tensor \eqref{O(N)_scalar_stress_tensor} of the free scalar $\phi_A$.
These operators are not new defect primaries of dimension $d$ and they can be  written as
\begin{align}
    \hat{\CT}
        &=
            \frac{1}{2(d-1)}\left[ \,p\,\normord{\hatt_1\cdot\hatt_1} + q\,[\hphi\,\hphi]_{1,0}\right] \ , \label{ST_decomposition1} \\
    \hat{\CT}^{\hat a\hat b}
        &=
            - \frac{d}{2(d-1)}\,[\hphi\,\hphi]_{0,2}^{\hat a\hat b} \ ,\label{ST_decomposition2} \\
    \hat{\CT}^{ij}
        &=
           \frac{1}{2(d-1)}\left[\, d\,\normord{\hatt_1^{\langle i}\,\hatt_1^{j\rangle}} - (d-2)\,\normord{\hphi\,\hatt_2^{ij}}\right] \ ,\label{ST_decomposition3}  \\
    \hat{\CT}^{\hat a i}
        &=
           - \frac{1}{d-1}\,[\hphi\,\hatt_1]_{0,1}^{\hat a i} \ .\label{ST_decomposition4}
\end{align}
The bilinear primaries on the right hand sides are the $\O(N)$ singlet part of table \ref{tab:bilinear_defect_primary}.

\subsection{Bulk-defect operator expansion}
\label{sec:GFF_DOE}

The DOE can be extracted from the two-point function of a bulk local operator with defect local operators.
The only non-vanishing two-point functions of $\phi_A$ with the defect primaries are those with $\hatt_{s, A}$,
\begin{align}
    \langle\,\phi_A(\hat x, x_\perp)\,\hatt_{s, B}(\hat y; w)\,\rangle
        =
            \delta_{AB}\,2^s\,(\Delta_\phi)_s\,\Cphi\,\frac{(w\cdot x_\perp)^s}{\left( |\hat x - \hat y|^2 + |x_\perp|^2\right)^{\Delta_\phi + s}} \ ,
\end{align}
and hence the DOE of $\phi_A$ is given by
\begin{align}\label{GFF_phi_DOE}
    \phi_A(x)
        =
            \sum_{s=0}^\infty\,\frac{1}{\Gamma(s+1)}\,\hatt_{s, A}(\hat x; x_\perp) + (\text{descendants})\ .
\end{align}
Note that the same expansion can be obtained by Taylor expanding the free field $\phi_A(\hat x, x_\perp)$ around $|x_\perp| = 0$.
In particular, the defect identity operator $\hat{\bm{1}}$ does not appear in \eqref{GFF_phi_DOE} as $\langle\,\phi_A(x)\,\rangle = 0$.

The DOE of the bulk flavor-singlet operator $\normord{\phi^2}$ is fixed by either calculating the bulk-defect two-point functions with the defect bilinear primaries or Taylor expanding the composite operator $\normord{\phi^2}$ around $|x_\perp| = 0$:
\begin{align}\label{GFF_phi2_DOE}
    \begin{aligned}
        \normord{\phi^2}(x)
            ={}&
                \hphi^2
                +
                2\,x_{\perp i}\,\normord{\hphi\,\hatt_{1}^i}
                 \\
            &
            ~
            +
                x_{\perp i}x_{\perp j}
                \left[\normord{\hatt_{1}^{\langle i}\,\hatt_{1}^{j\rangle}} + \normord{\hphi\,\hatt_{2}^{ij}}\right]
            +
                |x_\perp|^2\left[\,\frac{1}{q}\,\normord{\hatt_1\cdot\hatt_1} - \frac{1}{2\Delta_{\phi} + q}\,[\hphi\,\hphi]_{1,0}  \right] + \cdots\ ,
    \end{aligned}
\end{align}
where only defect primaries (up to $d$ for the bilinears) are displayed explicitly and the bilinears are the $\O(N)$ singlets.
The defect primary $\normord{\phi_{\langle A}\phi_{B\rangle}}$ in the $\O(N)$ symmetric traceless representation has a similar expansion as that of $\normord{\phi^2}$ with identical coefficients.
The bulk $\O(N)$ current \eqref{O(N)_current} expands into the defect operators in the $\O(N)$  anti-symmetric representation,
\begin{align}\label{GFF_J_DOE}
    J^{\hat a}_{AB}(x)
        &=
            \frac{1}{\Delta_\phi}\,\hat J^{\,\hat a}_{AB}
            + \cdots \ , \\
    \qquad
    J^{i}_{AB}(x)
        &=2
            ([\hphi\,\hatt_1]^i_{0,0})_{[AB]}
            + \cdots \ ,
\end{align}
with no identity channel, as required by the unbroken $\O(N)$ symmetry.
Similarly, the DOE of the stress tensor reads
\begin{align}\label{GFF_T_DOE}
    \begin{aligned}
        T^{\mu\nu}(x)
            &=
                \frac{1}{p\,q}\,\sfs_1^{\mu\nu}\,\hat\CT
                +
                \delta_{\hat a}^{\mu}\,\delta_{\hat b}^{\nu}\,\hat\CT^{\hat a \hat b}
                +
                2\, \delta_{\hat a}^{(\mu}\,\delta_{i}^{\nu)}\,\hat\CT^{\hat a i}
                +
                 \delta_{i}^{\mu}\,\delta_{j}^{\nu}\,\hat\CT^{ij}
                +
                \cdots \ ,
    \end{aligned}
\end{align}
where the defect primaries of dimension $d$ in \eqref{defect_stress_tensor_primaries} are displayed explicitly.

\subsection{Thermal one-point functions}
\label{sec:GFF_thermal_one_point}

Since the bulk theory is free, thermal one-point functions can be calculated straightforwardly by applying the Wick's theorem and the method of images.
The thermal two-point function of $\phi_A$ is
\begin{align}\label{GFF-phi-two-point}
    \langle\,\phi_A(x)\,\phi_B(y)\,\rangle_{\beta}
        =
            \delta_{AB}\,\sum_{n\in \BZ}\,\frac{C_{\phi}}{\left[\,(\tau_x - \tau_y + n\,\beta)^2 + |\vec{x} - \vec{y}|^2\,\right]^{\Delta_\phi}} \ ,
\end{align}
from which the thermal one-point function of the bulk composite operator $\normord{\phi^2}(x)$ follows by subtracting the $n=0$ term and taking the coincident limit:
\begin{align}\label{GFF-phi2}
    \langle\,\normord{\phi^2}(x)\,\rangle_{\beta}
        =
            \frac{b_{\phi^2}}{\beta^{2\Delta_\phi}} \ ,
\end{align}
with 
\begin{align}
    b_{\phi^2}
        =
        2\,N\,\zeta(d-2)\,C_{\phi} \ .
\end{align}
By taking the derivative of the two-point function \eqref{GFF-phi-two-point}, we obtain the thermal one-point function of the stress tensor \eqref{O(N)_scalar_stress_tensor}:
\begin{align}\label{free_scalar_bT}
    \langle\,T^{\mu\nu}(x)\,\rangle_{\beta}
        =
            \frac{b_T}{\beta^d}\,\left[\, e^{\mu}e^{\nu} - \frac{\delta^{\mu\nu}}{d}\,\right]     \ ,
            \quad
    b_{T}
        =
            -2\,N\,d(d-2)\,\zeta(d)\,C_{\phi} \ ,
\end{align}
where $e^\mu$ is the unit vector along the thermal circle.
Both \eqref{GFF-phi2} and \eqref{free_scalar_bT} are independent of the position $x$, as they must be in the absence of a non-trivial defect.

The thermal one-point functions of $\phi_A$ and of the $\O(N)$ non-singlet bulk primaries, such as $\normord{\phi_{\langle A}\phi_{B\rangle}}$ and $J^{\mu}_{AB}$, vanish by the $\O(N)$ symmetry, as summarized in the last column of table \ref{tab:GFF_bulk_primary}.

Turning to the defect local operators, the thermal one-point functions of odd powers of $\hphi_A$ vanish, $\langle\,\hphi_{A_1}\cdots \hphi_{A_{2k+1}}\,\rangle_{\beta} = 0$, since the defect primary $\hphi_A$ behaves as a generalized free field.
Also, the thermal one-point functions of the operators with transverse spins vanish due to the rotational invariance in the transverse direction to the defect.
Similarly, those of the $\O(N)$ non-singlet (T and A) vanish due to the $\O(N)$ symmetry.

The thermal one-point functions of the defect primaries bilinear in $\hphi_A$ can be read off from \eqref{GFF-phi-two-point} by acting with derivatives and taking the limits $y \to x$ and $|x_\perp| \to 0$.
For instance, 
\begin{align}\label{GFF_t1t1_thermal}
    \langle\,\normord{\hatt_{1\, A}^i\,\hatt_{1, B}^j}\,(\hat x)\,\rangle_{\beta}
        =
            \delta^{ij}\,\delta_{AB}\,\frac{4\,\Delta_\phi\,C_\phi\,\zeta(d)}{\beta^{d}} \ ,
\end{align}
from which we can read off the coefficients ${\hat b}_{\hO}$ for the trace part and symmetric traceless part in the transverse indices and in the flavor indices as shown in table \ref{tab:bilinear_defect_primary}.
The other non-vanishing coefficients are calculated similarly.
For later reference, we list thermal one-point coefficients of \eqref{ST_decomposition1}--\eqref{ST_decomposition4}:
\begin{align}\label{ST_decompose_b_coefficient}
    \begin{aligned}
        \hat b_{\hat{\CT}}
            &=
                -\frac{q}{d}\,b_T \ , \\
        \hat b_{\hat{\CT}^{\hat a\hat b}}
            &=
                b_T \ , \\
        \hat b_{\hat{\CT}^{ij}}
            &=
                \hat b_{\hat{\CT}^{\hat a i}}
            =
                0 \ ,
    \end{aligned}
\end{align}
where the coefficient $b_T$ is given by \eqref{free_scalar_bT}.

\subsection{Consistency of the low temperature expansion}
\label{sec:GFF_low_temperature}

For the bulk scalar operator $\phi_A$, the defect primaries appearing in the DOE \eqref{GFF_phi_DOE} are linear in $\hphi_A$ and vanish in the thermal expectation value, hence $\langle\,\phi_A(x)\,\rangle_{\beta} = 0$ as expected.

For the composite operator $\normord{\phi^2}$ of dimension $\Delta_\CO = d-2$, the only defect primary of dimension $d-2$ in the DOE \eqref{GFF_phi2_DOE} is $\hphi^2$, and the terms of dimension $d-1$ and the $x_{\perp i}x_{\perp j}$ term of dimension $d$ have transverse spin so that their thermal one-point functions vanish.
The remaining term of dimension $d$ is
\begin{align}
    |x_\perp|^2\left[\,\frac{1}{q}\,\frac{\hat b_{\normord{\hatt_1\cdot\hatt_1}}}{\beta^d} - \frac{1}{2\Delta_\phi + q}\,\frac{\hat b_{[\hphi\,\hphi]_{1,0}}}{\beta^d}  \right]
        =
            \frac{|x_\perp|^2}{\beta^d}\,\left[\,4N\,\Delta_\phi\,\zeta(d)\,C_\phi - 4N\,\Delta_\phi\,\zeta(d)\,C_\phi\,\right]
        =
            0 \ ,
\end{align}
where we substituted the thermal coefficients $\hat b_{\hO}$ in table \ref{tab:bilinear_defect_primary}.
The two dimension-$d$ defect scalars thus cancel each other and we are left with
\begin{align}
    \langle\,\normord{\phi^2}(x)\,\rangle_{\beta}
        =
            \frac{\hat b_{\hphi^2}}{\beta^{d-2}}
            =
            \frac{2\,N\,\zeta(d-2)\,C_\phi}{\beta^{d-2}} \ ,
\end{align}
which reproduces the exact result \eqref{GFF-phi2}.

For the stress tensor $T^{\mu\nu}$ of dimension $\Delta_\CO = d$, the DOE \eqref{GFF_T_DOE} yields
\begin{align}
    \begin{aligned}
        \langle \, T^{\mu\nu}(x)\,\rangle_{\beta}
            &=
                \frac{1}{p\,q}\,\sfs_1^{\mu\nu}
                \,\frac{\hat b_{\hat\CT}}{\beta^d}
             -
                \frac{1}{p}\,\sfs_3^{\mu\nu}\,\frac{\hat b_{\hat\CT^{\hat a \hat b}}}{\beta^d}
                \\
            &=
                \frac{b_T}{\beta^d}\left( e^\mu e^\nu - \frac{\delta^{\mu\nu}}{d}\right)
                \ ,
    \end{aligned}
\end{align}
where we used the coefficients \eqref{ST_decompose_b_coefficient} in the second line.
This agrees with the exact result \eqref{free_scalar_bT}, and the contributions of the defect primaries of dimension $\hD_{\hO} > d$ must cancel as they come with positive powers of $|x_\perp|$.
These results may be seen as a non-trivial check of the validity of the low temperature expansion.

\section{Free theories with a boundary}\label{sec:free_BCFT}

In this section, we consider a boundary CFT (BCFT), which is a special case with $q=1$ of the general setup in section \ref{sec:thermal_DCFT}.
We consider the free scalar theory in section \ref{ss:free scalar BCFT} and the free (Dirac) fermion in section \ref{sec:fermion_BCFT}.
In what follows, the boundary is located at $x_{\perp} = 0$ and the bulk theory is defined in $x_{\perp} \geq 0$.

\subsection{Free scalar field}\label{ss:free scalar BCFT}
The first example is a single free scalar $\phi$ (i.e., $N=1$ in \eqref{eq:bulk_free_scalar_action}) with the following conformal boundary conditions:
\begin{alignat}{3}
\begin{aligned}
    \text{Neumann}&:&\quad
    \partial_{\perp}\phi\left(x_{\perp}=0\right)
        &=
            0 \ ,
    \\
    \text{Dirichlet}&:&\quad
    \phi\left(x_{\perp}=0\right)
        &=
            0  \ .
\end{aligned}
\end{alignat}
The thermal bulk two-point function can be computed by the method of images in both the time and normal directions:
\begin{align}\label{eq:scalar_BCFT_bulk_2pt}
\begin{aligned}
    \langle\,\phi(x_1)\,&\phi(x_2)\,\rangle_{\pm,\,\beta} \\
        &=
            \sum_{n\in \BZ}\left(\frac{C_{\phi}}{\left[(\tau_1 - \tau_2 + n\,\beta)^2 + |\vec{x}_1 - \vec{x}_2|^2\right]^{\Delta_{\phi}}} \pm \frac{C_{\phi}}{\left[(\tau_1 - \tau_2 + n\,\beta)^2 + |\vec{x}_1 - \vec{\bar{x}}_2|^2\right]^{\Delta_{\phi}}}\right) \\
        &=
            \langle\,\phi(x_1)\,\phi(x_2)\,\rangle_{\beta}
            \pm
            \sum_{n\in \BZ}\frac{C_{\phi}}{\left[(\tau_1 - \tau_2 + n\,\beta)^2 + |\vec{x}_1 - \vec{\bar{x}}_2|^2\right]^{\Delta_{\phi}}}
    \ , 
\end{aligned}   
\end{align}
where $\vec{\bar{x}} \coloneqq (x_{\parallel}, -x_{\perp})$, and the upper and lower signs correspond to the Neumann and Dirichlet boundary conditions, respectively.
Based on this result, we calculate the one-point functions of $\phi^2$ and $T^{\mu\nu}$.

\subsubsection{Boundary local operators}\label{sec:scalar_bdy_op}
Before computing the bulk one-point functions, we summarize the boundary local operators appearing in the BOE of $\phi$, $\phi^2$ and $T^{\mu\nu}$ (see \cite{Liendo:2012hy,Herzog:2017xha} for the derivation of BCFT data from the conformal block decomposition of bulk two-point functions).

\begin{table}[ht]
    \centering
    \begin{tabular}{ccccc}
    \toprule
       bulk primary & $\hat{\Delta}_{\hat{\CO}}$ & $j$ & $b^2_{\CO\hat{\CO}}C_{\hat{\CO}}/C_{\CO}$ & $\hat{b}^2_{\hat{\CO}}/C_{\hat{\CO}}$ \\
    \midrule
    \multirow{2}{*}{$\phi$} & $\frac{d-2}{2}$ (Neumann) & $0$ & $2$ & $0$ \\
    & $\frac{d}{2}$ (Dirichlet) & $0$ & $2(d-2)$ & $0$ \\
    \cmidrule(lr){1-5}
    \multirow{3}{*}{$\phi^2$} & $0$ & $0$ & $2^{3-2d}$ & $1$ \\
    & $d-2$ & $0$ & $2(1\pm1)$ & $2\,\zeta(d-2)^2$ (Neumann) \\
    & $d-2+2\,m$ & $0$ & \eqref{eq:BOE_coeff_phi2} & \eqref{eq:bdy_thermal_1pt_scalar} \\
    \cmidrule(lr){1-5}
    \multirow{5}{*}{$T^{\mu\nu}$} & $d$ & $0$ & $\frac{2}{d(d-1)}$ & $2\,\zeta(d)^2$ \\
    & $d$ & $2$ & $2(1\pm1)$ & $4d(d-1) \zeta (d)^2$ (Neumann) \\
    & $d+2$ & $2$ & $\frac{4 d(d-1)(d+2)}{d+1}$ & $4(d+1)(d+2)\zeta(d+2)^2$\\
    & $d+4$ & $2$ & $\frac{4d(d-1)(d+2)^2(d+4)}{3(d+7)}$ & $12 (d+4) (d+7) \zeta (d+4)^2$\\
    & $d+2\,m$ & $2$ & \eqref{eq:BOE_coeff_stress_fscalar} & \eqref{eq:scalarBCFT_thermal_1pt_spin2} \\
    \bottomrule
    \end{tabular}
    \caption{The BCFT data appearing in the BOE of $\phi$, $\phi^2$ and $T^{\mu\nu}$.
    The scaling dimension, spin, the BOE coefficient, and the thermal one-point function.
    The upper and lower signs correspond to the Neumann and Dirichlet boundary conditions, respectively.
    The two-point coefficients of $\phi^2$ and $T^{\mu\nu}$ in the CFT without a boundary are given by $C_{\phi^2}=2\,C_{\phi}^2$, $C_{T}=\frac{d(d-2)^2}{(d-1)}C_{\phi}^2$, respectively.}
    \label{tab:bdy_primary_scalar}
\end{table}

We begin with the BOE of $\phi$, which contains only one operator, $\hat{\phi}$ or $\hat{t}_1 \coloneqq \widehat{\partial_{\perp}\phi}$ with dimension $\hat{\Delta}_{\pm} = \frac{d-1}{2} \mp \frac{1}{2}$
\begin{align}\label{scalar_BCFT_phi_BOE}
    \phi(\hat{x},x_{\perp})
        =
            \begin{dcases}
                b_{\phi\hat{\phi}}\,\hat{\phi} + (\text{descendants}) \quad &(\text{Neumann})\ ,\\
                b_{\phi\hat{t}_1}\,x_{\perp}\,\hat{t}_1 + (\text{descendants}) \quad &(\text{Dirichlet})\ ,
            \end{dcases}
\end{align}
where $b_{\phi\hat{\phi}} = b_{\phi\hat{t}_1} = 1$.
Note that $\langle\phi\rangle_{\pm,\,\beta}$ vanishes due to the $\BZ_2$-symmetry.
We will refer to the BOE \eqref{scalar_BCFT_phi_BOE} in section \ref{ss:Free_O(N)_ep1}.

In the BOE of $\phi^2$, the boundary double-twist operator $[\hat{\CO}\hat{\CO}]_{n,0}$ appears, where $\hat{\CO}=\hat{\phi}$ for the Neumann and $\hat{\CO}=\hat{t}_1$ for the Dirichlet boundary condition:
\begin{align}
    \phi^2(\hat{x},x_{\perp})
        =
            \frac{a_{\phi^2}}{x_{\perp}^{d-2}}\,\hat{\bm{1}} + \sum_{n=0}^{\infty}\frac{b_{\phi^2[\hat{\CO}\hat{\CO}]_{n,0}}}{x_{\perp}^{\Delta_{\phi^2}-2\hat{\Delta}_{\pm}-2\,n}}[\hat{\CO}\hat{\CO}]_{n,0} + (\text{descendants}) \ .
\end{align}
The first term is the zero-temperature BCFT one-point function \cite{McAvity:1993ue}
\begin{align}\label{eq:a_phi2_BCFT}
    a_{\phi^2}
        =
            b_{\phi^2\hat{\bm{1}}}
        =
            \pm\frac{C_{\phi}}{2^{d-2}} \ .
\end{align}
The BOE coefficients of $[\hat{\CO}\hat{\CO}]_{n,0}$ with dimension $\hat{\Delta} = d-2+2\,m~(m=0,1,2,\cdots)$ in the second term are given by\footnote{For BOE of bulk scalar operators, our $b^2_{\CO\hat{\CO}}C_{\hat{\CO}}/C_{\CO}$ is the same as $\mu_{l}^2\,4^{\Delta_{\hat{\CO}}-\Delta_{\CO}}$ in \cite{Liendo:2012hy}.}
\begin{align}\label{eq:BOE_coeff_phi2}
    b^2_{\phi^2[\hat{\CO}\hat{\CO}]_{n,0}}\frac{C_{[\hat{\CO}\hat{\CO}]_{n,0}}}{C_{\phi^2}}
        &=
            (1\pm\delta_{m,0})\,\frac{4^{1+m}}{(2\,m)!}\frac{\Gamma\left(\frac{d-1}{2}+m\right)\Gamma\left(\frac{d}{2}+m-1\right)\Gamma\left(d+2\,m-3\right)}{\Gamma\left(\frac{d}{2}-1\right)\Gamma\left(d-2\right)\Gamma\left(\frac{d+4m-3}{2}\right)} \ ,       
\end{align}
where $n=m~(m=0,1,\cdots)$ for the Neumann and $n=m-1~(m=1,2,\cdots)$ for the Dirichlet boundary condition.
In particular, the boundary scalar with $\hat{\Delta}=d$ corresponds to the displacement operator \eqref{displacement_BCFT_definition}:
\begin{align}\label{eq:scalar BCFT displacement}
    \hat{D}
        =
            \begin{dcases}
                \frac{1}{2(d-1)}\,[\hat{\phi}\hat{\phi}]_{1,0} \quad &(\text{Neumann})\ ,\\
                \frac{1}{2}\,[\hat{t}_1\hat{t}_1]_{0,0} \quad &(\text{Dirichlet}) \ ,
            \end{dcases}
\end{align}
where we used the normalization of the double-twist operators in section \ref{sec:GFF_bilinears}.

The BOE of the stress tensor only includes the displacement operator $\hat{D}$ and boundary spin-two symmetric traceless tensors. 
The latter operators also correspond to double-twist operators $[\hat{\CO}\hat{\CO}]_{n,2}$.
Thus, the BOE of the stress tensor is written as
\begin{align}\label{eq:BOE_stress_BCFT}
    T^{\mu\nu}(\hat{x},x_{\perp})
        =
            \frac{d}{d-1}\left(\delta^{\mu n}\delta^{\nu n} - \frac{\delta^{\mu\nu}}{d}\right)\hat{D}(\hat{x}) + \sum_{n=0}^{\infty}\frac{b_{T[\hat{\CO}\hat{\CO}]_{n,2}}}{x_{\perp}^{d-2\hat{\Delta}_{\pm}-2\,n-2}}\,\delta^{\mu}_{\hat{a}}\delta^{\nu}_{\hat{b}}\,[\hat{\CO}\hat{\CO}]_{n,2}^{\hat{a}\hat{b}} + (\text{descendants}) \ .
\end{align}
In the Neumann case, we identify $[\hat{\CO}\hat{\CO}]_{0,2}=[\hat{\phi}\hat{\phi}]_{0,2}$ with $\hat{\CT}^{\hat{a}\hat{b}}$ in \eqref{defect_stress_tensor_primaries},\footnote{The operator $\hat{\CT}^{\hat{a}n}$ in \eqref{defect_stress_tensor_primaries} is absent for the physical boundary condition with no energy flow across the boundary, including our case. If extra degrees of freedom are present on the boundary, $\hat{\CT}^{\hat{a}n}$ can exist as a descendant rather than a primary operator \cite{Herzog:2017xha}.}
\begin{align}
    \hat{\CT}^{\hat{a}\hat{b}}(\hat{x})
        \coloneqq
            T^{\hat{a}\hat{b}} (\hat{x},x_{\perp})\big|_{x_{\perp} \rightarrow 0} + \frac{\delta^{\hat{a}\hat{b}}}{d-1}\hat{D}(\hat{x})
        =   
            -\frac{d}{2(d-1)}\,[\hat{\phi}\hat{\phi}]_{0,2} 
            \quad (\text{Neumann}) \ ,
\end{align}
while $\hat{\CT}^{\hat{a}\hat{b}}$ vanishes by the equation of motion for the Dirichlet case.
The BOE coefficients of the second term in \eqref{eq:BOE_stress_BCFT} are given by
\begin{align}
    \frac{b^2_{T[\hat{\phi}\hat{\phi}]_{0,2}} C_{[\hat{\phi}\hat{\phi}]_{0,2}}}{C_{T}}
        &=
            4 \ ,
    \\
    \frac{b^2_{T[\hat{\CO}\hat{\CO}]_{n,2}} C_{[\hat{\CO}\hat{\CO}]_{n,2}}}{C_{T}}
        &=
            \frac{\pi^{\frac{1}{2}} \, 2^{1-d} \,\Gamma (d+2\,m-1)\,\Gamma (d+2 m+2)}{(d+1)\,(2 m)!\,\Gamma \left(\frac{d}{2}+1\right) \Gamma (d-1)\,\Gamma \left(\frac{d+1}{2}+2\,m\right)} \ ,\label{eq:BOE_coeff_stress_fscalar}
\end{align}
where $n=m~(m=1,2\cdots)$ for the Neumann and $n=m-1~(m=1,2,\cdots)$ for the Dirichlet case, respectively.
Several values of the BOE coefficients for lower-dimensional boundary operators are listed in table \ref{tab:bdy_primary_scalar}.

\subsubsection{Thermal one-point function of $\phi^2$}\label{sec:free_scalar_BCFT_thermal_1pt_phi2}
The one-point function of $\phi^2$ is simply obtained by taking the coincident limit and removing the divergent zero-mode in the above two-point function: 
\begin{align}\label{eq:BCFTphi2}
\begin{aligned}
    \langle\,\phi^2(x)\,\rangle_{\pm,\,\beta}
        &=
            \langle\,\phi^2(x)\,\rangle_{\beta}
            \pm
            \frac{C_{\phi}}{\beta^{d-2}}\,
            \sum_{n\in \BZ}\frac{1}{\left( n^2 + 4\,z^2 \right)^{\frac{d-2}{2}}} \ ,
\end{aligned}
\end{align}
where the first term represents the thermal one-point function \eqref{GFF-phi2} in the CFT without a boundary.

\subsubsection*{(i) Low temperature limit $z \to 0$}

In the low temperature limit $z \rightarrow 0$, performing the binomial expansion in the summand yields
\begin{align}\label{eq:scalar_BCFT_scalar_1pt_lowT}
\begin{aligned}
    \langle\,\phi^2(x)\,\rangle_{\pm,\,\beta}
        &=
            \frac{C_{\phi}}{\beta^{d-2}}\left[2\,\zeta(d-2) \pm
            \left(\frac{1}{\left(2z\right)^{d-2}} + \sum_{m=0}^{\infty} \frac{2^{1+2\,m}\,\left(\frac{d-2}{2}\right)_{m}\,\zeta(d-2+2\,m)}{m!}(-z^2)^m\right)\right] \\[1.5ex]
        &= 
            \begin{dcases}
                \frac{C_{\phi}}{\left(2\,x_{\perp}\right)^{d-2}} + \frac{4\,C_{\phi}\,\zeta(d-2)}{\beta^{d-2}} - \frac{4\,(d-2)\,C_{\phi}\,\zeta(d)\,x_{\perp}^{2}}{\beta^{d}} + \beta^{-d+2}\,O(z^4) \ ,
                \\
                -\frac{C_{\phi}}{\left(2\,x_{\perp}\right)^{d-2}} + \frac{4\,(d-2)\,C_{\phi}\,\zeta(d)\,x_{\perp}^{2}}{\beta^{d}} + \beta^{-d+2}\, O(z^4) \ ,
            \end{dcases}
\end{aligned}
\end{align}
for the Neumann and Dirichlet boundary condition, respectively.
In terms of the BOE, the first terms in both cases come from the identity contributions or equivalently the one-point function \eqref{eq:a_phi2_BCFT} at zero temperature.
The second term for the Neumann boundary condition can be interpreted as the one-point function of the boundary operator $\hat{\phi}^2(\hat{x})$ 
\begin{align}\label{eq:scalar_BCFT_defect_phi2}
    \langle\,\hat{\phi}^2(\hat{x})\,\rangle_{+,\,\beta}
        =
            \frac{\hat{b}_{\hat{\phi}^2}}{\beta^{d-2}} \ ,\qquad
    \hat{b}_{\hat{\phi}^2}
        =
            4\,C_{\phi}\,\zeta(d-2)
    \ .
\end{align}
The Dirichlet boundary condition does not allow the existence of such a boundary operator of dimension $d-2$ as seen from the above expansion.
The other terms correspond to higher-dimensional operators with $\hat{\Delta}=d-2+2\,m~(m=1,2,\cdots)$ listed in table \ref{tab:bdy_primary_scalar}.
We can extract the thermal one-point function coefficient of all the operators for $m\ge 1$ appearing in the BOE of $\phi^2$ from \eqref{eq:BOE_coeff_phi2} and \eqref{eq:scalar_BCFT_scalar_1pt_lowT}:
\begin{align}\label{eq:bdy_thermal_1pt_scalar}
    \frac{\hat{b}^2_{[\hat{\CO}\hat{\CO}]_{n,0}}}{C_{[\hat{\CO}\hat{\CO}]_{n,0}}}
        =
            \frac{4^m \,\Gamma \left(\frac{d-1}{2}\right)\,\Gamma \left(m+\frac{1}{2}\right)\, \Gamma \left(\frac{d+4 m-3}{2}\right)\,\zeta (d+2 m-2)^2}{\pi^{\frac{1}{2}}\, \Gamma (m+1)\,\Gamma \left(\frac{d+2\,m-3}{2}\right)\, \Gamma \left(\frac{d-1}{2}+m\right)} \ ,
\end{align}
where $\hat{\CO}$ and the relation between $n$ and $m$ are defined in section \ref{sec:scalar_bdy_op}.
For example, the displacement operator, corresponding to $m=1$, has the following one-point function
\begin{align}\label{eq:D_thermal_1pt}
    \frac{\hat{b}^2_{\hat{D}}}{C_{\hat{D}}}
        =
            2\,\zeta(d)^2 \ ,
\end{align}
which can also be derived from the definition of the displacement operator.

\subsubsection*{(ii) High temperature limit $z \to \infty$}
Applying the identity \eqref{eq:app_poisson_sum} obtained through the Poisson resummation formula,
we find the following form at $z \rightarrow \infty$:
\begin{align}\label{eq:scalar_BCFT_scalar_1pt_highT}
\begin{aligned}
    \langle\,\phi^2(x)\,\rangle_{\pm,\,\beta}
        &=
            \frac{C_{\phi}}{\beta^{d-2}}\left[2\,\zeta(d-2) \pm
            \frac{\pi^{\frac{1}{2}}\,\Gamma\left(\frac{d-3}{2}\right)}{\Gamma\left(\frac{d-2}{2}\right)\,(2z)^{d-3}}
            + O\left(e^{-4\pi z}\right)\right] \\
        &=
            \frac{1}{\beta^{d-2}}\left[\,b_{\phi^2} + \frac{\tilde{a}_{\phi^2}}{z^{d-3}} + O\left(e^{-4\pi z}\right)\right]  \ ,
\end{aligned}
\end{align}
where $\tilde{a}_{\phi^2}$ is the dimensional reduction of the zero temperature one-point function \eqref{eq:a_phi2_BCFT}, i.e.\footnote{When $d=3$, both $b_{\phi^2}$ and $\tilde{a}_{\phi^2}$ are divergent. This is the IR divergence caused by the gapless zero mode. More details will be investigated in section \ref{ss:Free_O(N)_p1}.}
\begin{align}
    \tilde{a}_{\phi^2}
        \coloneqq
            a_{\phi^2}|_{d \to d-1}
        =
            \pm 2^{1-d}\,\pi ^{\frac{1-d}{2}}\,\Gamma \left(\frac{d-3}{2}\right) \ .
\end{align}

This high temperature behavior can be understood by the Kaluza-Klein (KK) dimensional reduction on the thermal circle.
Expanding the field $\phi$ in KK-modes 
\begin{align}
    \phi(x)
        =
            \frac{1}{\sqrt{\smash[b]{\mathstrut \beta}}}\,\sum_{n \in \BZ}\,e^{-\i\,\omega_{n}\tau}\phi_{n}(\vec{x}) 
        \ , \qquad
    \omega_n
        =
            \frac{2\pi n}{\beta} \ ,
\end{align}
and integrating over the time direction, the action reduces to
\begin{align}
    I
        =
            \int_{\BR^{d-1}_{+}}\d^{d-1}x\left[\,\frac{1}{2}\left(\vec{\partial}\phi_0\right)^2 + \sum_{n=1}^{\infty}\left(\,|\vec{\partial}\phi_n|^2 + \omega_n^2\,|\phi_n|^2\,\right)\,\right] \ ,
\end{align}
where we used $\phi_n = \phi_{-n}^{\ast}$ with the boundary conditions
\begin{alignat}{3}
\begin{aligned}
    \text{Neumann}&:&\quad
    \partial_{\perp}\phi_n\left(x_{\perp}=0\right)
        &=
            0 \ ,
    \\
    \text{Dirichlet}&:&\quad
    \phi_n\left(x_{\perp}=0\right)
        &=
            0  \ ,
\end{aligned}
\end{alignat}
for all $n \in \BZ$.
The zero mode $\phi_0$ is nothing but a $(d-1)$-dimensional massless free scalar at zero temperature, while the other modes $\phi_{n\neq0}$ are massive with mass $m_n^2=\omega_n^2$.
In terms of the KK modes, the original one-point function is written as
\begin{align}
    \langle\,\phi^2(x)\,\rangle_{\pm,\,\beta}
        =
            \frac{1}{\beta}\,\sum_{n,m\in\BZ}\,e^{-\i(\omega_n+\omega_m)\tau}\,\langle\!\langle\,\phi_n(\vec{x})\,\phi_{m}(\vec{y})\,\rangle\!\rangle_{\pm}\big|_{\vec{y}\to\vec{x}}
        =
            \frac{1}{\beta}\,\sum_{n\in\BZ}\,\langle\!\langle\,\phi_n(\vec{x})\,\phi_{-n}(\vec{y})\,\rangle\!\rangle_{\pm}\big|_{\vec{y}\to\vec{x}} \ ,
\end{align}
where $\langle\!\langle\rangle\!\rangle$ indicates the correlation functions in the $(d-1)$-dimensional theory, and the propagators of the KK modes are given by
\begin{align}
    \langle\!\langle\,\phi_n(\vec{x})\,\phi_{-n}(\vec{y})\,\rangle\!\rangle_{\pm}
        =
            \int\,\frac{\d^{d-2}p_{\parallel}}{(2\pi)^{d-2}}\,\frac{e^{\i\,p_{\parallel}\cdot(x_{\parallel}-y_{\parallel})}}{2\sqrt{p_{\parallel}^2+\omega_n^2}}\,\left(e^{-\sqrt{p_{\parallel}^2+\omega_n^2}\,|x_{\perp}-y_{\perp}|} \pm e^{-\sqrt{p_{\parallel}^2+\omega_n^2}\,|x_{\perp}+y_{\perp}|}\right) \ .
\end{align}
Since the first term is a purely bulk contribution, summing over all modes yields $\langle\,\phi^2(x)\,\rangle_{\beta}$ in the coincident limit $\vec{y} \to \vec{x}$.
On the other hand, the image term includes the boundary effects.
Among all modes, the zero mode gives the leading contribution which is just $\langle\,\phi^2(x)\,\rangle_{\pm}|_{d \to d-1}$ as expected, while the contributions from the massive modes are exponentially suppressed at high temperature.
For completeness, we perform the momentum integral for the image term in the coincident limit:
\begin{align}
    \int\,\frac{\d^{d-2}p_{\parallel}}{(2\pi)^{d-2}}\,\frac{e^{-2\sqrt{p_{\parallel}^2+\omega_n^2}\,x_{\perp}}}{2\sqrt{p_{\parallel}^2+\omega_n^2}}
        =
            \frac{1}{\pi^{\frac{1}{2}}(4\pi)^{\frac{d-2}{2}}}\,\left(\frac{|\omega_n|}{x_{\perp}}\right)^{\frac{d-3}{2}}\,K_{\frac{d-3}{2}}\left(2|\omega_n|\,x_{\perp}\right) \ .
\end{align}
In particular, we can reproduce \eqref{eq:scalar_BCFT_scalar_1pt_highT} from the asymptotic form of the modified Bessel function under $|\omega_n\,x_{\perp}| \to 0$.

\subsubsection{Thermal one-point function of $ T^{\mu\nu}$}
We turn to the thermal one-point function of the stress tensor given by \eqref{O(N)_scalar_stress_tensor} at $N=1$.
The one-point function takes the form as \eqref{eq:BCFT_stress_1pt} with 
\begin{align}\label{eq:scalar BCFT stress thermal 1-pt}
    F_3(z)
        =
           \pm C_{\phi}\frac{2\,d\,(d-2)}{d-1}\,\sum_{n=1}^{\infty}\left[\frac{1}{\left( n^2 + 4\,z^2 \right)^{\frac{d}{2}}} -\frac{4\,z^2}{\left( n^2 + 4\,z^2 \right)^{\frac{d+2}{2}}}\right] \ .
\end{align}
We again evaluate the asymptotic behavior in the two regimes: (i) $z \rightarrow 0$ and (ii) $z \rightarrow \infty$.
Both limits were studied for $d=4$ in \cite{Kennedy:1979ar}, whose results are reproduced in our calculation.

\subsubsection*{(i) Low temperature limit $z \to 0$}

In the low temperature limit, we obtain\footnote{
We use the following identity:
\begin{align}
    \sum_{n=1}^{\infty}\left[\frac{1}{\left( n^2 + 4\,z^2 \right)^{\frac{d}{2}}} - \frac{4\,z^2}{\left( n^2 + 4\,z^2 \right)^{\frac{d+2}{2}}}\right]
            &=
                 \sum_{m=0}^{\infty}\frac{(-4)^{m}\,\Gamma\left(\frac{d}{2}+m+1\right)\,\zeta(d+2\,m)}{\Gamma\left(\frac{d}{2}+1\right)\,\Gamma\left(m+1\right)}z^{2\,m} \ .    
\end{align}
}
\begin{align}\label{eq:stress_free_scalar_BCFT}
\begin{aligned}
    \langle\,T^{00}(x)\,\rangle_{\pm,\,\beta}
        &=
            \frac{b_T}{\beta^d}\, \frac{d-1}{d}
            \mp 
            C_{\phi}\frac{2\,d\,(d-2)^2}{d-1}\frac{1}{\beta^d}\,\left[\,\zeta(d) - 2\,(d+2)\,\zeta(d+2)\,z^2 + O(z^4)\,\right] \ ,
    \\
    \langle\,T^{ab}(x)\,\rangle_{\pm,\,\beta}
        &=
            -\frac{b_T}{\beta^d}\,\frac{\delta^{ab}}{d}
            \pm
            C_{\phi}\frac{2\,d\,(d-2)}{d-1}\frac{\delta^{ab}}{\beta^d}\,\left[\,\zeta(d) - 2\,(d+2)\,\zeta(d+2)\,z^2 + O(z^4)\right] \ ,
    \\
    \langle\,T^{nn}(x)\,\rangle_{\pm,\,\beta}
        &=
            -\frac{b_T}{\beta^d}\,\frac{1}{d} \ .
\end{aligned}
\end{align}
For example, in $d=4$, the time component becomes
\begin{align}
    \langle\,T^{00}(x)\,\rangle_{\pm,\,\beta}
        =
            \begin{dcases}
                -\frac{17\,\pi^2}{270\,\beta^4} + \frac{32\,\pi^4\, z^2}{945\,\beta^4} + \beta^{-4}\,O(z^4) \quad \text{(Neumann)} \ , \vspace{4pt}\\
                -\frac{\pi^2}{270\,\beta^4} - \frac{32\,\pi^4\, z^2}{945\,\beta^4} + \beta^{-4}\,O(z^4) \quad \text{(Dirichlet)} \ ,
            \end{dcases}
\end{align}
which reproduces eqs.\,(3.24a,b) in \cite{Kennedy:1979ar}, up to the Wick rotation.

In this regime, we can compare with the BOE of the stress tensor \eqref{eq:BOE_stress_BCFT} by taking the thermal expectation value.
From the normal components of \eqref{eq:stress_free_scalar_BCFT}, we extract the one-point function of the displacement  as
\begin{align}\label{Dirichlet displacement thermal 1-pt}
    \langle\,\hat{D}(\hat{x})\,\rangle_{\pm,\,\beta}
        &=
            \frac{\hat{b}_{\hat{D}}}{\beta^d}
    \ , \qquad
    \hat{b}_{\hat{D}}
        =
            -\frac{b_T}{d} \ ,
\end{align}
which agrees with the result in \eqref{eq:D_thermal_1pt}, where the corresponding two-point function of the displacement operator is
\begin{align}\label{eq:D_2pt_scalar_BCFT}
    C_{\hat{D}}
        =
            2\,(d-2)^2\,C^2_{\phi} \ .
\end{align}
After subtracting the contribution of the displacement operator from \eqref{eq:stress_free_scalar_BCFT}, only the spin-two symmetric traceless tensors $[\hat{\CO}\hat{\CO}]_{n,2}$ remain.
In the Neumann case, the leading term after the subtraction is proportional to $\beta^{-d}$, from which the one-point function of $\hat{\CT}^{\hat{a}\hat{b}} \propto [\hat{\phi}\hat{\phi}]_{0,2}$ is determined as
\begin{align}\label{eq:scalarBCFT_bdy_T_1pt}
    \langle\,\hat{\CT}^{\hat{a}\hat{b}}(\hat{x})\,\rangle_{+,\,\beta}
        &=
            \frac{\hat{b}_{\hat{\CT}^{\hat{a}\hat{b}}}}{\beta^d}\left(e^{\hat{a}}e^{\hat{b}} - \frac{\delta^{\hat{a}\hat{b}}}{d-1}\right)
    \ , \qquad
    \hat{b}_{\hat{\CT}^{\hat{a}\hat{b}}}
        =
            2\,b_{T} 
    \ .
\end{align}
In contrast, the leading term behaves as $z^2\,\beta^{-d}$ in the Dirichlet case, which is consistent with the absence of $\hat{\CT}^{\hat{a}\hat{b}}$, as we mentioned in section \ref{sec:scalar_bdy_op}.
The one-point functions of higher dimensional operators with $\hat{\Delta}=d+2\,m~(m=1,2,\cdots)$ are encoded in the higher-order terms of $z$ in \eqref{eq:stress_free_scalar_BCFT}
\begin{align}
    b_{T[\hat{\CO}\hat{\CO}]_{n,2}}\hat{b}_{[\hat{\CO}\hat{\CO}]_{n,2}}
        =
            \mp\frac{(-1)^{m}\,2^{2\,m+1}\,\Gamma \left(\frac{d}{2}+m+1\right)\zeta (d+2 m)}{\pi ^{\frac{d}{2}}\,\Gamma (m+1)} \ ,
\end{align}
where $n=m$ for the Neumann, $n=m-1$ Dirichlet boundary condition, respectively.
Then, combining with the BOE coefficients \eqref{eq:BOE_coeff_stress_fscalar}, we can read off the thermal one-point function
\begin{align}\label{eq:scalarBCFT_thermal_1pt_spin2}
    \frac{\hat{b}^2_{[\hat{\CO}\hat{\CO}]_{n,2}}}{C_{[\hat{\CO}\hat{\CO}]_{n,2}}}
        =
            \frac{2^{d+4m+1}\,\Gamma \left(\frac{d+3}{2}\right) \Gamma \left(m+\frac{1}{2}\right) \Gamma \left(\frac{d}{2}+m+1\right) \Gamma \left(\frac{d+1}{2} +2\,m\right) \zeta (d+2 m)^2}{\pi\,m!\,\Gamma \left(\frac{d+3}{2}+m\right) \Gamma (d+2 m-1)} \ .
\end{align}

\subsubsection*{(ii) High temperature limit $z \to \infty$}

We again apply the Poisson summation formula in the same manner as in the derivation of \eqref{eq:scalar_BCFT_scalar_1pt_highT}. 
Since the calculation proceeds analogously, we present only the final results:
\begin{align}
\begin{aligned}
    \langle\,T^{00}(x)\,\rangle_{\pm,\,\beta}
        &=
            \frac{b_T}{\beta^d}\, \frac{d-1}{d}
            \mp 
            C_{\phi}\frac{(d-2)^2}{d-1}\frac{1}{\beta^d}\,\left[\frac{\pi^{\frac{1}{2}}\,\Gamma\left(\frac{d-1}{2}\right)}{\Gamma\left(\frac{d}{2}\right)\,(2z)^{d-1}} + O\left(e^{-4\pi z}\right)\right]
    \\
    \langle\,T^{ab}(x)\,\rangle_{\pm,\,\beta}
        &=
            -\frac{b_T}{\beta^d}\,\frac{\delta^{ab}}{d}
            \pm
            C_{\phi}\frac{d-2}{d-1}\frac{\delta^{ab}}{\beta^d}\,\left[\frac{\pi^{\frac{1}{2}}\,\Gamma\left(\frac{d-1}{2}\right)}{\Gamma\left(\frac{d}{2}\right)\,(2z)^{d-1}} + O\left(e^{-4\pi z}\right)\right]
    \\
    \langle\,T^{nn}(x)\,\rangle_{\pm,\,\beta}
        &=
            -\frac{b_T}{\beta^d}\,\frac{1}{d} \ .
\end{aligned}
\end{align}
For $d=4$, we find a perfect agreement with eq.\,(3.22a,c) in \cite{Kennedy:1979ar}, up to the Wick rotation.

\subsubsection{Thermal two-point function of boundary operators}
Since the boundary theory behaves as a $(d-1)$-dimensional CFT without a boundary, the boundary two-point function admits a decomposition analogous to that in \eqref{eq:thermal_2pt}, which allows us, in principle, to extract infinitely many thermal BCFT data.
In our example, the boundary theory is just a generalized free field, which has already been well studied in \cite{Iliesiu:2018fao}.

For a scalar primary operator $\hat{\CO}$ with dimension $\hat{\Delta}$ in a generalized free field, the thermal two-point function takes the following form \cite{Iliesiu:2018fao}:
\begin{align}
\begin{aligned}
        &\frac{\langle\,\hat{\CO}(\hat{x})\,\hat{\CO}(0)\,\rangle_{\CD,\,\beta}}{C_{\hat{\CO}}} \\
        &=
            \sum_{m\in \BZ}\frac{1}{\left[(\tau + m\beta)^2 + |x_{\parallel}|^2\right]^{\hat{\Delta}}} \\
        &=
            \frac{1}{|\hat{x}|^{2\hat{\Delta}}} + \sum_{n=0}^{\infty}\sum_{j = 0,2,\cdots}\frac{2\,\zeta(2\,\hat{\Delta}+2\,n+j)\left(j + \hat{\nu}\right)(\hat{\Delta})_{n+j}\left(\hat{\Delta}-\hat{\nu}\right)_n}{n!\left(\hat{\nu}\right)_{n+j+1}}\,C^{\left(\hat{\nu}\right)}_{j}\left(\frac{\tau}{|\hat{x}|}\right)\,\frac{|\hat{x}|^{2\,n+j}}{\beta^{2\,\hat{\Delta}+2\,n+j}} \ ,
\end{aligned}
\end{align}
where we defined $\hat{\nu} = \frac{d-3}{2}$ and $C_j^{(\nu)}$ is the Gegenbauer polynomial.
This is precisely the form of the thermal conformal block decomposition \eqref{eq:thermal_2pt} for a $(d-1)$-dimensional CFT, which has contributions from the identity and the double-twist operators $[\hat{\CO}\hat{\CO}]_{n,j}$
with even spin $j$ and dimensions $\hat{\Delta}_{n,j} = 2\,\hat{\Delta} + 2\,n + j$. 
More explicitly, the one-point functions of these operators satisfy
\begin{align}
    \frac{f_{\hat{\CO}\hat{\CO} [\hat{\CO}\hat{\CO}]_{n,j}}}{C_{\hat{\CO}}\,\sqrt{C_{[\hat{\CO}\hat{\CO}]_{n,j}}}}\,\frac{\hat{b}_{[\hat{\CO}\hat{\CO}]_{n,j}}}{\sqrt{C_{[\hat{\CO}\hat{\CO}]_{n,j}}}}\,\frac{j!}{2^j(\hat{\nu})_j}
        =
            \frac{2\,\zeta(2\,\hat{\Delta}+2\,n+j)\left(j + \hat{\nu}\right)(\hat{\Delta})_{n+j}\left(\hat{\Delta}-\hat{\nu}\right)_n}{n!\left(\hat{\nu}\right)_{n+j+1}} \ .
\end{align}
Since the OPE coefficients of the boundary operators were computed in \cite{Fitzpatrick:2011dm} as\footnote{The factor $2^{j}$ is multiplied to the result of \cite{Fitzpatrick:2011dm}, which is needed in our normalization.}
\begin{align}
    \left(\frac{f_{\hat{\CO}\hat{\CO} [\hat{\CO}\hat{\CO}]_{n,j}}}{C_{\hat{\CO}}\,\sqrt{C_{[\hat{\CO}\hat{\CO}]_{n,j}}}}\right)^2
        =
            \frac{(1+(-1)^j)\,2^{j}\, \left(\hat{\Delta} -\frac{d-3}{2}\right)_n^2 (\hat{\Delta} )_{n+j}^2}{j!\,n! \left(\frac{d-1}{2}+j\right)_n (2\,\hat{\Delta}+n-d +2)_n (2\,\hat{\Delta}+2\,n+j-1)_j \left(2\,\hat{\Delta}+n+j-\frac{d-1}{2} \right)_n} \ ,
\end{align}
the thermal one-point functions of the double-twist operators are given by
\begin{align}\label{eq:DT_1pt_scalar_BCFT}
    \frac{\hat{b}^2_{[\hat{\CO}\hat{\CO}]_{n,j}}}{C_{[\hat{\CO}\hat{\CO}]_{n,j}}}
        &=
            \frac{\left(\frac{d-1}{2}+j\right)_n (2\,\hat{\Delta}+n-d+2)_n (2\,\hat{\Delta}+2\,n+j-1)_j \left(2\,\hat{\Delta}+n+j-\frac{d-1}{2} \right)_n}{j!\,n!} \notag \\
            &\qquad\cdot
            \frac{2^{j+1} (j+\hat{\nu})^2 \left(\hat{\nu}\right)_j^2 \,\zeta (2\,\hat{\Delta}+2\,n+j)^2}{\left(\hat{\nu}\right)_{n+j+1}^2} \ .
\end{align}
This result includes the cases considered in the previous section, by setting $\hat{\CO}=\hat{\phi}$ for the Neumann and $\hO = \hat{t}_1$ for the Dirichlet boundary condition, agreeing with the corresponding results \eqref{eq:scalar_BCFT_defect_phi2}, \eqref{eq:bdy_thermal_1pt_scalar}, \eqref{eq:scalarBCFT_bdy_T_1pt}, and \eqref{eq:scalarBCFT_thermal_1pt_spin2}.

\subsection{Free fermion}\label{sec:fermion_BCFT}
We consider the free fermion theory on $\BS^1_\beta\times\BR^{d-1}_+$ with the action
\begin{align}
    I
        =
            \int \d^d x\,\bar{\psi}\gamma\cdot\oap\psi\ ,
\end{align}
where $\oap_\mu=\frac{1}{2}\left(\partial_{\mu}-\overleftarrow\partial_{\mu}\right)$.
The stress tensor is
\begin{align}
    T_{\mu\nu}
        =
            \frac{1}{2}\,\bar{\psi}\left(\gamma_\mu\oap_\nu+\gamma_\nu\oap_\mu\right)\psi\ .
\end{align}
The boundary condition is
\begin{align}\label{fermion boundary condition}
    (1-U)\,\psi\,\big|_{x_\perp=0} = 0\ ,
    \qquad 
    \bar{\psi}\,(1-\bar{U})\,\big|_{x_\perp=0}=0\ ,
\end{align}
where $U,\bar{U}$ are constant matrices that satisfy \begin{align}
    U\,\gamma^{d-1} = -\gamma^{d-1}\, \bar{U}\ ,
    \qquad 
    U\,\gamma^i = \gamma^i\,\bar{U}\ ,\qquad U^2 = \bar{U}^2 = 1\ .
\end{align}
When $d$ is even, the matrix $\gamma_*=\gamma_0\cdots \gamma_{d-1}$ is not an identity matrix, and a convenient choice for $U,\bar{U}$ is $U=\gamma^{d-1}\exp(\i\,\theta\,\gamma'_*)$ and $\bar{U} = -\gamma^{d-1}\exp(-\i\,\theta\,\gamma'_*)$, where $\gamma'_* = \i^\frac{d}{2}\gamma_*$ \cite{Luckock:1990xr}.
In the free massless theory, the chiral symmetry can be used to set $\theta = 0$.
When $d$ is odd, the matrix $\gamma_*$ is proportional to the identity matrix, and we simply choose $U = \gamma^{d-1}$ and $\bar{U} = -\gamma^{d-1}$.
We thus take $U=\gamma^{d-1}$ and $\bar{U}=-\gamma^{d-1}$ for all $d$.
At zero temperature, the propagator is 
\begin{align}
    \langle\,\psi(x)\,\bar{\psi}(y)\,\rangle
        =
            \frac{1}{\Omega_{d-1}}\,\left[\,
                \frac{\gamma_{\ha}\,(\hx-\hy)^{\ha}+\gamma_\perp(x_\perp-y_\perp)}{[(\hx-\hy)^2+(x_\perp-y_\perp)^2]^{\frac{d}{2}}}
                +
                \gamma_\perp\,\frac{\gamma_{\ha}\,(\hx-\hy)^{\ha}-\gamma_\perp(x_\perp+y_\perp)}{[(\hx-\hy)^2+(x_\perp+y_\perp)^2]^{\frac{d}{2}}}\right]\ .
\end{align}

For later convenience, we define the parity transformation as
\begin{align}
    \psi(x^a_\parallel,\,\cdots )\quad \to\quad S\,\psi(-x^a_\parallel,\,\cdots )\ ,
\end{align}
where $S=\gamma_a$ for $d$ odd, and $S=\gamma_*\gamma_a$ for $d$ even.
Under these transformations, in $d=$even the boundary condition is invariant, while in $d=$odd the boundary condition changes as
\begin{align}
    \begin{aligned}
        (1-U)\,\psi\,\big|_{x_\perp=0}&=0 \quad \to \quad (1+U)\,S\,\psi|_{x_\perp=0} = 0\  .\\
    \end{aligned}
\end{align} 
Hence only in $d=$odd the boundary condition breaks the parity invariance explicitly.

\subsubsection{Boundary local operators}
As in section \ref{sec:scalar_bdy_op}, we first list the boundary local primaries that appear in the BOE for $\bar{\psi}\psi$ and  $T^{\mu\nu}$ using the same method as in \cite{Liendo:2012hy}.
For the fermionic theory, it is more difficult to construct a basis of bilinear boundary primary than  in the scalar theory as the gamma matrices inserted between $\bar{\psi}$ and $\psi$ make it complicated to check whether the resulting operator is primary or not.
Thus, we do not bother to write down the explicit expressions of the boundary operators, except for the displacement operator $\hat{D}$ and the spin-two operator $\hat{\CT}^{\ha\hb}$ constructed from the bulk stress tensor.
See \cite{Herzog:2022jlx} for the BOE of fermionic operators. 

\begin{table}[ht]
    \centering
    \begin{tabular}{ccccc}
    \toprule
       bulk primary & $\hat{\Delta}_{\hat{\CO}}$ & $j$ & $b^2_{\CO\hat{\CO}}C_{\hat{\CO}}/C_{\CO}$ & $\hat{b}^2_{\hat{\CO}}/C_{\hat{\CO}}$ \\
    \midrule
    \multirow{3}{*}{$\bar{\psi}\psi$} & $0$ & $0$ & $2^{ \left \lfloor\frac{d}{2} \right \rfloor+2-2\,d}$ & $1$ \\
    & $d$ & $0$ & $4\,(d-1)$ & $\displaystyle \frac{2^{ \left \lfloor \frac{d}{2}\right \rfloor+2}}{d-1}\,\eta^2(d)$  \\
    & $d+2\,n$ & $0$ & \eqref{eq:BOE for barpsipsi} & \eqref{eq:thermal 1-pt of BOE barpsipsi} \\
    \cmidrule(lr){1-5}
    \multirow{3}{*}{$T^{\mu\nu}$} & $d$ & $0$ & $\displaystyle\frac{2}{d(d-1)}$ & $\displaystyle \frac{2^{ \left \lfloor \frac{d}{2}\right \rfloor+2}}{d-1}\,\eta^2(d)$ \\
    & $d$ & $2$ & $2$ & $2^{ \left \lfloor \frac{d}{2}\right \rfloor+2}\,d\,\eta^2(d)$ \\
    & $d+2\,n~(n\geq1)$ & $2$ & \eqref{eq:BOE_coeff_stress_fermion} & \eqref{eq:thermal_1pt_fermion_bilinear} \\
    \bottomrule
    \end{tabular}
    \caption{The BCFT data appearing in the BOE of $\bar{\psi}\psi$ and $T^{\mu\nu}$.
    The scaling dimension, spin, the BOE coefficient, and the thermal one-point function are tabulated.
    The two-point functions of $\bar{\psi}\psi$ and $T^{\mu\nu}$ in the CFT without a boundary are $C_{\bar{\psi}\psi}=\frac{2^{ \left \lfloor\frac{d}{2} \right \rfloor}}{\Omega^2_{d-1}},~C_T=d\,\frac{2^{\left\lfloor \frac{d}{2}\right\rfloor-1}}{\Omega_{d-1}^2}$.
    When $d=3$, the coefficient $\hat b_{\hO}$ in the second to the last line have an additional contribution due to the parity violating term.
    }
    \label{tab:BOE_free_fermion}
\end{table}

Let us first consider the BOE of the $\bar{\psi}\psi$ operator.
Since this is a scalar operator that effectively satisfies the Dirichlet boundary condition as in section \ref{ss:free scalar BCFT}, the two-point function becomes
\begin{align}\label{eq:psibarpsi_two_point}
    \langle\,\bar{\psi}\psi(x)\,\bar{\psi}\psi(y)\,\rangle_\text{con}
        =
            \frac{2^{ \left \lfloor \frac{d}{2}\right \rfloor}}{\Omega^2_{d-1}}\left[\,
                \frac{1}{[\,(\hx-\hy)^2+(x_\perp-y_\perp)^2\,]^{d-1}}-\frac{1}{[\,(\hx-\hy)^2+(x_\perp+y_\perp)^2\,]^{d-1}}
                \,\right]\ .
\end{align}
We then read off the BOE coefficients for the boundary scalars $\hO_n$ with $\hD =d+2\,n$ from \eqref{eq:psibarpsi_two_point} by using the conformal block expansion in BCFT \cite{Liendo:2012hy} as
\begin{align}\label{eq:BOE for barpsipsi}
    b^2_{\bar{\psi}\psi\,\hO_n}\,\frac{C_{\hO_n}}{C_{\bar{\psi}\psi}}
        =
            \frac{4^{n+1}(d-1)_{2\,n+1}\left(\frac d2\right)_n}{(2\,n+1)!\left(\frac{d+1}{2}+n\right)_n}\ .
\end{align}
The boundary scalar $\hat{\CO}_n$ is denoted as $\hat{\CO}_{-}[n,0]$, using one of the double-twist families with the schematic form \footnote{We use the notation in \cite{David:2023uya}.}
\begin{align}\label{eq:fermion_bilinear_minus}
    \hat{\CO}_{-}[n,j]
        =
            \bar{\psi}\,\gamma^{\hat{a}}\partial_{\hat{a}}\partial^{\hat{a}_1}\cdots\partial^{\hat{a}_j}\hat{\partial}^{2n}\,\psi\,\big|_{x_{\perp}=0} - \text{trace} \ .
\end{align}
Although, from the perspective of the boundary theory, the above operator should instead be expressed in terms of the boundary fermion, which is a Dirac (Weyl) fermion for even (odd) $d$, we adopt the above convention to keep the notation independent of the spacetime dimension.
The BOE reads
\begin{align}
    \bar{\psi}\psi(x)
        =
            \frac{2^{ \left \lfloor \frac{d}{2}\right \rfloor+1-d}}{\Omega_{d-1}}\frac{1}{x_\perp^{d-1}}
        +
        \sum_{n=0}^{\infty}\,b_{\bar{\psi}\psi\,\hO_n}\,|x_\perp|^{2\,n+1}\,\hO_n(\hx)\ +\text{descendants} \ .
\end{align}
For $n=0$, we choose the normalization $\hat{\CO}_0=\hat{D}=-\frac{1}{2}\partial_\perp(\bar{\psi}\psi)|_{x_{\perp}=0},~b_{\bar{\psi}\psi\,\hat D}=-2$.
Using the equation of motion and the boundary condition, this explicit expression for the displacement operator can be rewritten in the form given in \eqref{eq:fermion_bilinear_minus}.
The two-point function of the displacement operator $\hat{D}$ is \cite{Herzog:2021spv}
\begin{align}
    \langle\,\hat{D}(\hx)\,\hat{D}(\hat{y})\,\rangle_{\CD}
        =
            \frac{(d-1)2^{ \left \lfloor \frac{d}{2}\right \rfloor}}{\Omega^2_{d-1}}\frac{1}{|\hx-\hat{y}|^{2d}}.
\end{align}

For the bulk stress tensor, the BOE reads
\begin{align}\label{eq:BOE_stress_fermion}
    T^{\mu\nu}(x)
        =
            \frac{d}{d-1}\left(\delta^{\mu n}\delta^{\nu n} 
            -
            \frac{\delta^{\mu\nu}}{d}\right)\hat{D}(\hx) 
            +  \sum_{\hat{\CO},~n=0}^{\infty}b_{T\hat{\CO}}\,x_{\perp}^{2\,n}\,\delta^{\mu}_{\hat{a}}\,\delta^{\nu}_{\hat{b}}\,\hat{\CO}^{\hat{a}\hat{b}}_{\hat{\Delta}=d+2\,n}(\hx) \ +\text{descendants} ,
\end{align}
where $\hat{\CO}^{\hat{a}\hat{b}}_{\hat{\Delta}=d}=\hat{\CT}^{\hat{a}\hat{b}}$ is defined by \eqref{defect_stress_tensor_primaries}.
For $n \geq 1$, the sum over $\hat{\CO}$ comes from the two distinct double-twist families: $\hat{\CO}_{-}[n-1,2]$ defined in \eqref{eq:fermion_bilinear_minus}, and $\hat{\CO}_{+}[n,2]$ defined by
\begin{align}\label{eq:fermion_bilinear_plus}
    \hat{\CO}_{+}[n,j]
        =
            \left(\bar{\psi}\,\gamma^{\hat{a}_{1}}\partial^{\hat{a}_2}\cdots\partial^{\hat{a}_j}\hat{\partial}^{2n}\,\psi\,\big|_{x_{\perp}=0} + \text{cyclic}\right) - \text{trace} \ .
\end{align}
At $n=0$, there is the unique dimension-$d$ operator $\hat{\CO}^{\hat{a}\hat{b}}_{\hat{\Delta}=d}=\hat{\CT}^{\hat{a}\hat{b}}=\hat{\CO}_{+}[0,2]$.
The sum of the BOE coefficients for the two families can be read off from the stress tensor two-point function and the conformal block expansion as
\begin{align}\label{eq:BOE_coeff_stress_fermion}
    b^2_{T\hat{\CO}_{+}[n,2]}\,\frac{C_{\hat{\CO}_{+}[n,2]}}{C_T} + b^2_{T\hat{\CO}_{-}[n-1,2]}\,\frac{C_{\hat{\CO}_{-}[n-1,2]}}{C_T}
        =
            \frac{\pi^\frac{1}{2}\,2^{1-d}\, \Gamma(d+2\,n-1)\,\Gamma (d+2\,n+2)}{(d+1)\,\Gamma \left(\frac{d}{2}+1\right)\, \Gamma(d-1)\,\Gamma(2\,n+1)\, \Gamma\left(\frac{d+1}{2} +2\,n\right)} \ ,
\end{align} 
where the second term on the left-hand side is absent for $n=0$.

\subsubsection{Thermal one-point function of $\bar{\psi}\psi$}
The thermal one-point function of $:\bar{\psi}\psi:$ is 
\begin{align}
    \begin{aligned}
        \langle\,\bar{\psi}\psi(x)\,\rangle_{\CD,\,\beta}
            =
                \frac{1}{\beta^{d-1}}\sum_{n\in \BZ}\frac{(-1)^n\,2^{\lfloor \frac{d}{2} \rfloor+1}}{\Omega_{d-1}}\frac{z}{(n^2+4\,z^2)^{\frac{d}{2}}}\ .
    \end{aligned}
\end{align}
The low temperature expansion is
\begin{align}
    \begin{aligned}
        \langle\,\bar{\psi}\psi(x)\,\rangle_{\CD,\,\beta}
            =
            \frac{2^{\lfloor \frac{d}{2}\rfloor+1-d}}
            {\Omega_{d-1}\,|x_\perp|^{d-1}}
            -
            \frac{2^{\lfloor \frac{d}{2}\rfloor +2}}
            {\Omega_{d-1}\,\beta^{d-1}}\sum_{k=0}^{\infty}
            \frac{(-1)^k\, 4^k\, (\frac{d}{2})_k}{k!}\,
            \eta(d+2\,k)\,z^{2\,k+1}
            \ ,
    \end{aligned}
\end{align}
where $\eta(s)=(1-2^{1-s})\,\zeta(s)$ is the Dirichlet eta function.
Combining this with \eqref{eq:BOE for barpsipsi}, we can read off the thermal one-point function coefficient of each boundary primary that appears in the BOE, 
\begin{align}\label{eq:thermal 1-pt of BOE barpsipsi}
    \frac{\hb^2_{\hO}}{C_{\hO}}
        =
            \frac{2^{ \left \lfloor\frac{d}{2} \right \rfloor+2\,n+2}\,(\frac{d}{2})_n\,(2\,n+1)!(\frac{d+1}{2}+n)_n}{(n!)^2\,(d-1)_{2\,n+1}}\,\eta^2\,(d+2\,n)\ .
\end{align}
For example, the coefficient for the displacement operator corresponds to $n=0$, \begin{align}\label{eq:fermion BCFT displace thermal 1-pt}
    \hb_{\hat{D}}=\frac{2^{ \left \lfloor \frac{d}{2}\right \rfloor+1}\eta(d)}{\Omega_{d-1}}\ .
\end{align}
As in section \ref{ss:free scalar BCFT}, the thermal one-point function can be expressed as
\begin{align}
    \langle\,\bar{\psi}\psi(x)\,\rangle_{\CD,\,\beta}
        =
            \frac{2^{\lfloor \frac{d}{2} \rfloor+1}\,z}{\Omega_{d-1}\,\beta^{d-1}}\sum_{k\in \BZ}\,\int^{\infty}_{-\infty}\d y\,\frac{e^{-2 \pi\,\i\, y\left(k+\frac{1}{2}\right)}}{(y^2+4\,z^2)^{\frac{d}{2}}}\ .
\end{align}
Unlike the scalar case, the fermion has no Matsubara zero mode due to the anti-periodicity along the thermal circle.
All the modes are suppressed at high temperature exponentially,
\begin{align}
    \langle\,\bar{\psi}\psi(x)\,\rangle_{\CD,\,\beta}
        =
        \frac{2^{\lfloor \frac{d}{2}\rfloor+2-d}}{\beta^{d-1}}
        z^{1-\frac d2}\,e^{-2\pi z}
        \left[\,
        1+\frac{d\,(d-2)}{16\,\pi\, z}
        +O(z^{-2})\,\right]
        +
        O\left(z^{-\frac{d}{2}}\,e^{-6\pi z}\right)
        \ .
\end{align}

\subsubsection{Thermal one-point function of $T^{\mu\nu}$}
In $d\neq3$, as discussed in section \ref{ss:parity odd}, there are no parity violations in the thermal one point function, and one  finds
\begin{align}
    b_T=-2^{ \left \lfloor \frac{d}{2}\right \rfloor+1}\frac{d}{\Omega_{d-1}}\eta(d)\ , \qquad F_i(z)=0\ .
\end{align}
The existence of the boundary cannot be seen from this thermal one-point function.
In terms of the BOE \eqref{eq:BOE_stress_fermion}, we interpret this result as
\begin{align}
    &\hat{b}_{\hat{D}}
        =
            -\frac{b_T}{d} \ ,\qquad
    \hat{b}_{\hat{\CT}^{\hat{a}\hat{b}}}
        =
            b_T \ ,\\
    &b_{T\hat{\CO}_{+}[n,2]}\,\hat{b}_{\hat{\CO}_{+}[n,2]} + b_{T\hat{\CO}_{-}[n-1,2]}\,\hat{b}_{\hat{\CO}_{-}[n-1,2]}
        =
            0 \qquad(n \geq 1)\ . \label{eq:thermal_1pt_fermion_bilinear}
\end{align}

In $d=3$, the thermal one-point function receives an extra parity violating term as shown in section \ref{ss:parity odd}, 
\begin{align}
    \begin{aligned}
        F_5(z)
            =
                \frac{3\,\i}{\pi}\sum_{n=1}^\infty \frac{(-1)^n\, n^2}{(n^2+4\,z^2)^\frac{5}{2}}\ ,
    \end{aligned}
\end{align}
where $F_5(z)$ is defined in \eqref{parity violating of tensors}. 

Expanding this result at low temperature and using \eqref{eq:fermion BCFT displace thermal 1-pt} for the boundary operators of dimension $d$, we obtain 
\begin{align}
     \langle \,\hat{\CT}^{\hat{a}\hat{b}}\,\rangle_{\CD,\,\beta}
        =
            \frac{b_T}{\beta^d}\left[\left(e^{\hat{a}} e^{\hat{b}}-\frac{\delta^{\hat{a}\hat{b}}}{d-1}\right) + \delta_{d,3}\,\i\,\epsilon^{\hat{c}(\hat{a}}\,e^{\hat{b})}\,e_{\hat{c}}\right] \ .
\end{align}
For the boundary parallel spin-two operators with $\hD > d$, the thermal one-point functions vanish for $d\neq 3$.

\section{Free scalar theories with localized $\phi$ deformation}\label{sec:scalar_Wilson}

One of the simplest examples of DCFTs that have a Lagrangian description is the free massless scalar theory with a localized $\phi$ deformation:
\begin{align}\label{eq:free_scalar_defect_action}
    I
        =
            \frac{1}{2}\int\d^d x\,(\partial \phi)^2
            +
            h\int\d^p \hat x\,\phi
            \ .
\end{align}
The localized $\phi$ deformation is conformal and $h$ is the exactly marginal coupling  when $p$ equals the conformal dimension of $\phi$:
\begin{align}\label{eq:marginality_condition}
    p=\Delta_{\phi}=\frac{d}{2}-1 \ .
\end{align}
Equivalently, the defect coupling $h$ is marginal when $\hphi$ of the undeformed theory is a marginal defect operator, $\hD_{\hphi} = p$.
It generalizes the scalar Wilson line ($p=1$) in $d=4$ \cite{Kapustin:2005py} to higher-dimensional defects in even $d$ dimensions (see, e.g., \cite{Billo:2016cpy,Lauria:2020emq,Nishioka:2021uef}).
Throughout this section we work at the marginal point \eqref{eq:marginality_condition} and use
\begin{align}\label{eq:marginal_kinematics}
    \Delta_\phi = p \ ,
    \qquad
    d = 2\,p+2 \ ,
    \qquad
    q = d - p = p+2 \ ,
\end{align}
to simplify the expressions of section \ref{sec:GFF_on_defect}.

It will be convenient to decompose the field $\phi$ to $\phi = \phi_0 + \varphi$ where $\phi_0$ is the solution to the equation of motion $\partial^2\phi = h\,\delta(x_\perp)$.
Since the action is Gaussian for $\varphi$, it can be dealt with as a free field describing quantum fluctuation around the defect.
The classical part $\phi_0$ encodes the defect data and depends on the spacetime structure, which we will fix in the following.
All the defect local operators are built out of $\varphi$ alone, and they are those of section \ref{sec:GFF_on_defect} with the free field there identified with $\varphi$.
Hence the defect operator content of the present theory is the generalized free field theory of the tower $\hatt_s$ described in section \ref{sec:GFF_on_defect}, specialized to $d=2\,p+2$.
Table \ref{tab:bilinear_defect_primary_marginal} lists the defect bilinears of table \ref{tab:bilinear_defect_primary} at the marginal point.
The coupling $h$ introduces the defect identity channel, generated by $\phi_0$, and the sector linear in $h$, which we work out below.

\begin{table}[ht]
    \centering
    \begin{tabular}{cccc}
    \toprule
       $\hO$  & $\hat \Delta_{\hO}$ &
       $C_{\hO}/C_{\phi}^2$ & $\hat b_{\hO}/C_\phi$\\
    \midrule
    $\hphi^{2} = [\hphi\,\hphi]_{0,0}$
      & $2p$ & $2$
      & $2\,\zeta(2p)$ \\
    \cmidrule(lr){1-4}
    $\normord{\hphi\,\hatt_1^{\,i}} = [\hphi\,\hatt_1]^{i}_{0,0}$
      & $2\,p+1$ & $2\,p$ & $0$ \\
    \cmidrule(lr){1-4}
    $[\hphi\,\hphi]_{1,0}$
      & \multirow{6}{*}{$2\,p+2$}  & $\displaystyle \frac{8\,p^3\,(3\,p+2)}{p+2}$ & $\displaystyle 4\,p\,(3\,p+2)\,\zeta(2\,p+2)$ \\
    $\normord{\hatt_1\!\cdot\!\hatt_1}
      = \delta_{ij}[\hatt_1\,\hatt_1]^{ij}_{0,0}$
      & & $8\,p^2\,(p+2)$
      & $4\,p\,(p+2)\,\zeta(2\,p+2)$ \\
    $\normord{\hatt_1^{\langle i}\hatt_1^{\,j\rangle}}
      = [\hatt_1\hatt_1]^{\langle ij\rangle}_{0,0}$
     & & $8\,p^2$ & $0$ \\
    $\normord{\hphi\,\hatt_2^{\,ij}} = [\hphi\,\hatt_2]^{ij}_{0,0}$
     & & $8\,p\,(p+1)$ & $0$ \\
    $[\hphi\,\hatt_1]^{\hat a i}_{0,1}$
     & & $4\,p^2(p+1)(2\,p+1)$ & $0$ \\
    $[\hphi\,\hphi]^{\hat a\hat b}_{0,2}$
     & & $\displaystyle\frac{8\,p^2(2\,p+1)}{p + 1}$ & $8\,p\,(2\,p+1)\,\zeta(2\,p+2)$ \\
    \bottomrule
    \end{tabular}
    \caption{Table \ref{tab:bilinear_defect_primary} at the marginal point $d=2\,p+2$, where $\Delta_\phi = p$ and $q=p+2$.
    At $p=1$, the last row is absent as $\hat\partial^{\langle\hat a}\hat\partial^{\hat b\rangle} = 0$.
    }
    \label{tab:bilinear_defect_primary_marginal}
\end{table}

\subsection{Defect CFT data at zero temperature}
Before delving into the thermal properties of the theory, we examine the defect CFT data at zero temperature such as defect operators and DOE that will be relevant to the subsequent subsections.

\subsubsection{Bulk local operators}
At zero temperature, the classical part $\phi_0$ is given by
\begin{align}
    \phi_0 = \frac{a_\phi}{|x_\perp|^{p}} \ ,
\end{align}
where the coefficient $a_\phi$ is given by
\begin{align}
    a_\phi
        =
            -h\,C_{\phi}\,\frac{\Omega_{2p-1}}{\Omega_{p-1}}
        =
            -h\,C_\phi\,\frac{\pi^{\frac{p+1}{2}}}{2^{p-1}\,\Gamma\left(\frac{p+1}{2}\right)}
            \ .
\end{align}
The correlation functions of the free part $\varphi$ are obtained by the Wick contraction using \eqref{eq:GFF_bulk_two_point}.
It follows that the one- and two-point functions of $\phi$ are
\begin{align}\label{eq:a_phi}
    \langle\,\phi(x)\,\rangle_h
        =
            \frac{a_\phi}{|x_\perp|^{p}} \ ,
\end{align}
and 
\begin{align}
    \langle\,\phi(x)\,\phi(y)\,\rangle_h
        =
            \frac{\Cphi}{|x-y|^{2\,p}}
            +
            \frac{a_\phi^2}{|x_\perp|^{p}\,|y_\perp|^{p}} \ ,
\end{align}
respectively.
The bulk composite operator $\phi^2(x)$ becomes
\begin{align}
        \langle\,\phi^2(x)\,\rangle_{h}
            =
             \frac{a_\phi^2}{|x_{\perp}|^{2\,p}}\ .
\end{align}
Similarly, the one-point function of the stress tensor takes the form of \eqref{bulk 1-pt stress} 
\begin{align}\label{bulk 1pt stress phi deformation zeroT}
    \langle\, T^{\mu\nu}(x)\,\rangle_{h}
        =
            \frac{a_T}{q\,|x_\perp|^{d}}\left(\frac{q-1}{d}\,\sfs^{\mu\nu}_1 + \sfs^{\mu\nu}_2\right) \ ,
\end{align}
with $a_T$ given by\footnote{Our definition of $a_T$ differs from that of \cite{Billo:2016cpy} by an overall sign.} 
\begin{align}
    a_T
        =
            \frac{h^2}{(2\,p+1)\,\Omega_{p+1}^2} \ .
\end{align}

\subsubsection{Defect local operators}
The spectrum of the defect operators is almost the same as that of the trivial defect in section \ref{sec:GFF_on_defect}.
The defect primary $\hphi$ and those with transverse spin $\hatt_s$ are not affected by the deformation as they are defined through the fluctuation $\varphi$, and given by
\begin{align}
    \hphi(\hat x)
        := 
            \lim_{|x_\perp|\to 0}\,\varphi(\hat x, x_\perp)
            \ ,
\end{align}
and
\begin{align}
    \hatt_s(\hat x; w)
    :=
            \lim_{|x_\perp|\to 0}(w\cdot\partial_{\perp})^s\,\varphi(\hat x, x_\perp)  \ .
\end{align}
They have dimension $\hD_{\hphi} = p$ and $\hD_{\hatt_s} = p+s$, respectively.

Since the $\phi$ deformation induces a non-trivial defect, there is the displacement operator, which can be read off from the equation $\partial_\mu T^{\mu i} = h\,\delta(x_\perp)\,\partial^i_\perp\phi$:
\begin{align}
    \hat D^i(\hat x) := h\,\hatt_1^i \ .
\end{align}
Thus, the coefficient $C_{\hat D}$ of the two-point function reads
\begin{align}
    C_{\hat D} = 2\,p\,h^2\,C_\phi\ .    
\end{align}
Note that $\hD_{\hatt_1} = p+1$ is indeed the protected dimension of the displacement operator, consistent with the marginality condition \eqref{eq:marginality_condition}.

The defect operators bilinear in $\hphi$ are constructed as the double-twist operators as in section \ref{sec:GFF_bilinears} and \ref{sec:GFF_defect_stress_tensor} by replacing $T^{\mu\nu}$ in the definitions \eqref{defect_stress_tensor_primaries} with the quantum part $T^{\mu\nu}[\varphi] := T^{\mu\nu}|_{\phi\to\varphi}$.

\subsubsection{Bulk-defect operator expansion}
The DOE can be extracted from the two-point function of a bulk  local operator with defect local operators.
For instance, the one-point function of $\phi$ in \eqref{eq:a_phi} implies the defect identity operator $\hat{\bm{1}}$ appears in the expansion.
Moreover, the only non-vanishing two-point functions are those of $\phi$ and $\hat t_s$: 
\begin{align}
    \langle\,\phi(\hat x, x_\perp)\,\hatt_s(\hat y; w)\,\rangle_h
        =
            2^s\,(\Delta_\phi)_s\,\Cphi\,\frac{(w\cdot x_\perp)^s}{\left( |\hat x - \hat y|^2 + |x_\perp|^2\right)^{\Delta_\phi + s}} \ .
\end{align}
Thus, the DOE of $\phi$ is given by
\begin{align}\label{Scalar_Wilson_phi_DOE}
    \phi(x)
        =
            \frac{a_\phi}{|x_\perp|^{p}}\,\hat{\bm{1}} 
            +
            \sum_{s=0}^\infty\,\frac{1}{\Gamma(s+1)}\,\hatt_s(\hat x; x_\perp) + (\text{descendants})\ ,
\end{align}
which is the DOE \eqref{GFF_phi_DOE} of the fluctuation supplemented by the identity channel generated by $\phi_0$.
Note that the same expansion can be obtained by Taylor expanding the free field $\varphi(\hat x, x_\perp)$ in $\phi = \phi_0 + \varphi$ around $|x_\perp| = 0$.

The DOE of the bulk operator $\phi^2$ is fixed by either calculating the bulk-defect two-point functions with the defect bilinear primaries or Taylor expanding the composite operator $\varphi^2$ around $|x_\perp| = 0$:
\begin{align}\label{Scalar_Wilson_phi2_DOE}
    \begin{aligned}
        \phi^2(x)
            ={}& \frac{a_\phi^2}{|x_\perp|^{d-2}}\;\hat{\bm{1}}
            + 
            \frac{2\,a_\phi}{|x_\perp|^{\frac{d}{2}-1}}\, 
                \sum_{s=0}^{\infty}\frac{1}{\Gamma(s+1)}\,\hatt_{s}(\hat x; x_\perp)
            \\[2pt]
            &+ 
                \hphi^2
                + 
                2\,x_{\perp i}\,\normord{\hphi\,\hatt_{1}^i} 
                 \\
            &
            ~
            + 
                x_{\perp i}x_{\perp j}
                \left[\normord{\hatt_{1}^{\langle i}\,\hatt_{1}^{j\rangle}} + \normord{\hphi\,\hatt_{2}^{ij}}\right]
            +    
                |x_\perp|^2\left[\,\frac{1}{q}\,\normord{\hatt_1\cdot\hatt_1} - \frac{1}{2\Delta_\phi + q}\,[\hphi\,\hphi]_{1,0}  \right] + \cdots\ ,
    \end{aligned}
\end{align}
where only defect primaries (up to $d$ for the bilinears) are displayed explicitly.
The last two lines are the DOE \eqref{GFF_phi2_DOE} of $\varphi^2$, while the first line consists of the identity channel and the sector linear in $\phi_0$.

To determine the DOE of the stress tensor $T^{\mu\nu}$, we decompose it using $\phi = \phi_0 + \varphi$ as
\begin{align}
    T^{\mu\nu}
        =
            T^{\mu\nu}[\phi_0] 
            + 
            T^{\mu\nu}_\text{lin} 
            +
            T^{\mu\nu}[\varphi] \ ,
\end{align}
where $T^{\mu\nu}_\text{lin}$ is the linearized part
\begin{align}\label{stress_tensor_linear}
    T^{\mu\nu}_\text{lin}
        =
            (\partial^{\mu}\phi_0)(\partial^\nu\varphi)
            +
             (\partial^{\nu}\phi_0)(\partial^\mu\varphi)
            -
            \delta^{\mu\nu}\,\partial\phi_0\cdot\partial\varphi
            -
            2\,\xi\,\left(\partial^\mu\partial^\nu - \delta^{\mu\nu}\,\partial^2\right)\left(\phi_0\,\varphi\right) \ .
\end{align}
The classical part $T^{\mu\nu}[\phi_0]$ yields the identity channel while the quantum part $T^{\mu\nu}[\varphi]$ has the DOE to the defect primaries \eqref{defect_stress_tensor_primaries}.
The linearized part \eqref{stress_tensor_linear} can be expanded with the DOE of $\varphi$ given by the non-identity channel of \eqref{Scalar_Wilson_phi_DOE}.
Collecting all terms together, we find
\begin{align}\label{Scalar_Wilson_T_DOE}
    \begin{aligned}
        T^{\mu\nu}(x)
            &=
                \frac{a_T}{q\,|x_\perp|^{d}}\left(\frac{q-1}{d}\,\sfs^{\mu\nu}_1 + \sfs^{\mu\nu}_2\right)\,\hat{\bm{1}} \\
            &\qquad
            +
                \frac{p}{2\,p+1}\,\frac{a_\phi}{|x_\perp|^{\frac{d}{2}+1}}\,\sum_{s=0}^{\infty}\frac{1}{\Gamma(s+1)}\,\BT^{\mu\nu}\,\hatt_{s}(\hat x; x_\perp) \\
            &\qquad~
                +
                \frac{\sfs_1^{\mu\nu}}{p\,q}\,\hat\CT
                +
                \delta_{\hat a}^{\mu}\,\delta_{\hat b}^{\nu}\,\hat\CT^{\hat a \hat b}
                +
                2\, \delta_{\hat a}^{(\mu}\,\delta_{i}^{\nu)}\,\hat\CT^{\hat a i}
                +
                 \delta_{i}^{\mu}\,\delta_{j}^{\nu}\,\hat\CT^{ij}
                +
                \cdots \ ,
    \end{aligned}
\end{align}
where $\BT^{\mu\nu}$ is the differential operator
\begin{align}
    \BT^{\mu\nu}
        :=
            p\,\sfs_2^{\mu\nu} + \delta^{\mu\nu}(x_\perp\cdot \partial_\perp) - (2\,p+2)\,x_\perp^{(\mu}\,\partial_\perp^{\nu)} - |x_\perp|^2\,\partial_\perp^{\mu} \partial_\perp^{\nu}  \ ,
\end{align}
and only defect primaries (up to $d=2\,p+2$ for the bilinears) are displayed explicitly.
The defect primaries $\hat{\CT}$, $\hat{\CT}^{\hat a\hat b}$, $ \hat{\CT}^{ij}$ and $\hat{\CT}^{\hat a i}$ are given by \eqref{ST_decomposition1}--\eqref{ST_decomposition4} with $d=2\,p+2$.

\subsection{Thermal one-point functions}
Since the action is Gaussian for $\varphi$, the thermal one-point functions can be computed exactly as we will see for the bulk operators $\phi$, $\normord{\phi^2}$ and $T^{\mu\nu}$, and for the defect bilinears.

\subsubsection{Bulk local operators}
As in section \ref{sec:GFF_thermal_one_point}, the thermal one-point functions can be calculated by applying the Wick's theorem and the method of images.
For instance, the one-point function of the bulk operator $\phi$ is 
\begin{align}\label{Scalar-Wilson-phi-one-point}
    \begin{aligned}
        \langle\,\phi(x)\,\rangle_{h,\,\beta}
            &=
                -h\int_{0}^{\beta}\,\d\tau_{y}\int\,\d^{p-1} y_\parallel\,\langle\,\phi(x)\,\phi(\tau_{y},y_\parallel)\,\rangle_{0,\,\beta}  \\
            &=
                \frac{a_\phi}{|x_{\perp}|^{p}}\ .
    \end{aligned}
\end{align}
Note that it takes the same form as the one at zero temperature \eqref{eq:a_phi} as it happens to be independent of $\beta$, $\langle\,\phi(x)\,\rangle_{h,\,\beta} = \langle\,\phi(x)\,\rangle_{h}$.

The thermal two-point function of $\phi$ is calculated in a similar way as follows:
\begin{align}\label{Scalar-Wilson-phi-two-point}
    \begin{aligned}
        \langle\,\phi(x)\,\phi(y)\,\rangle_{h,\,\beta}
            &=
                \langle\,\phi(x)\,\phi(y)\,\rangle_{0,\,\beta} + \langle\,\phi(x)\,\rangle_{h,\,\beta}\,\langle\,\phi(y)\,\rangle_{h,\,\beta}\, \\
            &=
                \sum_{n\in \BZ}\,\frac{C_{\phi}}{\left[\,(\tau_x - \tau_y + n\,\beta)^2 + |\vec{x} - \vec{y}|^2\,\right]^{p}} + \frac{a_\phi^2}{|x_{\perp}|^{p}\,|y_{\perp}|^{p}}\ ,
    \end{aligned}
\end{align}
where we used \eqref{Scalar-Wilson-phi-one-point} for the second term.
It follows that the thermal one-point function of the bulk composite operator $\phi^2(x)$ is given by
\begin{align}\label{Scalar-Wilson-phi2}
        \langle\,\phi^2(x)\,\rangle_{h,\,\beta}
            =
                \frac{2\,\zeta(d-2)\,C_{\phi}}{\beta^{2\,p}} + \frac{a_\phi^2}{|x_{\perp}|^{2\,p}}\ .
\end{align}
By taking the derivative of the two-point function \eqref{Scalar-Wilson-phi-two-point}, we obtain the thermal one-point function of the stress tensor:
\begin{align}\label{Scalar-Wilson-ST}
    \begin{aligned}
        \langle\,T^{\mu\nu}(x)\,\rangle_{h,\,\beta}
            =
                \langle\,T^{\mu\nu}(x)\,\rangle_{\beta} 
                + 
                \langle\,T^{\mu\nu}(x)\,\rangle_{h} \ ,
    \end{aligned}
\end{align}
where the first term is the thermal one-point function \eqref{free_scalar_bT} of the free scalar theory without defect, and the second term is the one-point function at zero temperature given by \eqref{bulk 1pt stress phi deformation zeroT}.
In each of \eqref{Scalar-Wilson-phi2} and \eqref{Scalar-Wilson-ST}, the first term is the thermal one-point function of section \ref{sec:GFF_thermal_one_point} and the second is the one-point function generated by $\phi_0$ at zero temperature.
There is no cross term because the mixed contribution is linear in $\varphi$ which vanishes in the one-point functions.

\subsubsection{Defect local operators}
Since the defect primary $\hphi$ behaves as a generalized free field, the thermal one-point functions of odd powers of $\hphi$ vanish, $\langle\,\hphi^{2k+1}\,\rangle_{h,\, \beta} = 0$.
Also, the thermal one-point functions of the operators with transverse spins vanish due to the rotational invariance in the transverse direction to the defect.

The thermal one-point functions of the defect primaries bilinear in $\hphi$ can be read off from \eqref{Scalar-Wilson-phi-two-point} by acting with derivatives and taking the limits $y \to x$ and $|x_\perp| \to 0$.
For instance, the operator $\normord{\hatt_1^i\,\hatt_1^j}$ has the thermal expectation value:
\begin{align}
    \langle\,\normord{\hatt_1^i\,\hatt_1^j}\,(\hat x)\,\rangle_{h,\,\beta}
        =
            \frac{4\,\Delta_\phi\,C_\phi\,\zeta(d)}{\beta^{d}}\,\delta^{ij} \ ,
\end{align}
from which we can read off the coefficients ${\hat b}_{\hO}$ for the trace part and symmetric traceless part shown in table \ref{tab:bilinear_defect_primary_marginal}.
The other non-vanishing coefficients are calculated similarly.

\subsection{High and low temperature limits}
We examine the high and low temperature limits of the thermal one-point functions obtained in the previous section, and compare them with the general analysis in section \ref{ss:general_high_lowT}.

\paragraph{High temperature limit.}
Taking the high-temperature limit $\beta\to 0$ at fixed distance $|x_\perp|$ from the defect is equivalent to taking $|x_\perp|\to\infty$ at fixed $\beta$.
In this limit, the thermal one-point functions \eqref{Scalar-Wilson-phi-one-point}, \eqref{Scalar-Wilson-phi2} and \eqref{Scalar-Wilson-ST} approach those in the bulk free scalar theory without defect as expected.

\paragraph{Low temperature limit.}

The low temperature limit of the thermal one-point functions may be compared with the DOE.

For the bulk scalar operator $\phi$, the one-point function \eqref{Scalar-Wilson-phi-one-point} is consistent with the DOE \eqref{Scalar_Wilson_phi_DOE} as the defect primaries appearing there are linear in $\hat\phi$ and vanish in the thermal expectation value.

For the composite operator $\phi^2$, the DOE \eqref{Scalar_Wilson_phi2_DOE} implies
\begin{align}
    \begin{aligned}
        \langle\,\phi^2\,\rangle_{h,\,\beta}
            &=
                \frac{a_\phi^2}{|x_\perp|^{2\,p}}
                +
                \frac{\hat b_{\hphi^2}}{\beta^{2\,p}}
                +
                |x_\perp|^2\left[\,\frac{1}{p+2}\,\frac{\hat b_{\normord{\hatt_1\cdot\hatt_1}}}{\beta^{2\,p+2}} - \frac{1}{3\,p+2}\,\frac{\hat b_{[\hphi\,\hphi]_{1,0}}}{\beta^{2\,p+2}}  \right] + O\left(\beta^{-(d+2)}\right)  \\
            &=
                \frac{a_\phi^2}{|x_\perp|^{2\,p}}
                +
                \frac{2\,\zeta(2p)\,C_\phi}{\beta^{2\,p}}
                + O\left(\beta^{-(d+2)}\right)  \ ,
        \end{aligned}
\end{align}
where we substituted the thermal coefficient $\hat b_{\hO}$ in table \ref{tab:bilinear_defect_primary_marginal}.
This reproduces the exact result \eqref{Scalar-Wilson-phi2} up to the order $O\left(\beta^{-(d+2)}\right)$.

For the stress tensor $T^{\mu\nu}$, the DOE \eqref{Scalar_Wilson_T_DOE} yields
\begin{align}
    \begin{aligned}
        \langle \, T^{\mu\nu}(x)\,\rangle_{h,\,\beta}
            &=
                \frac{a_T}{q\,|x_\perp|^{d}}\left(\frac{q-1}{d}\,\sfs^{\mu\nu}_1 + \sfs^{\mu\nu}_2\right)
            +
                \frac{\sfs_1^{\mu\nu}}{p\,q}\,\frac{\hat b_{\hat\CT}}{\beta^{d}} 
            -
                \frac{\sfs_3^{\mu\nu}}{p}\,\frac{\hat b_{\hat\CT^{\hat a \hat b}}}{\beta^{d}}
            +
                O\left(\beta^{-(d+2)}\right)  \\
            &=
                \frac{a_T}{q\,|x_\perp|^{d}}\left(\frac{q-1}{d}\,\sfs^{\mu\nu}_1 + \sfs^{\mu\nu}_2\right)
                +
                \frac{b_T}{\beta^{d}}\left( e^\mu e^\nu - \frac{\delta^{\mu\nu}}{d}\right)
                +
                O\left(\beta^{-(d+2)}\right)
                \ ,
    \end{aligned}
\end{align}
where we used the coefficients \eqref{ST_decompose_b_coefficient}.
This agrees with the exact result \eqref{Scalar-Wilson-ST} up to $O\left(\beta^{-(d+2)}\right)$.

As in section \ref{sec:GFF_low_temperature}, these results may be seen as a non-trivial check of the low temperature expansion.

\section{Free $\O(N)$ models with localized $\phi^2$ deformation}\label{sec:O(N)_phi2_bulk_free}

Another simple class of DCFTs is the free $\O(N)$ model with an $\O(N)$-invariant localized mass term:
\begin{align}\label{O(N)_phi2_action}
    I
        =
            \frac12\int \d^dx\,(\partial\phi_A)^2 
            +
            h_0\int \d^p\hat x\, \phi_A^2 \ .
\end{align}
This model describes $N$ copies of the free massless scalar field with a localized mass.
Thus, the theory remains Gaussian even for $h_0\neq 0$.
The action is classically conformal when $d=p+2$.
In what follows, we will work in $d=p+2-\epsilon$ dimensions.

A part of the DCFT data of this model has been obtained in \cite{Ge:2025fsm} in the context of a study of the long-range Ising model.
There, the renormalized defect coupling $h$ has been derived by calculating the one-point function $\langle\,\phi_A^2\,\rangle$ to all orders and requiring it to be finite:
\begin{align}\label{Free_O(N)_h_0_h}
    h_0
        =
            M^\epsilon\,\frac{h}{1 - \frac{h}{\pi\,\epsilon}} \ ,
\end{align}
where $M$ is the renormalization scale.
It follows that the exact beta function is
\begin{align}
    \beta_h 
        = 
            -\epsilon\,h + \frac{h^2}{\pi} \ .
\end{align}
Hence, this model has a non-trivial IR fixed point at
\begin{align}\label{Free_O(N)_h_ast}
    h_\ast = \pi\,\epsilon \ .
\end{align}

The rest of this section is organized as follows.
In section \ref{ss:Free_O(N)_zeroT}, we examine the defect CFT data of the localized $\phi^2$-deformed $\O(N)$ model \eqref{O(N)_phi2_action} at zero temperature.
In section \ref{ss:Free_O(N)_finiteT} and \ref{ss:Free_O(N)_high_lowT}, we assume $p\ge 2$ and study the exact thermal one-point functions of the bulk primaries $\phi^2$ and $T^{\mu\nu}$ and their limiting behaviors in the high and low temperature limits.
We confirm that the low temperature expansions are consistent with the DOE.
The $p=1$ case is treated separately in section \ref{ss:Free_O(N)_p1}, where we observe the IR divergence that makes the asymptotic expansion of the thermal one-point functions break down.
We also examine the $\epsilon = 1$ case in section \ref{ss:Free_O(N)_ep1}, where the defect becomes an interface, and show that the theory factorizes to two copies of the Dirichlet BCFTs studied in section \ref{ss:free scalar BCFT}.

\subsection{Defect CFT data at zero temperature}\label{ss:Free_O(N)_zeroT}
We describe the DCFT data at the IR fixed point \eqref{Free_O(N)_h_ast} at zero temperature.
We focus on the defect operators bilinear in $\phi$ and determine their spectrum up to dimension $d+2\,\epsilon$.
We also fix the DOE for $\normord{\phi^2}$ and $T^{\mu\nu}$.

\subsubsection{Bulk local operators}
Since the bulk is free and the defect localized deformation does not change the bulk data, the bulk primaries are those of the $N$ free massless scalar fields in section \ref{sec:GFF_bulk_operators} and the bilinear primaries of lower dimensions are summarized in table \ref{tab:GFF_bulk_primary}.

The one-point functions of the bulk primaries can be calculated perturbatively.
Since the localized $\phi^2$ deformation in \eqref{O(N)_phi2_action} is $\O(N)$ invariant, the one-point functions of any operator in a non-trivial representation of $\O(N)$ such as $\phi_A$ vanish.
We thus focus on the $\O(N)$ singlet bilinear operators $\phi^2$ and $T^{\mu\nu}$ below.

\begin{figure}[!ht]
    \centering
    \begin{tikzpicture}[scale=0.9, every node/.style={font=\small}]
        \tikzset{defect/.style={line width=1.6pt}, prop/.style={line width=0.6pt}, vtx/.style={circle, fill=black, inner sep=1.6pt}, bulk/.style={circle, draw, fill=white, inner sep=1.4pt}}
        \begin{scope}[xshift=0cm]
            \draw[defect] (-1.3,0) -- (1.3,0);
            \node[vtx] (v1) at (0,0) {};
            \node[bulk] (x) at (-0.9,1.3) {};
            \node[bulk] (y) at (0.9,1.3) {};
            \draw[prop] (x) -- (v1) -- (y);
            \node[above left=-2pt] at (x) {$x$};
            \node[above right=-2pt] at (y) {$y$};
        \end{scope}
        \begin{scope}[xshift=4.2cm]
            \draw[defect] (-1.6,0) -- (1.6,0);
            \node[vtx] (v1) at (-0.6,0) {};
            \node[vtx] (v2) at (0.6,0) {};
            \node[bulk] (x) at (-1.2,1.3) {};
            \node[bulk] (y) at (1.2,1.3) {};
            \draw[prop] (x) -- (v1);
            \draw[prop] (v2) -- (y);
            \draw[prop] (v1) to[bend left=60] (v2);
            \node[above left=-2pt] at (x) {$x$};
            \node[above right=-2pt] at (y) {$y$};
        \end{scope}
        \begin{scope}[xshift=9cm]
            \draw[defect] (-2.1,0) -- (2.1,0);
            \node[vtx] (v1) at (-1.2,0) {};
            \node[vtx] (v2) at (0,0) {};
            \node[vtx] (v3) at (1.2,0) {};
            \node[bulk] (x) at (-1.8,1.3) {};
            \node[bulk] (y) at (1.8,1.3) {};
            \draw[prop] (x) -- (v1);
            \draw[prop] (v3) -- (y);
            \draw[prop] (v1) to[bend left=60] (v2);
            \draw[prop] (v2) to[bend left=60] (v3);
            \node[above left=-2pt] at (x) {$x$};
            \node[above right=-2pt] at (y) {$y$};
        \end{scope}
        \node at (12.2,0.6) {$\cdots$};
    \end{tikzpicture}
    \caption{The chain diagrams for the bulk two-point function $\langle\,\phi_A(x)\,\phi_B(y)\,\rangle_{h_0}$.
    The thick line represents the defect, black circles are the vertices $-h_0$, and every line is the free propagator $G_0(|\hat k|; x_\perp)$ at fixed parallel momentum $\hat k$.}
    \label{fig:chain_bulk}
\end{figure}
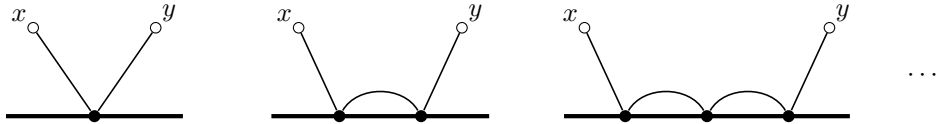

The one-point functions of $\phi^2$ and $T^{\mu\nu}$ follow from the bulk propagator $\langle\,\phi_A(x)\,\phi_B(y)\,\rangle_{h_0}$, which can be calculated exactly by summing up the chain diagrams in figure \ref{fig:chain_bulk} to all orders in $h_0$.
By taking Fourier transform along the parallel direction to the defect, the propagator becomes
\begin{align}\label{O(N)_phi2_bulk_propagator}
     \langle\,\phi_A(x)\,\phi_B(y)\,\rangle_{h_0}(\hat k)
        =
            \delta_{AB}\Big[\,G_0(|\hat k|; x_\perp - y_\perp)
            -
            \frac{2\,h_0}{1 +2\,h_0\,G_0(|\hat k|; 0)}\,G_0(|\hat k|; x_\perp)\,G_0(|\hat k|; y_\perp)\,\Big]\ ,
\end{align}
where $G_0(|\hat k|; x_\perp)$ is the free propagator at fixed parallel momentum $\hat k$:
\begin{align}
    G_0(|\hat k|; x_\perp)
        &:=
            \int\!\frac{\d^{q}k_\perp}{(2\pi)^{q}}\,
            \frac{e^{\,\i\,k_\perp\cdot x_\perp}}{\hat k^2+k_\perp^2}
        =
            \frac{1}{(2\pi)^\frac{q}{2}}
            \left(\frac{|\hat k|}{|x_\perp|}\right)^{\frac q2-1}
            K_{\frac q2-1}\!\left(|\hat k|\,|x_\perp|\right) \ .
\end{align}
On the defect, $G_0$ simplifies to
\begin{align}
    G_0(|\hat k|; 0)
        =
            (4\pi)^{\frac{\epsilon}{2}-1}\Gamma\left(\frac{\epsilon}{2}\right)\,|\hat k|^{-\epsilon} \ .
\end{align}

By Fourier transforming to the position space and setting $h_0$ to the IR fixed point value $h_0^{-1} = 0$, we obtain the exact bulk propagator:
\begin{align}
    \langle\,\phi_A(x)\,\phi_B(y)\,\rangle_{h_\ast}
        =
            \langle\,\phi_A(x)\,\phi_B(y)\,\rangle
            +
            \langle\,\phi_A(x)\,\phi_B(y)\,\rangle_{h_\ast}^\text{def} \ ,
\end{align}
where the first term is the free bulk propagator without defect given in \eqref{eq:GFF_bulk_two_point} while the second term represents the contribution from the defect to the propagator:
\begin{align}
    \begin{aligned}
    \langle\,\phi_A(x)\,\phi_B(y)\,\rangle_{h_\ast}^\text{def}
        =
            - \frac{\delta_{AB}}{\pi^\frac{q}{2}\,\Gamma\left(\frac{\epsilon}{2}\right)}\,\int &\frac{\d^{p}\hat k}{(2\pi)^{p}}\,e^{\i\,\hat k \cdot(\hat x - \hat y)}\, |x_\perp|^\frac{\epsilon}{2}\,|y_\perp|^\frac{\epsilon}{2}\,
            K_\frac{\epsilon}{2}\left( |\hat k|\,|x_\perp|\right)\,K_\frac{\epsilon}{2}\left( |\hat k|\,|y_\perp|\right) \ .
    \end{aligned}
\end{align}
The defect part can be rewritten by using the integral representation \eqref{eq:Bessel_K_int_rep} of the modified Bessel function $K_\nu$ and performing the $\hat k$ integral as
\begin{align}\label{Free_O(N)_phi2_propagator}
    \langle\,\phi_A(x)\,\phi_B(y)\,\rangle_{h_\ast}^\text{def}
        =
            -\frac{\delta_{AB}\,\Gamma\left(\frac{p}{2}\right)}{4\,\pi^\frac{d}{2}\,\Gamma\left(\frac{\epsilon}{2}\right)}\,|x_\perp|^\epsilon\,\int_0^1\d\xi\,\frac{\xi^{\Delta_\phi - 1}(1-\xi)^{\hD_{\hphi} - 1}}{\left[ \,(1-\xi)\,|x_\perp|^2 + \xi\,|y_\perp|^2 + \xi(1-\xi)\,\hat z^2\, \right]^{\frac{p}{2}}} \ ,
\end{align}
where $\hat z = \hat x - \hat y$.

By taking the coincident limit $y\to x$ of the bulk propagator \eqref{O(N)_phi2_bulk_propagator} and substituting $h_0 \to h_\ast$ with the relation \eqref{Free_O(N)_h_0_h}, we find the one-point coefficient of $\phi^2$:
\begin{align}\label{Free_O(N)_a_phi2}
    a_{\phi^2}
        =
            -\,\frac{N\;
                     \Gamma\!\left(\Delta_\phi\right)
                     \Gamma\!\left(\hD_{\hphi}\right)}%
                    {2^{\,p+1}\,\pi^{\frac{p+1-\epsilon}{2}}\,
                     \Gamma\!\left(\frac{\epsilon}{2}\right)
                     \Gamma\!\left(\frac{p+1}{2}\right)} \ ,
\end{align}
which is of order $\epsilon$.
Similarly, the one-point function of the stress tensor is obtained from the bulk propagator by acting with the derivatives followed by the coincident limit.
The one-point function takes the form \eqref{bulk 1-pt stress} with the coefficient
\begin{align}\label{Free_O(N)_a_T}
    a_T 
        =
             -\frac{\epsilon\;d\;\Delta_\phi}{2\,(d-1)\,(p+1)}\,a_{\phi^2} \ .
\end{align}
Since $a_{\phi^2} = O(\epsilon)$, it follows that $a_T$ vanishes at the leading order in $\epsilon$, i.e., $a_T = O(\epsilon^2)$.

Note that we do not use expansion in $\epsilon$ in the above derivations, and both $a_{\phi^2}$ and $a_T$ in \eqref{Free_O(N)_a_phi2} and \eqref{Free_O(N)_a_T} are exact.

\subsubsection{Defect local operators}
Since the localized mass term in the action \eqref{O(N)_phi2_action} is written as $h_0\,\int\d^p\hat x\,\hphi^2$, it can be regarded as a double-trace deformation of the generalized free field $\hphi_A$ of the undeformed theory with a trivial defect that we studied in section \ref{sec:GFF_on_defect}.
Indeed, the localized mass term is a slightly-relevant deformation with $\hD_{\hphi^2} = p-\epsilon$, which triggers an RG flow from the generalized free theory of $\hphi_A$ with $\hD_{\hphi} = \frac{p-\epsilon}{2}$ to the IR fixed point \eqref{Free_O(N)_h_ast} where $\hphi_A$ becomes the renormalized operator $[\hphi_A]$ of dimension
\begin{align}
    \hD_{[\hphi]} = p - \Delta_{\phi} = \frac{p+\epsilon}{2} \ ,
\end{align}
which may be seen as the ``shadow'' operator of $\hphi_A$ \cite{Klebanov:1999tb,Gubser:2002vv}.
Namely, $\hphi_A$ has the anomalous dimension $\gamma_{\hphi} = \hD_{[\hphi]} - \Delta_\phi = \epsilon$ under the flow.
This can be checked directly from the calculation of the two-point function of $[\hphi_A]$ to all orders in $h$ by introducing the renormalized operator in the minimal subtraction scheme through $[\hphi_A]_\text{min} := Z_{\hphi}^{-1}\,\hphi_A$ \cite{Ge:2025fsm,Giombi:2023dqs}.
The renormalization factor $Z_{\hphi}$ is given by $Z_{\hphi} = 1 - \frac{h}{\pi\,\epsilon}$, which yields the anomalous dimension $\gamma_{[\hphi]} = \partial \log Z_{\hphi}/\partial \log M|_{h=h_\ast} = \epsilon$ as expected.
Also, the two-point function takes the form:
\begin{align}
    \langle\,[\hphi_A]_\text{min}(\hat x)\,[\hphi_B]_\text{min}(0)\,\rangle_{h_\ast}
        =
            \delta_{AB}\,\frac{C_{\hphi}^\text{min}}{|\hat x|^{2\hD_{\hphi}}} \ ,
\end{align}
where $C_{\hphi}^\text{min}$ is calculated in the minimal subtraction scheme as
\begin{align}
    C_{\hphi}^\text{min}
        =
            \frac{\Gamma(\hD_{\hphi})}{\pi^{\epsilon}\,M^{2\,\epsilon}\,\Gamma(1+\frac{\epsilon}{2})\,\Gamma(1-\frac{\epsilon}{2})\,\Gamma(\Delta_\phi)}\,C_\phi  \ .
\end{align}
By redefining the renormalized operator as $[\hphi_A] = M^\epsilon\,[\hphi_A]_\text{min}$ with the finite renormalization, i.e., setting $M=1$ after the minimal subtraction, one can remove the $M$ dependence from the two-point function:
\begin{align}
    \langle\,[\hphi_A](\hat x)\,[\hphi_B](0)\,\rangle_{h_\ast}
        =
            \delta_{AB}\,\frac{C_{\hphi}}{|\hat x|^{2\hD_{\hphi}}} \ ,
\end{align}
where
\begin{align}
    C_{\hphi}
        =
            \frac{\Gamma(\hD_{\hphi})}{\pi^{\epsilon}
            \,\Gamma(1+\frac{\epsilon}{2})\,\Gamma(1-\frac{\epsilon}{2})\,\Gamma(\Delta_\phi)}\,C_\phi \ .  
\end{align}
We will use this renormalization scheme in the rest of this section.

On the other hand, the generalized free fields $\hatt_{s,A}$ with $s\ge 1$ in the undeformed theory do not get renormalized under the flow and continue to have dimension 
\begin{align}
    \hD_{\hatt_s} 
        = 
            \Delta_\phi + s \qquad (s\ge 1) \ ,
\end{align}
with the same two-point function as \eqref{eq:GFF_tower_two_point} at the IR fixed point.

The set of operators $\{\,[\hphi_A], \hatt_{s\ge 1, A}\,\}$ thus forms the basis of defect composite primaries at the IR fixed point and the spectrum follows from the Wick contraction.
For instance, the two-point functions of the composite operators $[\hphi_{A_1}\cdots \hphi_{A_n}]$ factorize to the product of $n$ copies of that of $\hphi$, hence they have the anomalous dimension $\gamma_{[\hphi^n]} = n\,\epsilon$ \cite{Ge:2025fsm}.
In general, a defect primary consisting of $k$ copies of $[\hphi]$ has the anomalous dimension $\gamma = k\,\epsilon$.

The displacement operator can be read off from the conservation law $\partial_\mu T^{\mu i} = h\,\delta(x_\perp)\,\partial_\perp^i \phi^2$ as
\begin{align}\label{displacement_epsilonOne}
    \hat D^i
        =
            2\,h\,\normord{[\hphi_A]\,\hatt_1^{iA}} \ ,
\end{align}
whose dimension is protected
\begin{align}
    \hD_{\hat D}
        =
            \hD_{[\hphi]} + \hD_{\hatt_1}
        =
            p + 1 \ ,
\end{align}
as required from the general consideration \cite{Billo:2016cpy}.
The normalization of the two-point function is
\begin{align}
    C_{\hat D}
        =
            8\,\pi^2\,N\,\Delta_\phi\,C_{\hphi}\,C_\phi\, \epsilon^2 \ .
\end{align}

For later convenience we also provide the normalization of $[\hphi^2],\,\normord{[\hphi]\,\hatt^i_1},\,\normord{\hatt_1\cdot\hatt_1}$,  \begin{align}\label{noramlizing factors}
    \begin{aligned}
        C_{[\hphi^2]}
        &=
            2\,N\,C^2_{\hphi} \ ,\\
    C_{\normord{[\hphi]\,\hatt^i_1}}
        &=
            2\,N\,\Delta_\phi\, C_\phi\, C_{\hphi} \ ,\\
    C_{\normord{\hatt_1\cdot\hatt_1}}
        &=
        8\,N\,q\,\Delta_\phi^2\,C^2_\phi \ .
    \end{aligned}
\end{align}

\subsubsection{Bulk-defect operator expansion}
The DOE for the field $\phi_A$ becomes
\begin{align}\label{surface defect OE of phi}
    \phi_A(x)
        =
            b_{\phi\hphi}\,|x_\perp|^\epsilon\, [\hphi_A] (\hat x)
            +
            \sum_{s\ge 1}\,\frac{1}{\Gamma(s+1)}\,\hatt_{s,A}(\hat x; x_\perp) + \text{(descendants)} \ ,
\end{align}
where the DOE coefficients of the $\hatt_{s\ge 1}$ are exactly the same as the free values as there are no corrections to the two-point function $\langle\,\phi\,\hatt_{s}\,\rangle$.
The coefficient $b_{\phi\hphi}$ can be read off from the two-point function $\langle\,\phi\,[\hphi]\,\rangle$.
In the minimal subtraction scheme, 
\begin{align}
    b_{\phi\hphi}
        =
            \pi^\frac{\epsilon}{2}\,\Gamma\left(1-\frac{\epsilon}{2}\right) \ .
\end{align}
We also record the scheme-independent combination of the coefficients:
\begin{align}
    b_{\phi\hphi}\,\left(\frac{C_{\hphi}}{C_\phi}\right)^\frac{1}{2}
        =
            \left[ \frac{\Gamma\left(1-\frac{\epsilon}{2}\right)\,\Gamma\left(\frac{p+\epsilon}{2}\right)}{\Gamma\left(1+\frac{\epsilon}{2}\right)\,\Gamma\left(\frac{p-\epsilon}{2}\right)}\right]^\frac{1}{2}  
            \ .
\end{align}

For the composite operator $\normord{\phi^2}$, the expansion takes a similar form:
\begin{align}\label{Free_O(N)_phi2_phi2_OE}
    \begin{aligned}
        \normord{\phi^2}(x)
            ={}&
                \frac{a_{\phi^2}}{|x_\perp|^{\Delta_{\phi^2}}}\,\hat{\bm{1}}
                +
                b_{\phi\hphi}^2\,|x_\perp|^{2\,\epsilon}\,[\hphi^2]
                +
                2\,b_{\phi\hphi}\,|x_\perp|^{1+\epsilon}\,n_{i}\,\normord{[\hphi]\,\hatt_{1}^i}
                 \\
            &
            ~
            +
                |x_\perp|^2\,n_i\,n_j\,
                \left[\normord{\hatt_{1}^{\langle i}\,\hatt_{1}^{j\rangle}} + b_{\phi\hphi}\,|x_\perp|^{\epsilon}\,\normord{\hphi\,\hatt_{2}^{ij}}\right]\\
            &
            ~~
            +
                |x_\perp|^2\left[\,\frac{1}{q}\,\normord{\hatt_1\cdot\hatt_1} 
                -
                \frac{b_{\phi \hphi}^2}{p+2+2\,\epsilon}\,
                |x_\perp|^{2\,\epsilon}\,[[\hphi]\,[\hphi]]_{1,0}  \right] + \cdots\ .
    \end{aligned}
\end{align}
Here, the DOE coefficients are determined by the bulk-defect two-point functions in the minimal subtraction scheme.
For instance, the coefficient for $[\hphi^2]$ follows from the two-point function $\langle\,\normord{\phi^2}\,[\hphi^2]\,\rangle = 2\,N\,\langle\,\phi\,[\hphi]\,\rangle^2$ and $C_{[\hphi^2]} = 2\,N\,C_{\hphi}^2$\ .
Note that this expansion is exact in $\epsilon$ and reduces to \eqref{GFF_phi2_DOE} in the $\epsilon\to 0$ limit.

For the stress tensor, the expansion has the following structure
\begin{align}\label{Free_O(N)_phi2_T_OE}
    T^{\mu\nu}(x)
        =
            \sum_{\hO^I}\,|x_\perp|^{\,\hD_{\hO}-d}\,b_{T\,\hO^I}\,\mathcal{S}^{\mu\nu}_{I}\;\hO^{I}(\hat x)
            \, +\, (\text{descendants})\ ,
\end{align}
where $\mathcal{S}^{\mu\nu}_{I}$ stands for the tensor structures and $I$ labels the transverse and parallel indices of the defect primaries $\hO^I$.
The tensor structure $\mathcal{S}^{\mu\nu}_{I}$ and the DOE coefficients $b_{T\,\hO^I}$ for the defect primaries of dimension $\hD_{\hO}\le d+2\,\epsilon$ that appear in the expansion are collected in table \ref{tab:phi2_T_OE}.
In the $\epsilon \to 0$ limit, it reduces correctly to \eqref{GFF_T_DOE} for the generalized free theory.
\begin{table}[ht]
    \small
    \renewcommand{\arraystretch}{1.05}
    \setlength{\tabcolsep}{4pt}
    \centering
    \begin{tabular}{ccccc}
    \toprule
        $\hO$ & $\hD_{\hO}$ & $\hat b_{\hO}$& $\mathcal{S}^{\mu\nu}_{I}$ & $b_{T\hO}$ \\
    \midrule
        $\hat{\bm{1}}$ & $0$ & $1$&$\displaystyle \frac{q-1}{q\,d}\,\sfs_1^{\mu\nu} + \frac{1}{q}\,\sfs_2^{\mu\nu}$
            & $a_T$ \\
    \cmidrule(lr){1-5}
        \multirow{2}{*}{$[\hphi^2]$} & 
        \multirow{2}{*}{$p+\epsilon$} &
        \multirow{2}{*}{ $2\,N\,\zeta(p+\epsilon)\,C_{\hphi}$}
            &$\sfs_1^{\mu\nu}$ & $\displaystyle b^{(1)}=\frac{\epsilon^2}{2\,q\,(d-1)}\,b_{\phi\hphi}^2$ \\
            & & & $\sfs_2^{\mu\nu}$& $\displaystyle b^{(2)}=-\frac{p\,\epsilon }{q\,(d-1)}\,b^2_{\phi\hphi}$ \\
    \cmidrule(lr){1-5}
        \multirow{4}{*}{$\normord{[\hphi]\,\hatt_1^{\,i}}$} & \multirow{4}{*}{$p+1$} &
            \multirow{4}{*}{$0$}
            &$\sfs_1^{\mu\nu}n_i$ & $b^{(1)}=\displaystyle \frac{\epsilon}{q\,(d-1)}\,b_{\phi\hphi}$ \\
            & & & $\sfs_2^{\mu\nu}n_i$&$b^{(2)}=\displaystyle -\frac{\epsilon\,(p-\epsilon)}{2(d-1)}\,b_{\phi\hphi}$ \\
            & & &$\delta_i^{(\mu}n^{\nu)}-\frac{h^{\mu\nu}_\perp}{q}n_i$ &$b^{(3)}=\displaystyle \frac{\epsilon\, d}{d-1}\,b_{\phi\hphi}$ \\
    \cmidrule(lr){1-5}
        $\normord{\hatt_1\!\cdot\!\hatt_1}$ & $d$
         & $2\,N\,q\,(p-\epsilon)\,\zeta(d)\,C_\phi$
            &$\sfs_1^{\mu\nu}$& $b^{(1)}=\displaystyle \frac{1}{2\,q\,(d-1)}$ \\
    \cmidrule(lr){1-5}
        $\normord{\hatt_1^{\langle i}\hatt_1^{\,j\rangle}}$ & $d$ & $0$
            &$\delta^\mu_i\delta^\nu_j$& $b^{(4)}=\displaystyle \frac{d}{2(d-1)}$ \\
    \cmidrule(lr){1-5}
        \multirow{5}{*}{$\normord{[\hphi]\,\hatt_2^{\,ij}}$} & \multirow{5}{*}{$d+\epsilon$}
           & \multirow{5}{*}{$0$}
            &$\sfs_1^{\mu\nu}n_in_j$& $\displaystyle b^{(1)}= \frac{\epsilon}{q\,(d-1)}\,b_{\phi\hphi}$ \\
            & & &$\sfs_2^{\mu\nu}n_in_j$& $b^{(2)}=\displaystyle -\frac{\epsilon\,(p-\epsilon)}{4\,(d-1)}\,b_{\phi\hphi}$ \\
            & & &$n_j(\delta^{(\mu}_{i}n^{\nu)}-\frac{h^{\mu\nu}_\perp}{q}n_i)$& $b^{(3)}=\displaystyle \frac{\epsilon\,d}{d-1}\,b_{\phi\hphi}$ \\
            & & &$\delta^\mu_i\delta^\nu_j$& $b^{(4)}=\displaystyle -\frac{p-\epsilon}{2(d-1)}\,b_{\phi\hphi}$ \\
    \cmidrule(lr){1-5}
        \multirow{2}{*}{$[[\hphi]\,\hatt_1]^{\hat a i}_{0,1}$} & 
        \multirow{2}{*}{$d+\epsilon$} &
        \multirow{2}{*}{$0$}
            &$\delta^{(\mu}_{\ha}n^{\nu)}n_i$& $b^{(1)}=\displaystyle \frac{2\,\epsilon}{p+1}\,b_{\phi\hphi}$ \\
            & & &$\delta^{(\mu}_{\ha}\delta^{\nu)}_i$& $b^{(2)}=\displaystyle -\frac{2}{p+1}\,b_{\phi\hphi}$ \\
    \cmidrule(lr){1-5}
        \multirow{3}{*}{$[[\hphi]\,[\hphi]]_{1,0}$} & \multirow{3}{*}{$d+2\,\epsilon$} &
        \multirow{3}{*}{$2\,N\,(p+\epsilon)\,(d+3\epsilon)\,\zeta(d+2\,\epsilon)\,C_{\hphi}$}
            &$\sfs_1^{\mu\nu}$& $b^{(1)}=\displaystyle \frac{(1-\epsilon)(2+\epsilon)(p+2)}{2\,p\,q\,(d-1)(d+3\epsilon)}\,b_{\phi\hphi}^2$ \\
            & & &$\sfs_2^{\mu\nu}$& $b^{(2)}=\displaystyle \frac{\epsilon\,(p+2)}{q\,(d-1)(d+3\epsilon)}\,b_{\phi\hphi}^2$ \\
    \cmidrule(lr){1-5}
        $[[\hphi]\,[\hphi]]^{\hat a\hat b}_{0,2}$ & $d+2\,\epsilon$
            & $4\,N\,(p+\epsilon)\,(p+1+\epsilon)\,\zeta(d+2\,\epsilon)\,C_{\hphi}$
            &$\delta^\mu_{\ha}\delta^\nu_{\hb}$& $\displaystyle -\frac{p+2+\epsilon}{2\,(p+1+\epsilon)}\,b_{\phi\hphi}^2$ \\
    \bottomrule
    \end{tabular}
    \caption{The DOE \eqref{Free_O(N)_phi2_T_OE} of the stress tensor for the defect primaries of dimension $\hD_{\hO}\le d+2\,\epsilon$ at the IR fixed point $h=h_\ast$, in the convention of \eqref{app:tensor OE structures}. All the coefficients are exact in $\epsilon$. We used $b^{(I)}$ for $b^{(I)}_{\CO\hO}$, and omitted vanishing coefficients.
    }
    \label{tab:phi2_T_OE}
\end{table}

\subsection{Thermal one-point functions}\label{ss:Free_O(N)_finiteT}
Having the calculation of the thermal one-point function of $\phi^2$ and $T^{\mu\nu}$ in mind, let us consider the bulk propagator at finite temperature, which can be obtained by replacing the momentum $\hat k$ in the free propagator \eqref{O(N)_phi2_bulk_propagator} with the thermal counterpart $(\omega_n, k_\parallel)$ with $\omega_n = 2\pi n/\beta$:
\begin{align}\label{O(N)_phi2_bulk_propagator_thermal}
     \langle\,\phi_A(x)\,\phi_B(y)\,\rangle_{h_0,\, \beta}(\omega_n, k_\parallel)
        =
            \delta_{AB}\Big[\,G_0(m_n; x_\perp - y_\perp)
            -
            \frac{2\,h_0}{1 +2\,h_0\,G_0(m_n; 0)}\,G_0(m_n; x_\perp)\,G_0(m_n; y_\perp)\,\Big]\ ,
\end{align}
where $m_n := \sqrt{\omega_n^2 + k_\parallel\cdot k_\parallel}$\ .
By Fourier transforming to the position space and setting $h_0$ to the IR fixed point value $h_0^{-1} = 0$, we obtain the thermal propagator:
\begin{align}\label{Free_O(N)_thermal_propagator}
    \langle\,\phi_A(x)\,\phi_B(y)\,\rangle_{h_\ast,\, \beta}
        =
            \langle\,\phi_A(x)\,\phi_B(y)\,\rangle_{\beta}
            +
            \langle\,\phi_A(x)\,\phi_B(y)\,\rangle_{h_\ast,\, \beta}^\text{def} \ .
\end{align}
Here, the first term is the bulk propagator without defect given in \eqref{GFF-phi-two-point} while the second term represents the contribution from the defect to the propagator:
\begin{align}\label{Free_O(N)_thermal_propagator_defect}
    \begin{aligned}
    \langle\,\phi_A(x)\,\phi_B(y)\,\rangle_{h_\ast,\, \beta}^\text{def}
        =
            - \frac{\delta_{AB}}{\pi^\frac{q}{2}\,\Gamma\left(\frac{\epsilon}{2}\right)}\,\frac{1}{\beta}\sum_{n\in\BZ}\int &\frac{\d^{p-1}k_\parallel}{(2\pi)^{p-1}}\,e^{\i\,\omega_n(\tau_x-\tau_y)}\,e^{\i\, k_\parallel\cdot(x_\parallel-y_\parallel)}\\
                &~\cdot |x_\perp|^\frac{\epsilon}{2}\,|y_\perp|^\frac{\epsilon}{2}\,K_\frac{\epsilon}{2}\left( m_n\,|x_\perp|\right)\,K_\frac{\epsilon}{2}\left( m_n\,|y_\perp|\right) \ .
    \end{aligned}
\end{align}
The defect part can be rewritten by using the integral representation \eqref{eq:Bessel_K_int_rep} of the modified Bessel function $K_\nu$ and performing the Poisson resummation,
as
\begin{align}
    \langle\,\phi_A(x)\,\phi_B(y)\,\rangle_{h_\ast,\, \beta}^\text{def}
        =
            -\frac{\delta_{AB}\,\Gamma\left(\frac{p}{2}\right)}{4\,\pi^\frac{d}{2}\,\Gamma\left(\frac{\epsilon}{2}\right)}\,|x_\perp|^\epsilon\,\sum_{w\in \BZ}\int_0^1\d\xi\,\frac{\xi^{\Delta_\phi - 1}(1-\xi)^{\hD_{\hphi} - 1}}{\left[ \,(1-\xi)|x_\perp|^2 + \xi\,|y_\perp|^2 + \xi(1-\xi)\,|\hat z_w|^2\, \right]^{\frac{p}{2}}} \ ,
\end{align}
where $\hat z_w := \left( \tau_x - \tau_y + w\beta,\, x_\parallel - y_\parallel\right)$.\footnote{We use the identity that follows from the resummation formula \eqref{eq:Poisson_resummation} and Gaussian integral:
\begin{align}
    \frac{1}{\beta}\sum_{n\in\BZ}\int \frac{\d^{p-1}k_\parallel}{(2\pi)^{p-1}}\,e^{\i\,(\omega_n\tau + k_\parallel\cdot x_\parallel)}\,e^{-a\,m_n^2}
        =
        (4\pi a)^{-\frac{p}{2}}\sum_{w\in \BZ}\,e^{-\frac{1}{4a}\left[ (\tau + w\beta)^2 + |x_\parallel|^2\right]} \ .
\end{align}
}

\paragraph{Thermal one-point functions of bilinear defect primaries.}
The thermal one-point functions of the bilinear defect operators are obtained from the bulk thermal propagator \eqref{Free_O(N)_thermal_propagator} by acting with derivatives and taking the defect limit.
Alternatively, there is a quicker derivation of the thermal one-point coefficient $\hat b_{\hO}$ that employs the fact that the defect theory is a generalized free field theory generated by $[\hphi_A]$ and $\hatt_{s\ge 1, A}$.
In section \ref{sec:GFF_bilinears}, the coefficients $\hat b_{\hO}$ for the bilinears in the generalized free theory for the trivial defect have been derived, and 
the coefficients of the present model can be determined similarly by applying the Wick contraction.
In either way, one obtains the coefficients $\hat b_{\hO}$ for the bilinear defect primaries listed in table \ref{tab:phi2_T_OE}.

\paragraph{Thermal one-point function of $\phi^2$.}

By taking the coincident limit $y\to x$ of the thermal propagator \eqref{Free_O(N)_thermal_propagator}, we find the thermal one-point function of $\normord{\phi^2}$:
\begin{align}\label{Free_ON_thermal_1pt_phi2}
    \langle\,\normord{\phi^2}(x)\,\rangle_{h_\ast,\,\beta}
        =
            \langle\,\normord{\phi^2}(x)\,\rangle_{\beta}
            +
            \langle\,\normord{\phi^2}(x)\,\rangle_{h_\ast,\,\beta}^\text{def} \ ,
\end{align}
where the first term is the thermal one-point function \eqref{GFF-phi2} without defect and the second term is
\begin{align}\label{Free_O(N)_1pt_phi2_defect}
    \langle\,\normord{\phi^2}(x)\,\rangle_{h_\ast,\,\beta}^\text{def} 
        =
            \frac{f(z)}{\beta^{\Delta_{\phi^2}}}\ ,
\end{align}
where
\begin{align}
    f(z) 
        :=              
            -\CN_p\,z^\epsilon\,
            \CI_{\frac{p}{2}}[\Delta_\phi - 1, \hD_{\hphi} - 1](z) \ ,
            \qquad
            \CN_p 
                :=
                \frac{N\,\Gamma\left(\frac{p}{2}\right)}{4\,\pi^\frac{d}{2}\,\Gamma\left(\frac{\epsilon}{2}\right)} \ ,
\end{align}
and 
\begin{align}
    \CI_\nu [a,b](z)
        :=
            \sum_{w\in \BZ}\int_0^1\d\xi\,\frac{\xi^{a}(1-\xi)^{b} }{\left[ \,z^2 + \xi(1-\xi)\,w^2\, \right]^{\nu}} \ .
\end{align}

\paragraph{Thermal one-point function of $T^{\mu\nu}$.}

The thermal one-point function of the stress tensor becomes
\begin{align}\label{Free_O(N)_thermal_1pt_stress_tensor}
    \langle\,T^{\mu\nu}\,\rangle_{h_\ast,\,\beta}
        =
            \langle\,T^{\mu\nu}\,\rangle_{\beta}
            +
            \langle\,T^{\mu\nu}\,\rangle_{h_\ast,\,\beta}^\text{def} \ ,
\end{align}
where the first term is the bulk part, i.e., the thermal one-point function \eqref{free_scalar_bT} without defect.
The second term is the contribution from the defect, which takes the following form from the general consideration \eqref{general_stress_tensor_thermal_1pt}:
\begin{align}\label{T_defect_part}
    \langle\,T^{\mu\nu}\,\rangle_{h_\ast,\,\beta}^\text{def}
        =
            \frac{1}{\beta^d}\,\sum_{i=1}^3\,F_i(z)\,\sfs_i^{\mu\nu} \ .
\end{align}
In the present case, the functions $F_i$ are given by
\begin{align}\label{Free_O(N)_T_thermal_1pt_F}
    \begin{aligned}
        F_1
            &=
                \frac{\CN_p\,z^\epsilon}{q\,(d-1)}\,
                \left[
                    \frac{\epsilon\,p}{2}\,\CI_{\frac{p}{2}+1}[\Delta_\phi, \hD_{\hphi}-1]  
                    -
                    (q+d-2)\,\CI_{\frac{p}{2}+1}[\Delta_\phi, \hD_{\hphi}] 
                    +
                    (q-1)(p+2)\,z^2\,\CI_{\frac{p}{2}+2}[\Delta_\phi, \hD_{\hphi}]               
                \right] \ , \\
        F_2
            &=
                -\frac{p\,\CN_p\,z^\epsilon}{q\,(d-1)}\,
                \left[ 
                    p\,\CI_{\frac{p}{2}+1}[\Delta_\phi, \hD_{\hphi}-1]  
                    -
                    (d-2)\,\CI_{\frac{p}{2}+1}[\Delta_\phi, \hD_{\hphi}] 
                    -
                    (p+2)\,z^2\,\CI_{\frac{p}{2}+2}[\Delta_\phi, \hD_{\hphi}]               
                \right]\ , \\
        F_3
            &=
                -(p+2)\,\CN_p\,z^\epsilon\,\left[\,\CI_{\frac{p}{2}+1}[\Delta_\phi, \hD_{\hphi}] -z^2\,\CI_{\frac{p}{2}+2}[\Delta_\phi, \hD_{\hphi}] \right] \ .
    \end{aligned}
\end{align}

\subsection{High and low temperature limits}\label{ss:Free_O(N)_high_lowT}
We examine the high and low temperature limits of the thermal one-point functions obtained in section \ref{ss:Free_O(N)_finiteT}.
In the high temperature limit, we read off the decay rate of the defect-induced part, while we reproduce the low temperature expansion from the DOE using the thermal data obtained in the previous section.
The following results are valid for $p\ge 2$, and we discuss the $p=1$ case in section \ref{ss:Free_O(N)_p1}.

\subsubsection{High temperature limit}
In the high temperature limit, $\beta \to 0$, the thermal one-point function of $\normord{\phi^2}$ given by \eqref{Free_ON_thermal_1pt_phi2} becomes
\begin{align}\label{Free_O(N)_phi2_highT}
    \langle\,\normord{\phi^2}(x)\,\rangle_{h_\ast,\,\beta}
        =
            \frac{1}{\beta^{\Delta_{\phi^2}}}\left[\,b_{\phi^2}
            +
            \frac{\tilde a_{\phi^2}}{z^{p-1-\epsilon}} 
            +
            O\left(e^{-4\pi z}\right)
            \right] \ ,
\end{align}
where we use the asymptotic form \eqref{eq:app_Inu_highT_general} of the function $\CI_\nu[a,b]$ around $z=\infty$ and $\tilde a_{\phi^2}$ is related to the one-point coefficient $a_{\phi^2}$ at zero temperature given in \eqref{Free_O(N)_a_phi2} as
\begin{align}
    \tilde a_{\phi^2}
    :=
        a_{\phi^2}|_{p\to p-1}
    =
        -
            \frac{N\,\Gamma(\Delta_\phi-\frac{1}{2})\,\Gamma(\hD_{\hphi}-\frac{1}{2})}{2^p\,\pi^\frac{p-\epsilon}{2}\,\Gamma(\frac{\epsilon}{2})\,\Gamma(\frac{p}{2})}\ .
\end{align}
The power-law behavior in \eqref{Free_O(N)_phi2_highT} and the reduction of the coefficient above are the consequence of the massless mode on $\BS_\beta \times \BR^{d-1}$ as seen for the free scalar theory with a boundary in section \ref{sec:free_scalar_BCFT_thermal_1pt_phi2}.

The one-point function of the stress tensor in the high temperature limit is obtained in a similar manner:
\begin{align}
    \begin{aligned}
        \langle\,T^{\mu\nu}\,\rangle_{h_\ast,\,\beta}
            =
                \frac{1}{\beta^d}\,&\left[\, b_T\left(e^\mu\,e^\nu - \frac{\delta^{\mu\nu}}{d} \right)\right. \\
            &~~+ 
                \frac{(d-3)\,\tilde a_{\phi^2}}{4\,p^2\,q\,z^{d-1}}
                \left(
                    \frac{(p-1)\,(q-1)^2}{d-1}\,\sfs_1^{\mu\nu}
                    +
                    p\,(q-1)\,\sfs_2^{\mu\nu}
                    +
                    q\,(p+1-q)\,\sfs_3^{\mu\nu}
                \right)\\
            &~~~~+ \left. 
                    O\left(e^{-4\pi z}\right)\,
                \right] \ .
    \end{aligned}
\end{align}

\subsubsection{Low temperature limit}

The low temperature limit $(\beta \to \infty)$ of $\langle\,\normord{\phi^2}\,\rangle_{h_\ast,\,\beta}$ is read off from \eqref{Free_ON_thermal_1pt_phi2} by using the asymptotic form \eqref{eq:app_Inu_lowT_general} of the function $\CI_\nu[a,b]$ around $z= 0$ as
\begin{align}\label{Free_O(N)_phi2_lowT}
    \begin{aligned}
    \langle\,\normord{\phi^2}(x)\,\rangle_{h_*,\,\beta}
        ={}&
            \frac{a_{\phi^2}}{|x_\perp|^{p-\epsilon}}
            +
            2\,N\,\zeta(p+\epsilon)\,b_{\phi\hphi}^2\,C_{\hphi}\,
            \frac{|x_\perp|^{2\,\epsilon}}{\beta^{p+\epsilon}}
            +
            4\,N\,\Delta_\phi\,\zeta(d)\,\Cphi\,\frac{|x_\perp|^2 }{\beta^{d}}\\
            &-
            4\,N\,\hD_{\hphi}\,\zeta(d+2\,\epsilon)\,b_{\phi\hphi}^2\,C_{\hphi}\,
            \frac{|x_\perp|^{\,2+2\,\epsilon}}{\beta^{\,d+2\,\epsilon}}
            +
            O\!\left(\frac{|x_\perp|^{4}}{\beta^{\,d+2}}\right) \ ,
    \end{aligned}
\end{align}
where the bulk part \eqref{GFF-phi2} is canceled by the $O(\beta^{2-d})$ term of the defect part.

For the stress tensor, the low temperature expansion becomes
\begin{align}\label{Free_O(N)_stress_tensor_lowT}
    \begin{aligned}
    \langle\,T^{\mu\nu}(x)\,\rangle_{h_*,\,\beta} 
        &=
            \frac{a_T}{q\,|x_\perp|^{\,d}}
            \left[\,\frac{q-1}{d}\,\sfs_1^{\mu\nu} + \sfs_2^{\mu\nu}\,\right] \\
            &~~
            +
            \frac{\epsilon\,b_{\phi\hphi}^2\,\hat b_{[\hphi^2]}}{q\,(d-1)}\,\frac{|x_\perp|^{\,2\,\epsilon-2}}{\beta^{\,p+\epsilon}}
            \left[\,\frac{\epsilon}{2}\,\sfs_1^{\mu\nu} - p\,\sfs_2^{\mu\nu}\,\right] 
            +
            \frac{\hat b_{\hatt_1\cdot\hatt_1}}{2\,q\,(d-1)\,\beta^{\,d}}
            \,\sfs_1^{\mu\nu}
        \\
            &~~ +
            \frac{b_{\phi\hphi}^2\,\hat b_{[[\hphi][\hphi]]_{1,0}}}{p+2+2\,\epsilon}\,\frac{|x_\perp|^{\,2\,\epsilon}}{\beta^{\,d+2\,\epsilon}}
            \left[\,
                \frac{p+2}{q\,(d-1)}\,\left(  
                \frac{(1-\epsilon)(2+\epsilon)}{2\,p}\,\sfs_1^{\mu\nu}
                + 
                \epsilon\,\sfs_2^{\mu\nu}\right)
                \right. \\
            &\hspace*{6.2cm}
                \left.
                +
                \frac{p+2+\epsilon}{p}\,\sfs_3^{\mu\nu}
            \,\right]
            +
            O\!\left(\frac{|x_\perp|^{2}}{\beta^{\,d+2}}\right) \ .
    \end{aligned}
\end{align}

Using the DOEs \eqref{Free_O(N)_phi2_phi2_OE} and \eqref{Free_O(N)_phi2_T_OE} and substituting the values listed in table \ref{tab:phi2_T_OE} reproduce the low temperature expansions \eqref{Free_O(N)_phi2_lowT} and \eqref{Free_O(N)_stress_tensor_lowT} as expected.

\subsection{$p=1$ case}\label{ss:Free_O(N)_p1}
For $p=1$, the defect becomes a line wrapping the thermal circle.
The rotation group $\SO(p)$ on the defect is trivial and there are no parallel spins.
Thus the symmetric traceless part of the defect primaries with parallel tensor structures such as $[\hphi\,\hphi]^{\hat a\,\hat b}_{0,2}$ no longer exists.
Moreover, the tensor structure $\sfs_3^{\mu\nu} = h^{\mu\nu} - p\,e^\mu e^\nu$ is absent as the parallel metric $h_{\mu\nu}$ becomes $h_{\mu\nu} = e_\mu e_\nu$ for line defects.
This modifies the thermal one-point function of the stress tensor \eqref{Free_O(N)_T_thermal_1pt_F} to
\begin{align}
    \langle\,T^{\mu\nu}\,\rangle^\text{def}_{h_\ast,\,\beta}
        =
            \frac{1}{\beta^d}\left[\,F_1\,\sfs_1^{\mu\nu} 
                +
                F_2\,\sfs_2^{\mu\nu} 
            \right] \ ,
\end{align}
with $F_1, F_2$ given by setting $p=1$ in \eqref{Free_O(N)_T_thermal_1pt_F}.

While most of the results for $p\ge 2$ appear to extend to $p=1$, there are some subtleties in this continuation, which are worth commenting on.
This is most clearly seen in the high temperature expansion of the thermal one-point function $\langle\,\normord{\phi^2}(x)\,\rangle_{h_\ast,\,\beta}$ in \eqref{Free_O(N)_phi2_highT}, where the defect contribution proportional to $z^{1+\epsilon - p}$ dominates over the bulk contribution of order $O(1)$ in the $z\to \infty$ limit.
This is in conflict with the expectation that the one-point function should approach the bulk part when the bulk operator is far away from the defect.
To understand the origin of this growing behavior of the defect contribution, we use the following representation
\begin{align}
    \begin{aligned}
    \langle\,\normord{\phi^2}(x)\,\rangle_{h_\ast,\, \beta}^\text{def}
        =
            - \frac{N}{\pi^\frac{q}{2}\,\Gamma\left(\frac{\epsilon}{2}\right)}\,\frac{1}{\beta}\sum_{n\in\BZ}\, |x_\perp|^{\epsilon}\,K_\frac{\epsilon}{2}\left( m_n\,|x_\perp|\right)^2 \ ,
    \end{aligned}
\end{align}
which can be obtained from \eqref{Free_O(N)_thermal_propagator_defect} by taking the coincident limit.
Since there are no longitudinal momenta $k_\parallel$ the thermal mass is linear in the Matsubara mode $n$, i.e., $m_n = |\omega_n| = |2\pi n/\beta|$, and there exists the zero mode at $n=0$.
The $n=0$ term proportional to $K_\frac{\epsilon}{2}(0)^2$ is infinite for any $\epsilon$.
This is an IR divergence arising from the zero mode.
A similar IR divergence also appears in the bulk one-point function, which takes the form in momentum space
\begin{align}\label{phi2_bulk_part_momentum}
    \langle\,\normord{\phi^2}(x)\,\rangle_{\beta}
        =
            \frac{N}{\beta}\,\sum_{n\in\BZ}\int\frac{\d^{d-1}\vec k}{(2\pi)^{d-1}}\,\frac{1}{\vec k^2 + \omega_n^2} 
            -
            N\,\int\frac{\d^{d}k}{(2\pi)^{d}}\,\frac{1}{k^2} \ .
\end{align}
When $p=1$, the momentum integral for the zero mode $(n=0)$ term is divergent around $\vec k =0$ as $d - 1 = 2-\epsilon < 2$.\footnote{The second term (the normal ordering term) in \eqref{phi2_bulk_part_momentum} is zero in dimensional regularization.}

To regularize the IR divergence, one may introduce a small mass $\mu$ to the thermal mass as $\omega_n^2 \to \omega_n^2 + \mu^2$ as a regulator.
Then the zero mode contribution to the thermal one-point function becomes
\begin{align}\label{phi2_zero_mode}
     \langle\,\normord{\phi^2}(x)\,\rangle_{h_\ast,\, \beta}\big|_{n=0}
        =
            \frac{N}{\beta}\left[\,\frac{\Gamma(\frac{\epsilon}{2})}{(4\,\pi)^{1-\frac{\epsilon}{2}}\,\mu^\epsilon} 
            -
            \frac{|x_\perp|^\epsilon}{\pi^{1-\frac{\epsilon}{2}}\,\Gamma(\frac{\epsilon}{2})}\,K_{\frac{\epsilon}{2}}(\mu\,|x_\perp|)^2
            \right] \ .
\end{align}
The bulk $\mu^{-\epsilon}$ term is canceled by the defect zero mode in the massless $\mu \to 0$ limit, and the zero mode contribution \eqref{phi2_zero_mode} reproduces the $z^{1+\epsilon - p}$ term \eqref{Free_O(N)_phi2_highT} in the high temperature limit.
Thus the growing behavior in $\langle\,\normord{\phi^2}(x)\,\rangle_{h_\ast,\, \beta}$ may be understood as the fluctuation of the zero mode.

\subsection{Interface case ($\epsilon = 1$)}\label{ss:Free_O(N)_ep1}
It has been hypothesized decades ago that an interface CFT factorizes into two copies of decoupled BCFTs in \cite{Bray:1977fvl}, where the factorization for the bulk-free interface theory   \begin{align}
    I=\frac12\int \d^dx\,(\partial\phi_A)^2 
            +h_0\int \d^{d-1}\hx\,\phi_A^2(\hx),
\end{align}
is shown at the IR fixed point $h_0\to \infty$.
The factorization is also hypothesized and verified using the large-$N$ and $\epsilon$-expansion for bulk interacting theories with an interface. 
See e.g., \cite{Bray:1977fvl, Krishnan:2023cff,Raviv-Moshe:2023yvq, Diatlyk:2024ngd, Popov:2025cha, Ge:2025fsm} for a detailed discussion.

Following the bulk-free discussion in \cite{Raviv-Moshe:2023yvq}, we analytically continue to $\epsilon=1$, where the theory becomes an interface CFT. We comment on the dictionary introduced in \cite{deSabbata:2024xwn} between the interface primaries and the boundary primaries in the Dirichlet BCFT investigated in section \ref{ss:free scalar BCFT}.
We then perform a test of the factorization property by comparing the thermal one-point functions in both theories.
To compare with section \ref{ss:free scalar BCFT}, we set $N=1$ in this section.

First, the factorization can be observed already from zero-temperature calculation.
After analytically continued to $\epsilon=1$, the two-point function simplifies to
\begin{align}
    \langle\,\phi(x)\,\phi(y)\,\rangle_{h_\ast}\big|_{\epsilon=1}
        =
            C_{\phi}\left[\frac{1}{[(\hx-\hy)^2+(x_\perp-y_\perp)^2]^{\frac{p-1}{2}}}-\frac{1}{[(\hx-\hy)^2+(x_\perp+y_\perp)^2]^{\frac{p-1}{2}}}\right],
\end{align}
which is precisely the bulk free Dirichlet BCFT propagator \eqref{eq:scalar_BCFT_bulk_2pt}.
Also, the one-point coefficient of $\phi^2$ at $\epsilon=1$ agrees with that in the Dirichlet BCFT \eqref{eq:BCFTphi2}, 
\begin{align}
    a_{\phi^2}|_{\epsilon=1}
        =
            -\frac{\Gamma(\frac{d-2}{2})}{2^d\,\pi^\frac{d}{2}}\ .
\end{align}
The bulk theories in the Dirichlet BCFT and the interface CFT are identical, hence we can compare the BOE and DOE of the bulk operators.
From a more elementary perspective, $\hat{\phi}$ and $\hat{t}_1$ become degenerate at $\epsilon=1$, which implies the following relation.
By comparing the BOE \eqref{scalar_BCFT_phi_BOE} of $\phi$ with DOE \eqref{surface defect OE of phi} \cite{deSabbata:2024xwn}
\begin{align}\label{eq:dic}
    \hat{t}^{\text{bdy}}_\text{L,R} 
        = 
            \hat\Phi_{\text{L,R}} \ ,
\end{align}
where the left-hand side represents the boundary operator (corresponding to $\hat{t}_1$ in section \ref{ss:free scalar BCFT}) with L,R denoting the Dirichlet BCFTs extending to $x_{\perp}<0,~x_{\perp}>0$, respectively, while the defect operators $\hat\Phi_{\text{L,R}}$ in the right-hand side are those in the interface CFT defined by
\begin{align}
    \hat\Phi_{\text{L,R}}
        \coloneqq
            \pi\,[\hphi] \mp \hatt_1 \ .
\end{align}
Note that the two-point function of $\hat\Phi_{\text{L}}$ and $\hat\Phi_{\text{R}}$ vanishes as is consistent with the conjectured factorization of the interface CFT into the two decoupled Dirichlet BCFTs.
It is then natural to expect analogous relations for the composite operators. For example, the following equation should hold
\begin{align}
    [\hatt^\text{bdy}_\text{L,R}\hatt^\text{bdy}_\text{L,R}]_{n,j}=[\hat\Phi_\text{L,R}\hat\Phi_\text{L,R}]_{n,j}\ ,
\end{align}
for the bilinear operators.
In particular, recall the displacement operators in the BCFTs and the interface CFT are given by 
\begin{align}
    \hat{D}^\text{bdy}_\text{L,R}
        &=
        \frac{1}{2}(\hatt^\text{bdy}_\text{L,R})^2 
        =
            \frac{\pi^2}{2}\,[\hphi^2]
            +
            \frac{1}{2}\,\normord{\hatt_1\cdot \hatt_1}
            \mp \pi\,\normord{[\hphi]\hatt_1} \ ,
        \\
    \hat{D}&=2\,\pi\,\normord{[\hphi]\hatt_1}\ ,
\end{align}
where we used \eqref{eq:dic} in the first line.
Then, we can confirm the standard relation (see \cite{Diatlyk:2024ngd})
\begin{align}\label{relation in Wang et.al}
    \hat{D}=\hat{D}^\text{bdy}_\text{R}-\hat{D}^\text{bdy}_\text{L}\ ,
\end{align}
together with the corresponding relation between the two-point-function normalization factors 
\begin{align}\label{eq:normalization relation}
    C^\text{bdy}_{\hat{D}}
        =
            \frac{\pi^4}{4}\,C_{[\hphi^2]}
            +
            \frac{1}{4}\,C_{\normord{\hatt_1\cdot \hatt_1}}
            +
            \pi^2\,C_{\normord{[\hphi]\hatt_1}}\ ,
\end{align}
which can be verified using the results in section \ref{sec:free_BCFT} and  \eqref{noramlizing factors}, analogous to the relation $C_{\hat{D}}=2\,C^\text{bdy}_{\hat{D}}$ obtained in \cite{Diatlyk:2024ngd}.
Furthermore, we find the factorization property at finite temperature.
In fact, setting $\epsilon=1$, the thermal one-point functions of $\phi^2$ and $T^{\mu\nu}$ become
\begin{align}
    \langle\,\normord{\phi^2}(x)\,\rangle_{h_\ast,\,\beta}^\text{def}\big|_{\epsilon=1}
        &=
            -\frac{C_\phi}{\beta^{d-2}}\sum_{w\in\BZ}\frac{1}{(w^2+4\,z^2)^{\frac{d-2}{2}}}\ ,\\
    \langle \,T^{\mu\nu}\,\rangle^\text{def}_{h_\ast,\,\beta}\big|_{\epsilon=1}
        &=
            \frac{d\,(d-2)\,C_\phi}{\beta^d}\sum_{w\in \BZ}\frac{w^2}{(w^2+4\,z^2)^{\frac{d+2}{2}}}\,
            \left(e^\mu e^\nu - \frac{h^{\mu\nu}}{d-1}\right) \ ,
\end{align}
which precisely coincide with those in \eqref{eq:BCFTphi2} and \eqref{eq:scalar BCFT stress thermal 1-pt}.

\section{Discussion}\label{sec:discussion}
In this paper, we studied the dynamics of DCFTs at finite temperature with a defect of $p$ dimensions wrapping the thermal circle.
We constrained the structures of the thermal one-point functions and their high and low-temperature expansions.
The relation between the low-temperature expansion of a bulk thermal one-point function and the thermal expectation value of the DOE was emphasized, which allows us to extract the thermal data $\hat b_{\hO}$ and to check the consistency. 

We examined the generalized free field theory on the trivial defect, listed the DCFT data and the spectrum of the defect primaries.
We then extended these data to several DCFT models with free bulk theories in general dimension, and calculated their thermal data exactly.
From the calculation in section \ref{ss:free scalar BCFT} and \ref{ss:Free_O(N)_ep1}, we verified the factorization of the interface CFT in the free $\O(N)$ model from a thermal perspective exactly, and obtained a dictionary between the interface and boundary primary operators.

The most direct extension of this work is to make the bulk theory or the defect theory interacting.
For instance, the interacting $\O(N)$ model in $d=4-\epsilon$ dimensions has a non-trivial fixed point even in the presence of the $\O(N)$-breaking localized $\phi$ deformation on a line \cite{Cuomo:2021kfm} and the $\O(N)$-symmetric localized $\phi^2$ deformation on a surface \cite{Shachar:2022fqk,Trepanier:2023tvb,Raviv-Moshe:2023yvq,Giombi:2023dqs,Diatlyk:2024ngd}.
In work in progress \cite{Li-Nakayama-Nishioka}, we calculate the DOEs of $\phi^2$ and $T^{\mu\nu}$ and the thermal correlation functions perturbatively in $\epsilon$ and extract the defect CFT data of these models.
The latter model is expected to factorize into two BCFTs when $\epsilon$ is continued to one \cite{Krishnan:2023cff,Diatlyk:2024ngd,Popov:2025cha}, which may be tested at finite temperature.

It would also be interesting to study the thermal aspects of DCFTs holographically.
The simplest approach is to introduce a probe brane in the AdS black hole background and calculate the thermal one-point function of the bulk operators from the backreaction of the brane \cite{Karch:2005ms} (see also \cite{Jensen:2013lxa,Kobayashi:2018lil} for the probe brane model for defects of general dimensions at zero temperature).
Another approach is to use the AdS/BCFT model \cite{Takayanagi:2011zk,Fujita:2011fp} and its generalization to CFTs with defects of general dimensions \cite{Nakayama:2025hqo}, where defects are holographically described by end-of-the-world branes.
At zero temperature, the one-point functions of bulk operators have been calculated and shown to reproduce the expected forms in DCFTs \cite{Fujita:2011fp,Kastikainen:2021ybu,Park:2024pkt,Nakayama:2025hqo}.
By replacing the background geometry with the AdS black hole in these models, one may calculate thermal one-point functions holographically and check if they take the general forms in section \ref{sec:thermal_DCFT}.

A $p$-dimensional defect wrapping the thermal circle that we considered in this paper reduces to a $(p-1)$-dimensional defect on $\BR^{d-1}$ at high temperature.
A part of the DCFT data may be captured by the thermal effective action in $(p-1)$ dimensions, which takes the form $I_{E,\, \CD}[\hat \gamma] = - \int \d^{p-1}x_\parallel\,\sqrt{\hat \gamma}\left[\,f_\CD\,\beta^{1-p} + c_\CD\,\beta^{3-p}\,\hat R + \cdots\right]$ where $\hat \gamma$ and $\hat R$ are the metric and the Ricci scalar on the defect.
The thermal one-point function of the stress tensor follows from the variation of the effective action.
Thus, the coefficient $f_\CD$ can be determined by the integrated one-point function.
Once $f_\CD$ is obtained through this relation, one may determine the DCFT data such as the density of states and the DOE coefficients as in the CFT case \cite{Benjamin:2023qsc,Allameh:2024qqp} (see \cite{Diatlyk:2024qpr,Kravchuk:2024qoh} for related works), and derive the $n\to 0$ limit of R\'enyi entropy for DCFTs from the effective action by following the argument of \cite{Kusuki:2025pgx}.
This is beyond the scope of this paper, and we hope to address it in the future.

A DCFT placed on thermal manifolds other than $\BS^1_\beta \times \BR^{d-1}$ is also worth studying.
The simplest example is $\BS^1_{\beta} \times \BS^{d-1}_{R}$ ($\BS^1_{\beta} \times \BH\BS^{d-1}_{R}$ for a BCFT). 
On this geometry, however, the thermal DCFT data are related to those at zero temperature through the state-operator correspondence in the large volume limit.
Nevertheless, the analysis of a DCFT on $\BS^1_{\beta} \times \BS^{d-1}_{R}$ may reveal new properties which are inaccessible on the flat space (e.g., the spectrum of higher-dimensional operators), as seen for a CFT without defects  \cite{Gobeil:2018fzy,Buric:2024kxo,David:2024pir,Buric:2025uqt,David:2025tqn,Buric:2026pes,David:2026yis}.

For line defects ($p=1$), the KMS relation, the bulk OPE and the DOE have been used to constrain the thermal one-point coefficients $\hat b_{\hO}$ by the bootstrap method \cite{Barrat:2024aoa}.
Also, away from the OPE regime, there has been recent progress of the thermal bootstrap in CFT at finite temperature \cite{Arnaudo:2026axe,Barrat:2026jfg}.
It would be interesting to extend these to the case with defects of $p\ge 2$ dimensions.

In this paper we have considered defects wrapping the thermal circle.
The complementary situation has been studied in the context of the measurement-induced physics, where local measurements are identified with the insertion of an interface or boundaries at a fixed time slice.
Such constructions have been explored at zero temperature in \cite{Garratt:2022ycp,Lee:2023fsk}, 
and used to study selective measurements of thermal states \cite{Najafi:2016kwb} and measurement-induced entanglement through the replicated path integrals \cite{Khanna:2025fac}.
These settings motivate us to extend our analysis to a DCFT with a defect at a fixed time slice and study the bulk one-point functions of operators outside the measured region, which can possibly reveal the effect of measurements on local observables.

\acknowledgments
We are grateful to C.\,P.\,Herzog and D.\,Ge for valuable discussions.
The work of T.\,N. was supported in part by the JSPS Grant-in-Aid for Scientific Research (B) No.\,24K00629, and
Grant-in-Aid for Transformative Research Areas (A) ``Extreme Universe''
No.\,21H05182 and No.\,21H05190.
T.\,N. would like to thank the Isaac Newton Institute for Mathematical Sciences, Cambridge, for support and hospitality during the programme ``Quantum Field Theory with Boundaries, Impurities, and Defects'' where work on this paper was undertaken.
This work was supported by EPSRC grant no EP/R014604/1. Y.L. gratefully acknowledges scholarship support from the Ito Foundation for International Education Exchange and support from the Multidisciplinary PhD Program for Pioneering Quantum Beam Application (PQBA) at The University of Osaka.
The work of H.N. was supported by JST SPRING, Grant Number JPMJSP2138.

\appendix

\section{Two-point functions in defect CFTs}\label{app:DCFT_zetoT_details}
In this appendix we list the two-point functions in a defect CFT at zero temperature that are relevant to the main text, and examine the relations between the bulk-defect two-point coefficients and those of the DOE.
More details can be found in e.g., \cite{Billo:2016cpy,Lauria:2018klo}.

\subsection{Tensor structures}

The inversion tensor $I^{\mu \nu}$ and its restriction $\hat{I}^{\ha\hb}(\hx)$ to the defect are defined by
\begin{align}\label{app:inversion tensor}
    I^{\mu\nu}(x)
        =
            \delta^{\mu\nu} - 2\,\frac{x^\mu\, x^\nu}{|x|^2}\ ,\qquad 
    \hat{I}^{\ha\hb}(\hx)
        =
            \delta^{\ha\hb} - 2\,\frac{\hx^{\ha}\,\hx^{\hb}}{|\hx|^2}\ .
\end{align}
We introduce the notation $\hx^\mu:= \delta^\mu_{\hat a}\,\hx^{\hat a}$, and denote the inversion tensor for $x - \hx'$ by
\begin{align}
    I^\mu_{~\nu}
        :=
            I^\mu_{~\nu}(x-\hx') 
        =
            \delta^\mu_\nu - 2\, \frac{(x-\hx')^\mu\,(x-\hx')^\nu}{|x-\hx'|^2} \ .
\end{align}

To simplify the notation, we contract the transverse indices of a defect operator by a polarization vector $w_i$ which is null, $w_i\,w^i = 0$ as
\begin{align}\label{app:polarization}
    \hO_{(s,j)}^{\hat a_1\cdots\hat a_j}(\hx;w)
        :=
            w_{i_1}\cdots w_{i_s}\,\hO_{(s,j)}^{i_1\cdots i_s, \hat a_1\cdots\hat a_j}(\hx)\ .
\end{align}

Besides $n^\mu$, $h^{\mu\nu}$, $h^{\mu\nu}_\perp$ and $\sfs^{\mu\nu}_{1,2}$ given in \eqref{tensor basis}, we introduce the following tensors for later convenience:
\begin{align}\label{app:tensor structures}
    \begin{aligned}
    N^{\mu\nu}
        &:=
            \frac{1}{q}\,\left(\frac{q-1}{d}\,\sfs^{\mu\nu}_1 + \sfs^{\mu\nu}_2\right)\ ,\\
    E^\mu
        &:=
            I^\mu_{~\nu}\,n^\nu
        =
            n^\mu - 2\,\frac{|x_\perp|}{|x-\hx'|^2}\,(x-\hx')^\mu\ , \\
    P^\mu
        &:=
            I^\mu_{~\nu}\,w^\nu - (n\cdot w)\,E^\mu
        =
            w^\mu - (n\cdot w)\,n^\mu \ ,
    \end{aligned}
\end{align}
where $w^\mu := \delta^\mu_i\,w^i$.
Note that $N^{\mu\nu}$ is the traceless tensor that appear in the bulk one-point function \eqref{bulk 1-pt stress} of the stress tensor, and $E^\mu$ is a unit vector, $E^\mu E_\mu=1$.
$P^\mu$ is orthogonal to $n^\mu$ and $E^\mu$ and satisfies $P\cdot P = - (n\cdot w)^2$.
Also $E^\mu$ and $I^\mu_{~\hat a}$ become
\begin{align}
    E^\mu \to n^\mu \ , \qquad 
    I^\mu_{~\hat a} \to \hat I^\mu_{~\hat a}(\hx - \hx') \ ,
\end{align}
in the $|x_\perp|\to 0$ limit.

We also denote symmetrization, anti-symmetrization and traceless symmetrization for a pair of indices $a, b$ by round, square and angle brackets.
For instance, given a tensor $M_{ab}$,
\begin{align}
    M_{(ab)}
        &:=
            \frac{1}{2}\,\left( M_{ab} + M_{ba}\right) \ , \\
    M_{[ab]}
        &:=
            \frac{1}{2}\,\left( M_{ab} - M_{ba}\right) \ , \\
    M_{\langle ab\rangle}
        &:=
            \frac{1}{2}\,\left( M_{ab} + M_{ba}\right)  - \frac{\delta_{ab}}{m}M^{c}_{~c}\ ,
\end{align}
where $m$ is the dimension of the metric $\delta_{ab}$. We also define
\begin{align}\label{app:A definition}
    A
        =
            (d-1)\,\hDO-d\,p \ ,
\end{align}
which frequently appears in the following.

\subsection{Defect-defect two-point functions}\label{app:defect-defect}
The defect-defect two-point functions are determined by the conformal symmetry $\SO(p+1, 1)\times \SO(q)$ on the defect. 
Thus, the two-point function of a pair of defect scalars is the same as that in a usual CFT,
\begin{align}
    \langle\,\hO(\hx)\,\hO'(\hx')\,\rangle
        =
            \delta_{\hO,\hO'}\,\frac{C_{\hO}}{|\hx - \hx'|^{2\hDO}}\ .
\end{align}
Likewise, the two-point function of two defect parallel-spin operators is 
\begin{align}
    \langle\,\hO^{\ha_1\cdots\ha_j}(\hx)\,\hO'_{\hb_1\cdots\hb_l}(\hx')\,\rangle
        =
            C_{\hO}\,\delta_{\hO,\hO'}\,\frac{\hat I^{(\ha_1}_{~~\hb_1}(\hx-\hx')\cdots \hat I^{\ha_j)}_{~~\hb_j}(\hx-\hx')-\text{traces}}{|\hx-\hx'|^{2\hDO}} \ .
\end{align}
The two-point function of two defect transverse-spin operators is \begin{align}
    \langle\,\hO^{i_1\cdots i_s}(\hx)\,\hO'_{j_1\cdots j_t}(\hx')\,\rangle
        =
            C_{\hO}\,\delta_{\hO,\hO'}\,\frac{\delta^{(i_1}_{(j_1}\cdots\delta^{i_s)}_{j_s)}-\text{traces}}{|\hx-\hx'|^{2\hDO}}.
\end{align}
For a pair of mixed-spin operators, the two-point function is 
\begin{align}
    \begin{aligned}
        &\langle\,\hO^{i_1\cdots i_s, \ha_1\cdots\ha_j}(\hx)\,\hO'_{j_1\cdots j_t,\hb_1\cdots\hb_l}(\hx')\,\rangle\\
            &\qquad
                =
                    C_{\hO}\,\delta_{\hO,\hO'}\frac{\left(\hat I^{(\ha_1}_{~~(\hb_1}(\hx-\hx')\cdots \hat I^{\ha_j)}_{~~\hb_j)}(\hx-\hx')-\text{traces}\right)\left(\delta^{(i_1}_{(j_1}\cdots\delta^{i_s)}_{j_s)}-\text{traces}\right)}{|\hx-\hx'|^{2\hDO}} \ .
    \end{aligned}
\end{align}

\subsection{Bulk-defect two-point functions}\label{app:bulk-defect}
We list the bulk-defect two-point functions of a bulk operator with spin $l\le 2$ and a defect operator with parallel spin $j\le 2$.
We use the symbol $(l,s,j)$ to denote the type of bulk and defect operators that we consider. 

We also introduce the structure for the two-point function:
\begin{align}
    X(x, \hx') 
        :=
        \frac{1}{|x_\perp|^{\Delta-\hDO}|x-\hx'|^{2\hDO}} \ .
\end{align}
Since the polarization vector $w$ multiplies with non-negative power in the two-point functions by definition, the terms with $(n\cdot w)^s$ should be removed when $s$ is negative in the following list.

\paragraph{Bulk scalar.}
\begin{itemize}
    \item $(0,s,0)$:
        \begin{align}\label{app:2pt (0,s,0)}
            \langle\,\CO(\hx,x_\perp)\,\hO_{(s,0)}(\hx';w)\,\rangle
                =
                    c_{\CO\hO}\,(n\cdot w)^s\,X(x,\hx') \ .
        \end{align}
\end{itemize}

\paragraph{Bulk spin-one.}
\begin{itemize}
    \item $(1,s,0)$:
        \begin{align}\label{app:2pt (1,s,0)}
            \langle\,\CO^\mu(\hx,x_\perp)\,\hO_{(s,0)}(\hx';w)\,\rangle
                =
                    \left[\,c^{(1)}_{\CO\hO}\,(n\cdot w)^{s-1}\,P^\mu+c^{(2)}_{\CO\hO}\,(n\cdot w)^{s}\,E^\mu\,\right]X(x,\hx')\ .
        \end{align}
        When $\CO^\mu$ is conserved, we have the constraint
        \begin{align}\label{app:2pt constraint (1,s,0)}
            (q+s-2)\,c^{(1)}_{\CO\hO}
                =
                    (\hDO-p)\,c^{(2)}_{\CO\hO}\ ,
        \end{align}
        which for $s=0$ reads $(\hDO-p)\,c_{\CO\hO}=0$.
    \item $(1,s,1)$:
        \begin{align}\label{app:2pt (1,s,1)}
            \langle\,\CO^\mu(\hx,x_\perp)\,\hO_{(s,1)}^{\ha}(\hx';w)\,\rangle
                =
                    c_{\CO\hO}\,(n\cdot w)^{s}\,I^{\mu\ha}\,X(x,\hx')\ .
        \end{align}
        There is no constraint from the conservation.
\end{itemize}

\paragraph{Bulk spin-two.}
\begin{itemize}
    \item $(2,s,0)$:
        \begin{align}\label{app:2pt (2,s,0)}
            \begin{aligned}
                &\langle\,\CO^{\mu\nu}(\hx,x_\perp)\,\hO_{(s,0)}(\hx';w)\,\rangle\\
                &\qquad
                    =
                    \left[\,c^{(1)}_{\CO\hO}\,(n\cdot w)^{s}\left(E^\mu E^\nu-\frac{\delta^{\mu\nu}}{d}\right)
                    +
                    c^{(2)}_{\CO\hO}\,(n\cdot w)^{s}\,N^{\mu\nu}
                    +
                    c^{(3)}_{\CO\hO}\,(n\cdot w)^{s-1}\,E^{(\mu}P^{\nu)}\right.\\
                &\qquad\qquad\left.
                    +
                    c^{(4)}_{\CO\hO}\,(n\cdot w)^{s-2}\left(P^\mu P^\nu+\frac{(n\cdot w)^2}{d}\,\delta^{\mu\nu}\right)\right]X(x,\hx')\ .
            \end{aligned}
        \end{align}
        When $\CO^{\mu\nu}$ is conserved, we have the constraints
        \begin{align}\label{app:2pt constraint (2,s,0)}
            \begin{aligned}
                A\,c^{(1)}_{\CO\hO}-(q-1)\,\hDO\,c^{(2)}_{\CO\hO}-\frac{d\,(q+s-2)}{2}\,c^{(3)}_{\CO\hO}+\hDO\,c^{(4)}_{\CO\hO}
                    &=0\ ,\\
                s\,c^{(1)}_{\CO\hO}-s\,(p+1)\,c^{(2)}_{\CO\hO}-\frac{d\,(\hDO-p)}{2}\,c^{(3)}_{\CO\hO}+\big(d\,(q+s-2)-s\big)\,c^{(4)}_{\CO\hO}
                    &=0\ .
            \end{aligned}
        \end{align}
        For $s=0$ where $c^{(3)}_{\CO\hO}$ and $c^{(4)}_{\CO\hO}$ are absent,  the first constraint reduces to $A\,c^{(1)}_{\CO\hO}=(q-1)\,\hDO\,c^{(2)}_{\CO\hO}$ while the second becomes trivial.
        For $s=1$ where $c^{(4)}_{\CO\hO}$ is absent, the two relations are independent unless $\hDO=p+1$, where they become proportional to each other.
        For $s\ge 2$, they are independent constraints for every $\hDO$.
    \item $(2,s,1)$:
        \begin{align}\label{app:2pt (2,s,1)}
            \langle\,\CO^{\mu\nu}(\hx,x_\perp)\,\hO_{(s,1)}^{\ha}(\hx';w)\,\rangle
                =
                    \left[\,c^{(1)}_{\CO\hO}\,(n\cdot w)^{s}\,I^{\ha(\mu}E^{\nu)}+c^{(2)}_{\CO\hO}\,(n\cdot w)^{s-1}\,I^{\ha(\mu}P^{\nu)}\,\right]X(x,\hx')\ .
        \end{align}
        When $\CO^{\mu\nu}$ is conserved, we have the constraint
        \begin{align}\label{app:2pt constraint (2,s,1)}
            (\hDO-p-1)\,c^{(1)}_{\CO\hO}
                =
                    (q+s-2)\,c^{(2)}_{\CO\hO}\ ,
        \end{align}
        which for $s=0$ reads $(\hDO-p-1)\,c_{\CO\hO}=0$.
    \item $(2,s,2)$:
        \begin{align}\label{app:2pt (2,s,2)}
            \langle\,\CO^{\mu\nu}(\hx,x_\perp)\,\hO_{(s,2)}^{\ha\hb}(\hx';w)\,\rangle
                =
                    c_{\CO\hO}\,(n\cdot w)^{s}\left[\,I^{\ha(\mu}I^{\nu)\hb}-\frac{\delta^{\ha\hb}}{p}\,I^{(\mu}{}_{\hat{c}}I^{\nu)\hat{c}}\,\right]X(x,\hx')\ .
        \end{align}
        There is no constraint from the conservation.
\end{itemize}

\subsection{Bulk-defect operator expansion}\label{app:DOE}
The DOE coefficients are determined by taking both sides of the DOE with a defect operator with the two-point functions in appendix \ref{app:bulk-defect} and comparing their structures.

For a bulk scalar, it follows from \eqref{app:2pt (0,s,0)} that only the defect operators $\hO_{(s,0)}$ with transverse spin can contribute to the DOE, hence the DOE is given as in \eqref{scalar DOE}.

\paragraph{Bulk spin-one.}
A bulk spin-one operator $\CO^\mu$ has non-vanishing two-point functions with defect primaries with parallel spin $j\le 1$ as in \eqref{app:2pt (1,s,0)} and \eqref{app:2pt (1,s,1)}.
The DOE can be read off as
\begin{align}\label{app:vector OE}
    \begin{aligned}
        \CO^\mu(\hx,x_\perp)
            =
            \frac{1}{|x_\perp|^{\Delta_\CO}}\Big[&\,n^\mu\sum_{\hO}b_{\CO\hO}\,|x_\perp|^{\hDO}\,\hO(\hx)+\sum_{\hO_{(0,1)}}b_{\CO\hO}\,|x_\perp|^{\hDO}\,\delta^\mu_{\ha}\,\hO_{(0,1)}^{\ha}(\hx)\\
            &+
            \sum_{\hO_{(s\ge 1,0)}}|x_\perp|^{\hDO}\left(b^{(1)}_{\CO\hO}\,\delta^\mu_{i_1}\,n_{i_2}\cdots n_{i_s}
                +
                b^{(2)}_{\CO\hO}\,n^\mu\, n_{i_1}\cdots n_{i_s}\right)\,\hO_{(s,0)}^{i_1\cdots i_s}(\hx)\\
            &+
            \sum_{\hO_{(s\ge 1,1)}}b_{\CO\hO}\,|x_\perp|^{\hDO}\,\delta^\mu_{\ha}\,n_{i_1}\cdots n_{i_s}\,\hO_{(s,1)}^{i_1\cdots i_s,\ha}(\hx)\Big] 
            +
            \text{(descendants)} \ ,
    \end{aligned}
\end{align}
where the defect identity does not contribute as the one-point function of $\CO^\mu$ vanishes (see section \ref{ss:DCFT_correlators_zeroT}).
The DOE coefficients $ b_{\CO\hO}$ are related to the two-point coefficients $c_{\CO\hO}$ in \eqref{app:2pt (1,s,0)} as follows:
\begin{align}\label{app:b from c spin-1}
    \begin{aligned}
        (0,0),\ (0,1),\ (s,1):&\quad b_{\CO\hO}\,C_{\hO} = c_{\CO\hO}\ ,\\
        (s\ge 1,0):&\quad b_{\CO\hO}^{(1)}\,C_{\hO} = c_{\CO\hO}^{(1)}\ ,\qquad b_{\CO\hO}^{(2)}\,C_{\hO} = c_{\CO\hO}^{(2)} - c_{\CO\hO}^{(1)}\ .
    \end{aligned}
\end{align}

\paragraph{Bulk spin-two.}
For bulk spin-two operator, the DOE reads from \eqref{app:2pt (2,s,0)}, \eqref{app:2pt (2,s,1)} and \eqref{app:2pt (2,s,2)} as
\begin{align}\label{app:tensor OE}
    \begin{aligned}
        \CO^{\mu\nu}(\hx,x_\perp)
            =
                \frac{a_\CO}{q\,|x_\perp|^{\Delta_\CO}}\left(\frac{q-1}{d}\sfs^{\mu\nu}_1+\sfs^{\mu\nu}_2\right)\,\hat{\bm{1}}
                +
                \sum_{\hO_{(s,j\le 2)}}|x_\perp|^{\hDO-\Delta_\CO}\,\Big[\sum_{I}b^{(I)}_{\CO\hO}\,\mathcal{S}^{\mu\nu}_I\Big]\,\hO_{(s,j)}(\hx)\\
                +
                \text{(descendants)} \ ,
    \end{aligned}
\end{align}
where the first term is the contribution from the defect identity and the tensor structures in the second term take the following forms:
\begin{align}\label{app:tensor OE structures}
    \begin{aligned}
        (0,0):&\quad b^{(1)}_{\CO\hO}\,\sfs_1^{\mu\nu} + b^{(2)}_{\CO\hO}\,\sfs_2^{\mu\nu}\ ,\\
        (s\ge 1,0):&\quad \left(\,b^{(1)}_{\CO\hO}\,\sfs_1^{\mu\nu} + b^{(2)}_{\CO\hO}\,\sfs_2^{\mu\nu}\,\right)\,n_{i_1}\cdots n_{i_s}
        +
        b^{(3)}_{\CO\hO}\,\left(\,n^{(\mu}\,\delta^{\nu)}_{i_1}- \frac{h^{\mu\nu}_\perp}{q}\, n_{i_1}\right)\,n_{i_2}\cdots n_{i_s}\\
        &\qquad
        +
        b^{(4)}_{\CO\hO}\,\delta^\mu_{i_1}\,\delta^\nu_{i_2}\,n_{i_3}\cdots n_{i_s}\ ,\\
        (0,1):&\quad b_{\CO\hO}\,n^{(\mu}\delta^{\nu)}_{\ha}\ ,\\
        (s \ge 1,1):&\quad b^{(1)}_{\CO\hO}\,\delta^{(\mu}_{\ha}\,n^{\nu)}\,n_{i_1}\cdots n_{i_s} + b^{(2)}_{\CO\hO}\,\delta^{(\mu}_{\ha}\,\delta^{\nu)}_{i_1}\,n_{i_2}\cdots n_{i_s}\ ,\\
        (s,2):&\quad b_{\CO\hO}\,\delta^\mu_{\ha}\,\delta^\nu_{\hb}\,n_{i_1}\cdots n_{i_s}\ ,
    \end{aligned}
\end{align}
where $b^{(4)}_{\CO\hO}\equiv0$ for $s=1$.
The coefficients $b^{(I)}_{\CO\hat{\CO}}$ are related to the two-point coefficients as follows:
\begin{align}\label{app:b from c spin-2}
    \begin{aligned}
        (s\ge 0,0):&\quad b^{(1)}_{\CO\hO}\,C_{\hO}=\frac{c^{(1)}_{\CO\hO}+(q-1)\,c^{(2)}_{\CO\hO}-c^{(4)}_{\CO\hO}}{q\,d}\ ,\qquad b^{(2)}_{\CO\hO}\,C_{\hO}=\frac{c^{(2)}_{\CO\hO}+c^{(3)}_{\CO\hO}-c^{(1)}_{\CO\hO}-c^{(4)}_{\CO\hO}}{q}\ ,\\
        &\quad b^{(3)}_{\CO\hO}\,C_{\hO}=c^{(3)}_{\CO\hO}-2\,c^{(4)}_{\CO\hO}\ ,\qquad b^{(4)}_{\CO\hO}\,C_{\hO}=c^{(4)}_{\CO\hO}\ ,\\
        (0,1),\ (s,2):&\quad b_{\CO\hO}\,C_{\hO}=c_{\CO\hO}\ ,\\
        (s\ge 1,1):&\quad b^{(1)}_{\CO\hO}\,C_{\hO}=c^{(1)}_{\CO\hO}-c^{(2)}_{\CO\hO}\ ,\qquad b^{(2)}_{\CO\hO}\,C_{\hO}=c^{(2)}_{\CO\hO}\ .
    \end{aligned}
\end{align}

\subsubsection{Constraints from conservation law}

\paragraph{Conserved current.}
When $\CO^\mu=J^\mu$ is a conserved current with $\Delta_J=d-1$, the conservation law imposes the constraints on the DOE coefficients:
\begin{align}\label{app:current OE constraints}
    \begin{aligned}
    (0,0):&\quad (\hDO-p)\,b_{J\hO}=0\ ,\\
    (s\ge 1,0):&\quad (q+s-2)\,b^{(1)}_{J\hO}=(\hDO-p)\big(b^{(1)}_{J\hO}+b^{(2)}_{J\hO}\big)\ .
    \end{aligned}
\end{align}
The first one forces the defect scalars that appear in the DOE \eqref{app:vector OE} to have $\hD_{\hO} = p$.

\paragraph{Stress tensor.}
When $\CO^{\mu\nu}=T^{\mu\nu}$ is the stress tensor with $\Delta_T=d$, the conservation law imposes the constraints on the DOE coefficients:
\begin{align}\label{stress DOE conservative coefficient}
    \begin{aligned}
            (0,0)&:\quad p\,(\hDO-d)\,b^{(1)}_{T\hO}=(q-1)(\hDO-p)\,b^{(2)}_{T\hO} \ ,\\
            (s \ge 1,0):&\quad 2\,p\,q\,(d-\hDO)\,b^{(1)}_{T\hO}+2\,q\,(q-1)(\hDO-p)\,b^{(2)}_{T\hO}\\
            &\qquad+\left[q\,(q+s-2)-2\,(q-1)(\hDO-p)\right]b^{(3)}_{T\hO} + 2\,q\,(d+s-2-\hDO)\,b^{(4)}_{T\hO} = 0\ ,\\
            &\quad 2\,q\,s\left(p\,b^{(1)}_{T\hO} + b^{(2)}_{T\hO}\right)+\left[q\,(\hDO-p)-2\,s\right]b^{(3)}_{T\hO} + 2\,q\,(\hDO-d-s+2)\,b^{(4)}_{T\hO}=0\ ,
    \end{aligned}
\end{align}
where $b^{(4)}_{T\hO}= 0$ for $s=1$, and 
\begin{align}
    \begin{aligned}
            (0,1)&:\quad (\hDO-p-1)\,b_{T\hO}=0\ , \\
            (s \ge 1,1)&:\quad (\hDO-p-1)\left(b^{(1)}_{T\hO}+b^{(2)}_{T\hO}\right)=(q+s-2)\,b^{(2)}_{T\hO}\ .
    \end{aligned}
\end{align}

\section{Conformal Ward identities at finite temperature}\label{app:Ward_identity}
In this appendix, we derive the modified version of the conformal Ward identities at finite temperature using the topological operator method. 
The effect of the defect under conformal transformation can be summarized by defining the total stress tensor $T^{\mu\nu}_\text{tot}(x) = T^{\mu\nu}(x) + \delta^{(q)}(x_\perp)\,\delta^\mu_{\hat a}\,\delta^\nu_{\hat b}\,\hat B^{\hat a \hat b}(\hx)$, where $\hat B^{\hat a \hat b}$ denotes the variation of the action with respect to the induced metric on the defect \cite{Billo:2016cpy}.
This satisfies the operator equations 
\begin{align}
    \partial_\mu T^{\mu a}_\text{tot} = 0\ ,
    \qquad 
    \partial_\mu T^{\mu i}_\text{tot} = \delta^{(q)}(x_\perp)\,\hat D^i\ ,
    \qquad
    (T_\text{tot})^\mu_{\,\,\mu}=0\ ,
\end{align}
which follow the diffeomorphism and Weyl invariance imposed on correlation functions.
Without the defect, $T^{\mu\nu}_\text{tot}$ reduces to $T^{\mu\nu}$ that satisfies $\partial_\mu T^{\mu\nu}=0$ and $T^\mu_{~\mu}=0$.
For a conformal Killing vector $\epsilon_\mu$ satisfying $\partial_{(\mu}\epsilon_{\nu)} = (\partial\cdot\epsilon)\,\delta_{\mu\nu}/d$, one finds
\begin{align}\label{app:current conservation}
    \partial_\mu\left(\epsilon_\nu\, T^{\mu\nu}_\text{tot}\right) 
        = 
        \delta^{(q)}(x_\perp)\,\epsilon_i\,\hat D^i \ ,
\end{align}
as long as it is away from other local operators.
We thus define the generator $Q_\epsilon(\Sigma)$ for the conformal transformation associated with $\epsilon^\mu$ by
\begin{align}
    Q_\epsilon(\Sigma)
        =
            -\int_\Sigma \d S_\mu\epsilon_\nu T^{\mu\nu}_\text{tot}(x)\ , 
    \qquad 
    [\,Q_\epsilon,\CO(x)\,] 
        =
            \delta_\epsilon\CO(x)\ ,
\end{align}
where $\Sigma$ is a codimension-one hypersurface, and the area element $\d S_\mu$ is oriented outward from the region surrounded by $\Sigma$.
Note that $Q_\epsilon(\Sigma)$ is topological and the hypersurface $\Sigma$ can be deformed freely as long as it does not cross a local operator or the defect.

\begin{figure}[h]
    \begin{center}
    \begin{tikzpicture}[
        point/.style={fill=black, circle, inner sep=1pt},
        defect/.style={orange, thick},
        main line/.style={thick, black},
        surface/.style={teal, densely dotted, thick},
        neighborhood/.style={teal, densely dotted, thick, radius=0.17},
        label text/.style={font=\small},
        scale=0.9
    ]
        \begin{scope}
            \draw[main line] (0,0) rectangle (7,3);
            \node[right, label text] at (7.1,3) {$\tau=\beta$};
            \node[right, label text] at (7.1,0) {$\tau=0$};
            \draw[defect] (3.5,0) -- (3.5,3);
            \node[above, orange] at (3.5,3) {$\Dp$};
            \coordinate (a1) at (1.2,1.2);
            \coordinate (a2) at (2.4,2.4);
            \coordinate (a3) at (5.8,2.5);
            \coordinate (a4) at (4.8,0.7);
            \node[point, label={[label text]right:$x_1$}] at (a1) {};
            \node[point, label={[label text]right:$x_2$}] at (a2) {};
            \node[point, label={[label text]right:$x_3$}] at (a3) {};
            \node[point, label={[label text]right:$x_4$}] at (a4) {};
            \draw[surface] (6.0,1.55) circle (0.35);
            \node[teal, label text, right] at (6.35,1.55) {$\Sigma$};
            \node[label text] at (3.5,-0.7) {(a)};
        \end{scope}

        \node[font=\Large] at (8.4,1.5) {$\Longrightarrow$};

        \begin{scope}[xshift=9.8cm]
            \draw[main line] (0,0) rectangle (7,3);
            \node[right, label text] at (7.1,3) {$\tau=\beta$};
            \node[right, label text] at (7.1,0) {$\tau=0$};
            \draw[defect] (3.5,0) -- (3.5,3);
            \node[above, orange] at (3.5,3) {$\Dp$};
            \coordinate (b1) at (1.2,1.2);
            \coordinate (b2) at (2.4,2.4);
            \coordinate (b3) at (5.8,2.5);
            \coordinate (b4) at (4.8,0.7);
            \draw[surface] (0.12,0.12) rectangle (3.2,2.88);
            \draw[surface] (3.8,0.12) rectangle (6.88,2.88);
            \draw[neighborhood] (b1) circle;
            \draw[neighborhood] (b2) circle;
            \draw[neighborhood] (b3) circle;
            \draw[neighborhood] (b4) circle;
            \node[point, label={[label text]right:$x_1$}] at (b1) {};
            \node[point, label={[label text]right:$x_2$}] at (b2) {};
            \node[point, label={[label text]right:$x_3$}] at (b3) {};
            \node[point, label={[label text]right:$x_4$}] at (b4) {};
            \node[teal, label text] at (1.6,-0.35) {$\Sigma$};
            \node[label text] at (3.5,-0.7) {(b)};
        \end{scope}
    \end{tikzpicture}
    \caption{Derivation of the modified conformal Ward identity on $\BS^1_\beta\times\BR^{d-1}$.
    The horizontal direction is a spatial direction and the vertical direction is $\tau$ with the periodicity $\beta$.
    The orange line is the $p$-dimensional defect wrapping the thermal circle.
    (a) The topological operator $Q_\epsilon$ is inserted on a small contractible surface $\Sigma$ (dotted), which gives zero. 
    (b) $\Sigma$ is deformed so that it wraps on the local operators, the defect, and the two hyperplanes at $\tau=0$ and $\tau=\beta$.}
    \label{fig:Ward identity}
    \end{center}
\end{figure}
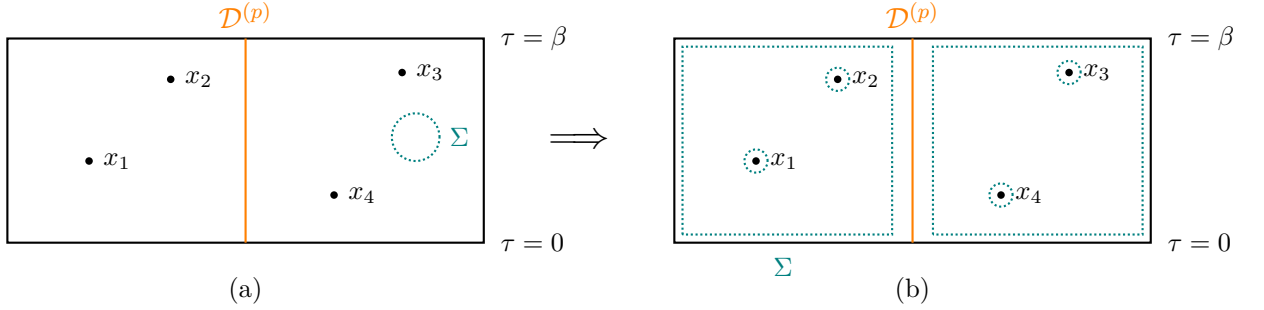

To derive the Ward identities at finite temperature, we place the defect CFT on $\BS^1_\beta\times\BR^{d-1}$ with periodicity $\tau \sim \tau + \beta$, and consider an $n$-point function $\langle\,\CO_1(x_1)\cdots \CO_n(x_n)\,\rangle_{\CD,\,\beta}$.
We insert the topological operator $Q_\epsilon$ on a small surface $\Sigma$ which is away from the defect and does not enclose any $x_m$ as in figure \ref{fig:Ward identity}(a).
Since $Q_\epsilon$ is contractible to a point, the insertion makes the total correlation function vanish.
We then deform $\Sigma$ so that it wraps on the operators at $x_m$, the defect and the two hyperplanes at $\tau=0$ and $\tau=\beta$ as in figure \ref{fig:Ward identity}(b).
By integrating $\partial_\mu(\epsilon_\nu T^{\mu\nu}_\text{tot}(x))$ in the region bounded by the dotted line in figure \ref{fig:Ward identity}(b) and using the Stokes theorem, we obtain
\begin{align}\label{modified_Ward_identity}
    \begin{aligned}
        &\sum_m\,\langle\,\CO_1(x_1)\cdots\delta_\epsilon\CO_m(x_m)\cdots\CO_n(x_n)\,\rangle_{\CD,\,\beta} \\
            &\qquad 
                =
                    \int_\CD \d^p x\,\epsilon^i(\hx)\,\langle\, \hat D_i(x)\,\CO_1(x_1)\cdots\CO_n(x_n)\,\rangle_{\CD,\,\beta,\,\text{con}}\\
            &\qquad\quad
                -
                    \int\d^{d-1}\vec{x}\,\left[\, \epsilon_\nu(\beta,\vec{x})-\epsilon_\nu(0,\vec{x})\,\right] \langle\, T^{0\nu}_\text{tot}(0,\vec{x})\,\CO_1(x_1)\cdots\CO_n(x_n)\,\rangle_{\CD,\,\beta,\,\text{con}} \ .
    \end{aligned}
\end{align}
The displacement term arises from the defect, while the second term is the boundary contribution at $\tau =0$ and $\tau=\beta$, where we used the periodicity of local operators.
The boundary term vanishes when $\epsilon^\mu$ is periodic, but it is not so for the dilatation $\epsilon^\mu = x^\mu$.
Also, the subscript in the boundary term means the connected correlation function defined by
\begin{align}
    \langle\, T^{\mu\nu}_\text{tot}\,\CO_1\cdots\CO_n\,\rangle_{\CD,\,\beta,\,\text{con}}
        :=
            \langle\, T^{\mu\nu}_\text{tot}\,\CO_1\cdots\CO_n\,\rangle_{\CD,\,\beta}
            -
            \langle\, T^{\mu\nu}_\text{tot}\,\rangle_{\CD,\,\beta}\,\langle\,\CO_1\cdots\CO_n\,\rangle_{\CD,\,\beta} \ .
\end{align}
Note that the identity \eqref{modified_Ward_identity} holds whether the defect wraps around the thermal circle or not.
By letting $\epsilon^\mu = x^\mu$ and noting that the displacement term vanishes as $x_\perp^i=0$ at the defect, we obtain
\begin{align}
    \left[\,\sum_m\left(\Delta_m+x^\mu_m\frac{\partial}{\partial x^\mu_m}\right)
        +
            \beta\,\frac{\partial}{\partial\beta}\,\right]\,\langle\,\CO_1(x_1)\cdots\CO_n(x_n)\,\rangle_{\CD,\,\beta} = 0 \ ,
\end{align}
where we have converted the boundary term to the derivative with respect to $\beta$ by noting that a change of the temperature is equivalent to the variation of the metric, $\delta g_{00} = 2\,\delta\beta /\beta$. 
This is the modified conformal Ward identity, originally derived in \cite{Barrat:2024aoa} based on the Lagrangian description of the defect for $p=1$.
The derivation here does not assume a Lagrangian description and works for any defect with $p\ge 1$.

\section{Useful formulas}\label{sec:app_useful}

\paragraph{Schwinger parametrization.}

\begin{align}
    \label{eq:app_schwinger}
    \frac{1}{A^{\nu}}
        =
            \frac{1}{\Gamma(\nu)}\int_0^\infty\d s\,s^{\nu-1}\,e^{-s\,A} \ .
\end{align}

\paragraph{Integral representation of the modified Bessel function of second kind.}
\begin{align}\label{eq:Bessel_K_int_rep}
    K_\nu(a\,x)
        =
        2^{-\nu-1}\,\left( \frac{x}{a}\right)^\nu\,\int_0^\infty \d t\,t^{-\nu-1}\,e^{- a^2\,t - \frac{x^2}{4\,t}} \ .
\end{align}

\paragraph{Poisson resummation formula.}
\begin{align}\label{eq:Poisson_resummation}
    \sum_{n\in \BZ}\,e^{-a\,n^2}
        =
        \sqrt{\frac{\pi}{a}}\,\sum_{w\in \BZ}e^{-\frac{\pi^2 w^2}{a}} \ .
\end{align}

\section{Properties of $\CI_\nu[a,b](z)$}\label{sec:app_Inu}
In this appendix, we record some properties and the asymptotic expansions of the following function:
\begin{align}\label{eq:app_J}
    \CI_{\nu}\!\left[a, b\right](z)
        :=
            \sum_{w\,\in\,\BZ}\int_0^1\d\xi\,
            \frac{\xi^{\,a}\,(1-\xi)^{\,b}}{\left[\,z^2 + \xi(1-\xi)\,w^2\,\right]^{\nu}}\ .
\end{align}

It is straightforward to see that the following identities hold:
\begin{align}
    \CI_{\nu}[a,b] 
        &= 
            \CI_{\nu}[b,a]\ ,
    \label{eq:app_I_sym}
    \\
    \CI_{\nu}[a,b] 
        &= 
            \CI_{\nu}[a+1,b] + \CI_{\nu}[a,b+1]\ ,
    \label{eq:app_I_recursion}
    \\
    \partial_z\CI_{\nu}[a,b] 
        &= 
            -2\,\nu\,z\,\CI_{\nu+1}[a,b]\ .
    \label{eq:app_I_deriv}
\end{align}

\subsection{Asymptotic expansion in $z\to \infty$}
To read off the asymptotic expansion of \eqref{eq:app_J} in the $z\to \infty$ limit, we use the identity
\begin{align}\label{eq:app_poisson_sum}
    \sum_{w\,\in\,\BZ}\frac{1}{\left(z^2 + c^2\,w^2\right)^{\nu}}
        =
            \frac{\sqrt\pi}{c\,\Gamma(\nu)}\left[\,
                \Gamma\!\left(\nu-\tfrac12\right)z^{\,1-2\,\nu}
                +
                4\,\sum_{n\ge1}\left(\frac{\pi\, n}{c\,z}\right)^{\nu-\frac12}
                K_{\nu-\frac12}\!\left(\frac{2\pi\, n\,z}{c}\right)
            \right] ,
\end{align}
which can be obtained by using the Schwinger parametrization \eqref{eq:app_schwinger}, Poisson resummation formula \eqref{eq:Poisson_resummation} and the integral representation of $K_\nu$ in \eqref{eq:Bessel_K_int_rep}.
By applying it to \eqref{eq:app_J} and performing the $\xi$-integral, one finds the asymptotic expansion in $z\to \infty$:
\begin{align}\label{eq:app_Inu_highT_general}
    \CI_{\nu}[a,b](z)
        =
            \frac{\sqrt\pi\,\Gamma\left(\nu-\frac12\right)\,\Gamma(a+\frac12)\,\Gamma(b+\frac12)}{\Gamma(\nu)\,\Gamma(a+b+1)}\,z^{\,1-2\nu}
            +
            O(e^{-4 \pi z}) \ .
\end{align}

\subsection{Asymptotic expansion in $z\to 0$}
\begin{figure}[h]
    \centering
    \begin{tikzpicture}[>=stealth
    ]
        \fill[gray!12] (-1.4,-1.6) rectangle (-0.8,1.6);
        \draw[->] (-4.4,0) -- (6.4,0) node[below right=-2pt] {$\mathrm{Re}\,\sigma$};
        \draw[->] (0,-1.9) -- (0,1.9); \node[anchor=west] at (0.15,1.75) {$\mathrm{Im}\,\sigma$};
        \draw[thick,->] (-1.1,-1.5) -- (-1.1,0.35);
        \draw[thick] (-1.1,0.3) -- (-1.1,1.5);
        \draw[thick, dashed] (-1.1,1.5) -- (4.6,1.5) arc[start angle=90, end angle=0, x radius=1.4, y radius=1.5] -- (6.0,0.0);
        \draw[thick, dashed] (6.0,0.0) arc[start angle=0, end angle=-90, x radius=1.4, y radius=1.5] -- (-1.1,-1.5);
        \node[below] at (-1.1,-1.95) {$\mathrm{Re}\,\sigma = c_0$};
        \foreach \x in {-1.4,-3.5}{\node[RoyalBlue] at (\x,0) {$\bullet$};}
        \node[RoyalBlue, below=2pt] at (-1.45,-0.1) {$-\nu$};
        \node[RoyalBlue, below=2pt] at (-3.4,-0.1) {$-\nu-1$};
        \node[RoyalBlue] at (-2.4,0.1) {$\cdots$};
        \foreach \x in {0,2,4}{\node at (\x,0) {$\bullet$};}
        \node[below=2pt] at (0.15,-0.1) {$0$};
        \node[below=2pt] at (2,-0.1) {$1$};
        \node[below=2pt] at (4,-0.1) {$2$};
        \node at (5.4,0.1) {$\cdots$};
        \foreach \x in {-0.8,1.2,3.2}{\node[BrickRed] at (\x,0) {$\bullet$};}
        \node[BrickRed, above=3pt] at (-0.8,0.1) {$A$};
        \node[BrickRed, above=3pt] at (1.2,0.1) {$A+1$};
        \node[BrickRed, above=3pt] at (3.2,0.1) {$A+2$};
        \foreach \x in {0.6,2.6,4.6}{\node[OliveGreen] at (\x,0) {$\bullet$};}
        \node[OliveGreen, below=2pt] at (0.6,-0.55) {$B$};
        \node[OliveGreen, below=2pt] at (2.6,-0.55) {$B+1$};
        \node[OliveGreen, below=2pt] at (4.6,-0.55) {$B+2$};
        \begin{scope}[yshift=0.4cm]
            \node[anchor=west, RoyalBlue] at (-4.3,3.5) {$\bullet\ \Gamma(\nu+\sigma)$};
            \node[anchor=west] at (-4.3,3.05) {$\bullet\ \Gamma(-\sigma)\ \to\ \CH_k \sim z^{2k}$};
            \node[anchor=west, BrickRed] at (-4.3,2.6) {$\bullet\ \Gamma(A-\sigma)\ \to\ \mathcal{K}_k(A,B) \sim z^{2A+2k}$};
            \node[anchor=west, OliveGreen] at (-4.3,2.15) {$\bullet\ \Gamma(B-\sigma)\ \to\ \mathcal{K}_k(B,A) \sim z^{2B+2k}$};
        \end{scope}        
    \end{tikzpicture}
    \caption{The contour used in the integral \eqref{eq:app_Inu_lowT}. }
    \label{fig:app_Inu_contour}
\end{figure}
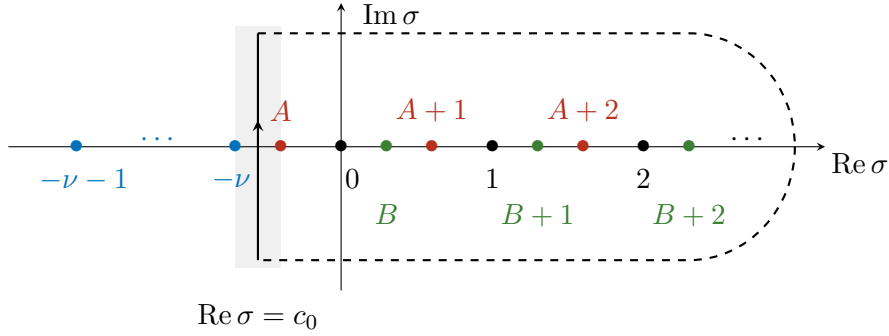

The $w=0$ term in \eqref{eq:app_J} becomes
\begin{align}\label{eq:app_I_zero}
    \CI_{\nu}\!\left[a, b\right]\big|_{w=0}
        =
            \frac{\Gamma(a+1)\,\Gamma(b+1)}{\Gamma(a+b+1)}\,\frac{1}{z^{2\,\nu}} \ .
\end{align}
For the term with $w\neq 0$, we apply the Mellin-Barnes representation and perform the integral over $\xi$ to obtain
\begin{align}\label{eq:app_Inu_lowT}
    \CI_{\nu}\!\left[a, b\right]\big|_{w\neq 0}
        =
            \frac{1}{w^{2\,\nu}\,\Gamma(\nu)}\,\frac{1}{2\,\pi\,\i}\,\int_{c_0-\i\,\infty}^{c_0+\i\,\infty}\d\sigma\,\frac{\Gamma(-\sigma)\,\Gamma(\nu +\sigma)\,\Gamma(A-\sigma)\,\Gamma(B-\sigma)}{\Gamma(A+B-2\,\sigma)}\,\left(\frac{z^2}{w^2}\right)^{\sigma} \ ,
\end{align}
where $A,B$ are defined by
\begin{align}
    A := a+1-\nu\ ,
    \qquad
    B := b+1-\nu\ .
\end{align}
The contour is chosen as $-\nu < c_0 < \text{min}(0, A, B)$ so that the $\sigma$-integral is convergent.
By closing the contour to the right as in figure \ref{fig:app_Inu_contour}, picking up the poles of the gamma functions and summing over the winding modes with the $w=0$ term \eqref{eq:app_I_zero}, 
\begin{align}\label{eq:app_Inu_lowT_general}
    \CI_{\nu}[a,b](z)
        =
            \frac{\Gamma(a+1)\,\Gamma(b+1)}{\Gamma(a+b+1)}\,\frac{1}{z^{2\,\nu}} 
            +
            \frac{2}{\Gamma(\nu)}\sum_{k\ge0}\frac{(-1)^k}{k!}
            \Big[\,\CH_k(A,B) + \mathcal{K}_k(A,B) + \mathcal{K}_k(B,A)\,\Big]\ ,
\end{align}
where
\begin{align}
    \begin{aligned}
        \CH_k(A,B)
            &:=
                \Gamma(\nu+k)\,\frac{\Gamma(A-k)\,\Gamma(B-k)}{\Gamma(A+B-2\,k)}\,
                \zeta(2\,\nu+2\,k)\,z^{2\,k}\ ,
            \\
    \CK_k(A,B)
            &:=
                \Gamma(-A-k)\,\Gamma(\nu+A+k)\,\frac{\Gamma(B-A-k)}{\Gamma(B-A-2k)}\,
                \zeta\!\left(2\,\nu+2\,A+2\,k\right)\,z^{2\,A+2\,k}\ .
    \end{aligned}
\end{align}

\bibliographystyle{JHEP}
\bibliography{DCFT}

\end{document}